\pdfoutput=1
\documentclass[11pt]{article}
\usepackage[final]{acl}
\usepackage{times}
\usepackage{latexsym}
\usepackage[T1]{fontenc}
\usepackage[utf8]{inputenc}
\usepackage[final,nopatch=footnote]{microtype}
\usepackage{inconsolata}
\usepackage{graphicx}
\usepackage[ruled,vlined,lined,commentsnumbered]{algorithm2e}
\usepackage{amsmath,amsfonts}
\usepackage{amssymb}
\usepackage{array}
\usepackage{booktabs}
\usepackage[skip=1pt,labelfont=bf]{caption}
\usepackage{calligra}
\usepackage{color, colortbl}
\usepackage{courier}
\usepackage{csvsimple}
\usepackage{enumitem}
\usepackage{fancybox}
\usepackage{listings}
\usepackage{longtable}
\usepackage{lscape}
\usepackage{makecell}
\usepackage{marvosym}
\usepackage{moreverb}
\usepackage{multicol}
\usepackage{multirow}
\usepackage{pifont}
\usepackage{rotating}
\usepackage{setspace}
\usepackage{caption}
\usepackage{subfigure}
\usepackage{parcolumns}
\usepackage[most]{tcolorbox}
\usepackage{threeparttable}
\usepackage{tikz}
\usepackage[normalem]{ulem}
\usepackage{url}
\usepackage{soul}
\usepackage{wasysym}
\usepackage{svg}
\usepackage{textcomp}
\usepackage{xcolor}
\usepackage{wrapfig}
\usepackage{pdflscape}
\usepackage{hyperref}

\usepackage{balance}
\colorlet{punct}{red!60!black}
\definecolor{background}{HTML}{EEEEEE}
\definecolor{delim}{RGB}{20,105,176}
\definecolor{yellow}{RGB}{254, 254, 0}
\definecolor{lightblue}{RGB}{0, 254, 254}
\definecolor{mygray}{gray}{0.98}
\colorlet{numb}{magenta!60!black}

\lstdefinestyle{sol}{
    language=Python,
    basicstyle=\ttfamily\small,
    keywordstyle=\color{blue!70},
    commentstyle=\color{green!50!black},
    stringstyle=\color{red!70},
    numbers=left,
    numberstyle=\tiny\color{gray},
    stepnumber=1,
    showspaces=false,
    showstringspaces=false,
    breaklines=true,
    frame=single,
    tabsize=4,
    morekeywords={function, address, external, returns, internal, assertEq},
    escapeinside=`` 
}

\lstdefinelanguage{json}{
    basicstyle=\normalfont\ttfamily,
    numbers=left,
    numberstyle=\scriptsize,
    stepnumber=1,
    numbersep=8pt,
    showstringspaces=false,
    breaklines=true,
    frame=lines,
    backgroundcolor=\color{background},
    literate=
     *{0}{{{\color{numb}0}}}{1}
      {1}{{{\color{numb}1}}}{1}
      {2}{{{\color{numb}2}}}{1}
      {3}{{{\color{numb}3}}}{1}
      {4}{{{\color{numb}4}}}{1}
      {5}{{{\color{numb}5}}}{1}
      {6}{{{\color{numb}6}}}{1}
      {7}{{{\color{numb}7}}}{1}
      {8}{{{\color{numb}8}}}{1}
      {9}{{{\color{numb}9}}}{1}
      {:}{{{\color{punct}{:}}}}{1}
      {,}{{{\color{punct}{,}}}}{1}
      {\{}{{{\color{delim}{\{}}}}{1}
      {\}}{{{\color{delim}{\}}}}}{1}
      {[}{{{\color{delim}{[}}}}{1}
      {]}{{{\color{delim}{]}}}}{1},
}

\newcommand{\mytitle}{RepoGenesis\xspace}
\newcommand{\datasetname}{RepoGenesis\xspace}
\newcommand{\agentname}{GenesisAgent\xspace}

\title{\mytitle: Benchmarking End-to-End Microservice Generation from Readme to Repository}

\author{
  Zhiyuan Peng\textsuperscript{1}\thanks{Equal contribution.},
  Xin Yin\textsuperscript{3}\footnotemark[1],
  Pu Zhao\textsuperscript{2},
  Fangkai Yang\textsuperscript{2},
  Lu Wang\textsuperscript{2}, \\
  \bfseries Ran Jia\textsuperscript{2},
  Xu Chen\textsuperscript{2},
  Qingwei Lin\textsuperscript{2},
  Saravan Rajmohan\textsuperscript{2},
  Dongmei Zhang\textsuperscript{2} \\
  \textsuperscript{1}Shanghai Jiao Tong University, \textsuperscript{2}Microsoft, \textsuperscript{3}Zhejiang University \\
  \texttt{pzy2000@sjtu.edu.cn}, \texttt{xyin@zju.edu.cn} \\
  \texttt{\{puzhao, fangkai.yang, wlu, Jia.Ran\}@microsoft.com} \\
  \texttt{\{xu.chen, qlin, sarr1, dongmeiz\}@microsoft.com}
}

\begin{document}
\maketitle

\begin{abstract}
Large language models and agents have achieved remarkable progress in code generation. 
However, existing benchmarks focus on isolated function/class-level generation (e.g., ClassEval) or modifications to existing codebases (e.g., SWE-Bench), neglecting complete microservice repository generation that reflects real-world 0-to-1 development workflows. 
To bridge this gap, we introduce \datasetname, the first multilingual benchmark for repository-level end-to-end web microservice generation, comprising 106 repositories (60 Python, 46 Java) across 18 domains and 11 frameworks, with 1,258 API endpoints and 2,335 test cases verified through a ``review-rebuttal'' quality assurance process.
We evaluate open-source agents (e.g., DeepCode) and commercial IDEs (e.g., Cursor) using Pass@1, API Coverage (AC), and Deployment Success Rate (DSR). 
Results reveal that despite high AC (up to 73.91\%) and DSR (up to 100\%), the best-performing system achieves only 23.67\% Pass@1 on Python and 21.45\% on Java, exposing deficiencies in architectural coherence, dependency management, and cross-file consistency. 
Notably, \agentname-8B, fine-tuned on \datasetname (train), achieves performance comparable to GPT-5 mini, demonstrating the quality of \datasetname for advancing microservice generation.
We release our benchmark at \url{https://github.com/microsoft/DKI_LLM/tree/main/RepoGenesis}.
\end{abstract}

\section{Introduction}
\label{sec:introduction}
The paradigm of software development is undergoing a fundamental shift, transitioning towards autonomous agents for code generation.
Coding agents have achieved widespread adoption~\cite{hong2023metagpt,qwenagent,li2024modelscope,deepcode2025}, with specialized tools such as OpenHands~\cite{wang2024openhands} and Cursor~\cite{cursor_ide} being increasingly used by developers to assist in or automate the coding process.
As these agents advance, the demand for more comprehensive benchmarks that reflect real-world development scenarios becomes increasingly critical.

In practice, modern software development often involves constructing complete systems rather than isolated code fragments.
Microservices have become a cornerstone of contemporary software architecture, widely adopted for building scalable, flexible, and independently deployable systems~\cite{richardson2018microservices,newman2021building}.
Developing microservices requires not only generating functional code but also designing coherent architectures, managing inter-service dependencies, and ensuring system-wide consistency. 
However, existing benchmarks fail to adequately evaluate the capability of agents to generate complete microservices repositories from scratch.
As shown in Table~\ref{tab:comparison}, most current benchmarks focus on three paradigms with inherent limitations: 
(1) \textit{isolated code generation}, 
evaluating function-level~\cite{chen2021evaluating,austin2021program,lai2023ds,cassano2023multipl} or class-level~\cite{du2023classeval} code generation; 
(2) \textit{codebase modification}, 
assessing the ability to resolve issues or complete code within existing repositories~\cite{jimenez2024swe,yu2024codereval,liu2023repobench,li2024evocodebench,ni2025gittaskbench,li2025fea}; and 
(3) \textit{repository-level generation}, which either focuses on non-microservice architectures~\cite{luo2025rpg,ding2025nl2repo} or offers limited scale and diversity~\cite{zan2024codes}.

\newcommand{\cmark}{\textcolor{green}{\checkmark}}
\newcommand{\xmark}{\textcolor{red}{\ding{55}}}
\newcommand{\halfcheck}{\textcolor{blue}{\ooalign{\hss\ding{55}\hss\cr\hss\checkmark\hss}}}
\begin{table*}[htbp!]
\centering
\caption{Comparison of \datasetname with existing code generation benchmarks. 
``Repo-Aware'' indicates repository-level context awareness. 
``Repo-Level'' denotes full repository generation capability. 
``NL2Repo.'' indicates natural language to repository generation.}
\label{tab:comparison}
\resizebox{\textwidth}{!}{
\begin{tabular}{llccccccc}
\toprule
Benchmark & Task & Repositories & Multilingual & Language & Repo-Aware & \makecell{Repo-Level} & \makecell{NL2Repo.} & Microservice \\
\midrule
Concode~\cite{iyer2018mapping} & Func-lvl. Gen. & $\approx$33k & \xmark & Java & \cmark & \xmark & \xmark & \xmark \\
HumanEval~\cite{chen2021evaluating} & Func-lvl. Gen. & \xmark & \xmark & Python & \xmark & \xmark & \xmark & \xmark \\
MBPP~\cite{austin2021program} & Func-lvl. Gen. & \xmark & \xmark & Python & \xmark & \xmark & \xmark & \xmark \\
DS-1000~\cite{lai2023ds} & Func-lvl. Gen. & \xmark & \xmark & Python & \xmark & \xmark & \xmark & \xmark \\
MultiPL-E~\cite{cassano2023multipl} & Func-lvl. Gen. & \xmark & \cmark & Multi & \xmark & \xmark & \xmark & \xmark \\
ClassEval~\cite{du2023classeval} & Class-lvl. Gen. & \xmark & \xmark & Python & \xmark & \xmark & \xmark & \xmark \\
RepoBench~\cite{liu2023repobench} & Completion & $\approx$25k & \cmark & Py/Java & \cmark & \xmark & \xmark & \xmark \\
CodeS~\cite{zan2024codes} & Full Repo. Gen. & 19 & \xmark & Python & \cmark & \cmark & \cmark & \xmark \\
CoderEval~\cite{yu2024codereval} & Func-lvl. Gen. & 53 & \cmark & Py/Java & \cmark & \xmark & \xmark & \xmark \\
EvoCodeBench~\cite{li2024evocodebench} & Func-lvl. Gen. & 25 & \xmark & Python & \cmark & \xmark & \xmark & \xmark \\
SWE-bench~\cite{jimenez2024swe} & Issue Resolution & 12 & \xmark & Python & \cmark & \halfcheck & \xmark & \xmark \\
SolEval~\cite{peng2025soleval} & Func-lvl. Gen. & 28 & \xmark & Solidity & \cmark & \xmark & \xmark & \xmark \\
GitTaskBench~\cite{ni2025gittaskbench} & Task Solving & 18 & \xmark & Python & \cmark & \halfcheck & \xmark & \xmark \\
RPG~\cite{luo2025rpg} & Full Repo. Gen. & 6 & \xmark & Python & \cmark & \cmark & \cmark & \xmark \\
NL2Repo-Bench~\cite{ding2025nl2repo} & Full Repo. Gen. & 104 & \xmark & Python & \cmark & \cmark & \cmark & \xmark \\
\midrule
\rowcolor{gray!10} \textbf{\datasetname (Ours)} & \textbf{Full Repo. Gen.} & \textbf{30 (test) + 76 (train)} & \cmark & \textbf{Py/Java} & \cmark & \cmark & \cmark & \cmark \\
\bottomrule
\end{tabular}
}
\end{table*}

To bridge this gap, we introduce \datasetname, the first benchmark specifically designed to evaluate repository-level web microservice generation.
\datasetname targets critical yet under-evaluated capabilities essential for real-world engineering: architectural design, dependency management, and system-level consistency.
Comprising 106 repositories (76 for training, 30 for evaluation) spanning 18 domains and 11 frameworks in both Python and Java, \datasetname provides a comprehensive testbed with 1,258 API endpoints and 2,335 test cases.
Each sample consists of a requirement document specifying service functionality, API endpoints, and input/output schemas, paired with rigorous black-box test suites that validate functional correctness and completeness.
The benchmark includes both real-world GitHub projects and expert-curated implementations, ensuring diverse architectural patterns and practical relevance.
To ensure benchmark reliability, we implement a rigorous quality assurance mechanism featuring a ``review-rebuttal'' loop with three LLM reviewers and a human Area Chair, guaranteeing high-quality test cases and reducing potential biases.

\begin{figure}[htbp!]
\centering
\includegraphics[width=.9\columnwidth]{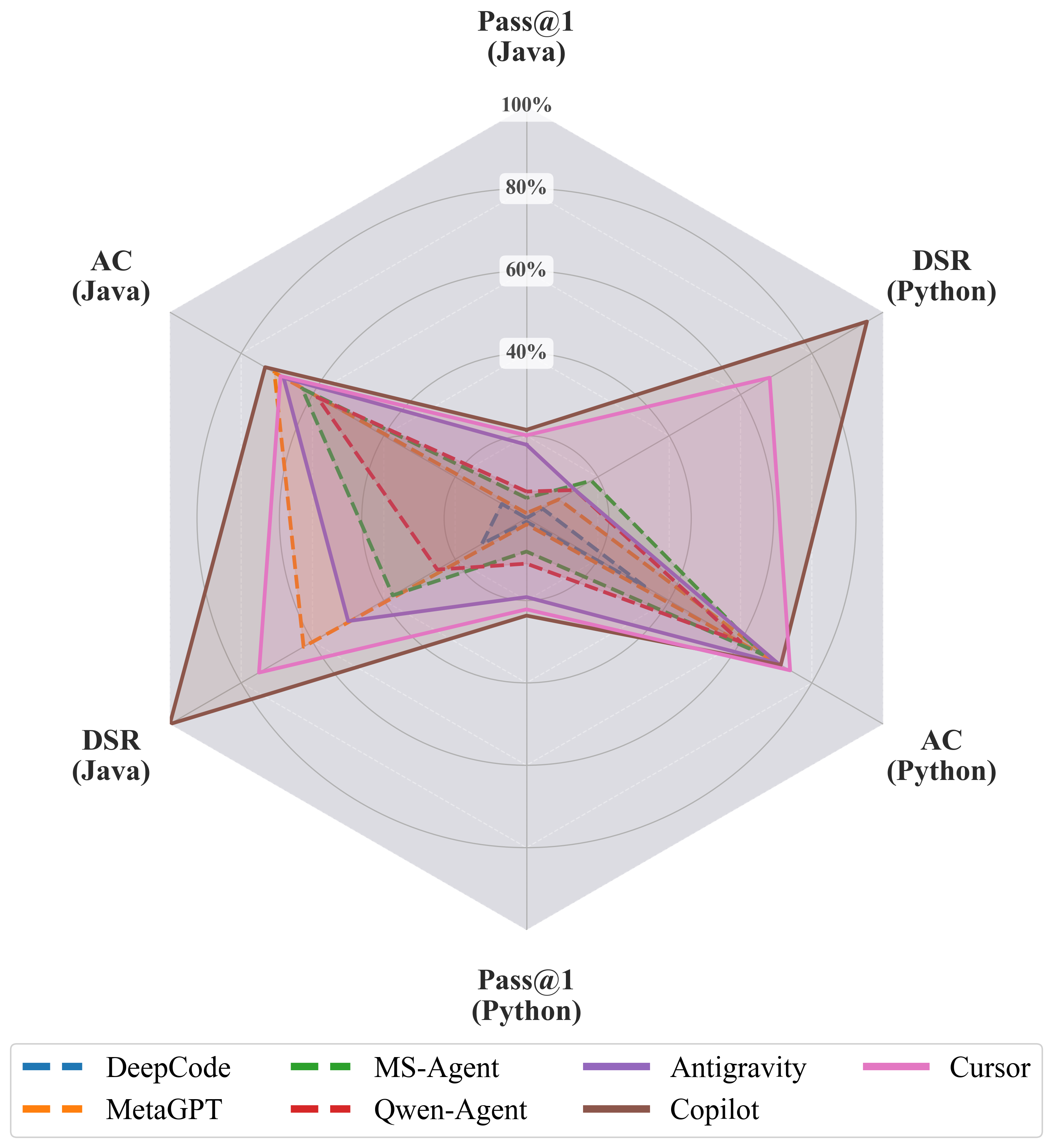}
\caption{Capability map of agents and IDEs}
\label{fig:capability_map}
\end{figure}

We establish a multi-dimensional evaluation methodology comprising Pass@1 for functional correctness, API Coverage (AC) for implementation completeness, and Deployment Success Rate (DSR) for deployability.
Evaluations on four open-source agents (i.e., DeepCode~\cite{deepcode2025}, MetaGPT~\cite{hong2023metagpt}, MS-Agent~\cite{li2024modelscope}, Qwen-Agent~\cite{qwenagent}) and three commercial IDEs (i.e., Antigravity~\cite{antigravity_ide}, Cursor~\cite{cursor_ide}, Copilot~\cite{github_copilot}) reveal a significant performance gap. 
As illustrated in \autoref{fig:capability_map}, even the best-performing agent achieves only 23.67\% Pass@1. 
Error analysis in Table~\ref{tab:error_analysis} reveals that failures stem primarily from cross-file consistency issues (50.2\%), architectural coherence problems (26.0\%), and dependency management errors (23.8\%), underscoring fundamental challenges in repository-level code generation.

\begin{table}[htbp!]
\centering
\caption{Distribution of critical deficiencies in model-generated repositories, categorized into three fundamental challenges in repository-level code generation.}
\label{tab:error_analysis}
\resizebox{0.95\columnwidth}{!}{
\begin{tabular}{lrrr}
\toprule
\textbf{Deficiency Type} & \textbf{Python} & \textbf{Java} & \textbf{Total} \\
\midrule
Architectural Coherence & 45 (25.9\%) & 27 (26.2\%) & 72 (26.0\%) \\
Dependency Management & 20 (11.5\%) & 46 (\textbf{44.7\%}) & 66 (23.8\%) \\
Cross-file Consistency & 109 (\textbf{62.6\%}) & 30 (29.1\%) & 139 (\textbf{50.2\%}) \\
\midrule
\rowcolor{gray!10} \textbf{Total Failures} & \textbf{174} & \textbf{103} & \textbf{277} \\
\bottomrule
\end{tabular}
}
\end{table}

In summary, our contributions are as follows:

\begin{itemize}[leftmargin=*]
    \item We introduce \datasetname, the first multilingual benchmark for repository-level web microservice generation from natural language requirements.
    \item We establish a rigorous quality assurance mechanism featuring a ``review-rebuttal'' loop with three LLM reviewers and a human Area Chair to ensure benchmark reliability.
    \item We propose a multi-dimensional evaluation methodology (i.e., Pass@1, AC, and DSR) and conduct comprehensive evaluations on state-of-the-art coding agents and commercial IDEs.
    \item Our evaluation reveals critical insights: (1) commercial IDEs excel in deployability but struggle with functional correctness, exposing a gap between infrastructure and logic generation; and (2) fine-tuned \agentname-8B achieves GPT-5 mini-level performance, demonstrating effective benchmark-driven model improvement.
\end{itemize}

\section{Benchmark -- \datasetname}
\label{sec:benchmark}
\begin{figure*}
    \centering
    \includegraphics[width=\textwidth]{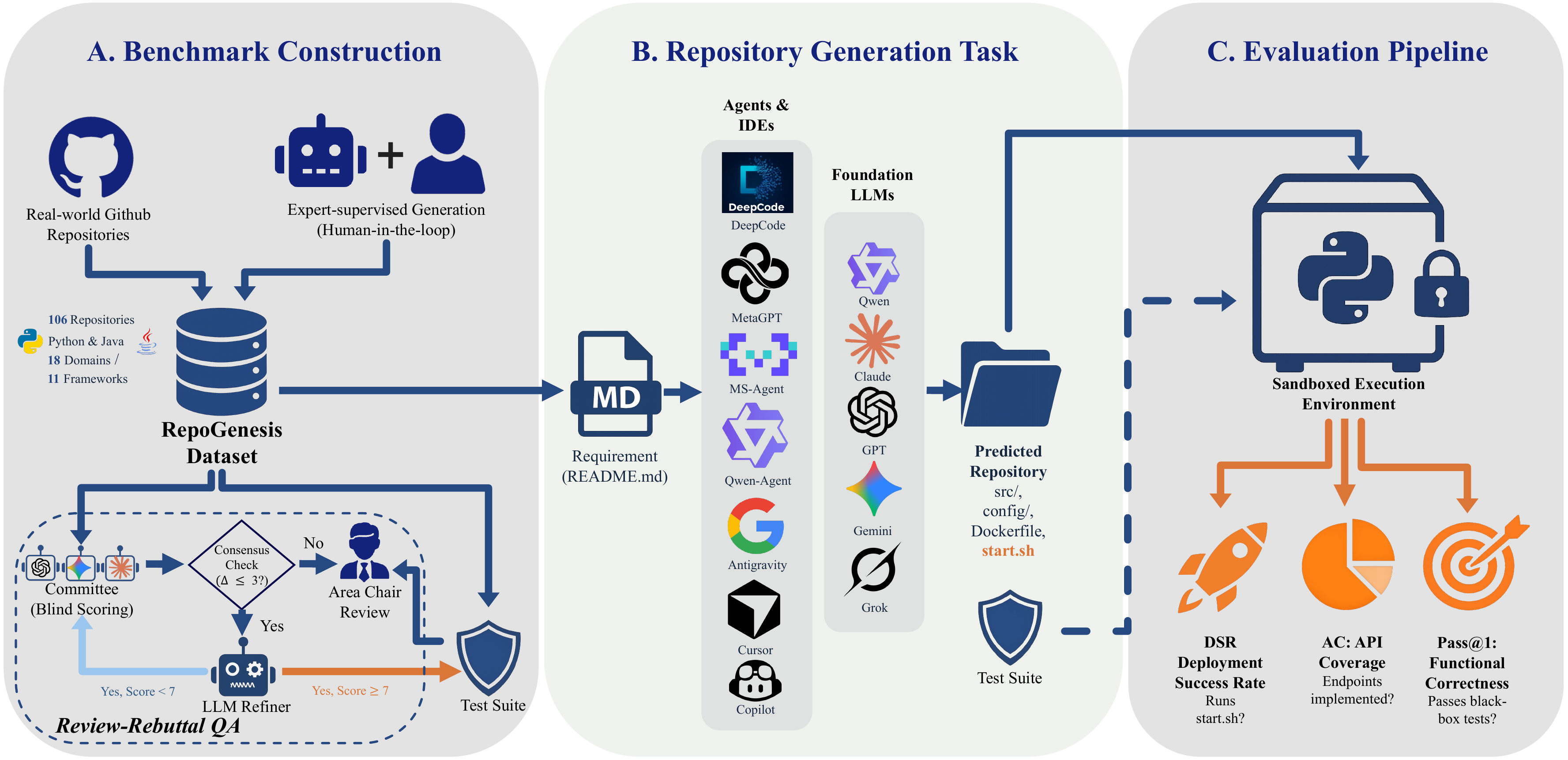}
    \caption{Overview of benchmark construction and evaluation pipeline, 
    which undergo rigorous quality assurance through a ``Review-Rebuttal QA''. 
    Given a requirement document (README.md), 
    various agents and IDEs generate repositories. 
    Repositories are executed in sandboxed environments and evaluated across DSR, AC, and Pass@1.}
    \label{fig:overview}
\end{figure*}

\subsection{Statistics of \datasetname}
As shown in Table~\ref{tab:dataset-statistics}, \datasetname (Verified) comprises 30 web microservice repositories spanning 18 domains and 11 frameworks, serving as our primary evaluation benchmark.

To balance domain coverage with quality control, we combine 6 real-world GitHub repositories with 24 expert-supervised repositories\footnote{Repositories with human-authored requirement documents and LLM-generated test suites under iterative human supervision.} targeting underrepresented domains (e.g., gaming backends, file management, and authentication services).
To mitigate potential bias from the expert-supervised subset (80\%), we implement: (1) human-authored requirement specifications, (2) a multi-model review-rebuttal quality assurance process with human Area Chair (\S\ref{sec:quality_assurance}), and (3) manual filtering to remove ambiguous requirements and flaky tests~\cite{jimenez2024swe}.
This design ensures domain diversity while maintaining evaluation rigor
Repositories are classified into Easy, Medium, and Hard difficulty levels based on code complexity metrics (see Appendix~\S\ref{sec:difficulty_classification} for more details).

To support model training, we additionally construct \datasetname (Train) with 76 repositories (\S\ref{sec:extending_dataset}).
In total, \datasetname encompasses 1,258 API endpoints and 2,335 test cases, with complete statistics in \autoref{appendix:dataset-statistics}.

\subsection{Task Definition and Challenges}
Given a requirement document (README.md) specifying service functionality, API endpoints with input/output schemas, authentication mechanisms, and operational constraints, models must generate a fully deployable repository including all source code, configuration files, and dependency specifications.
The generated repository must pass comprehensive black-box test suites validating functional correctness, API compliance, and error handling.
This task presents three fundamental challenges: (1) translating high-level requirements into concrete software architectures with proper code organization and module decomposition, (2) ensuring cross-file consistency across APIs, data models, and error handling, and (3) handling real-world system complexity including authentication, database interactions, and input validation.
For more details, see Appendix~\S\ref{appendix:task_definition}.

\subsection{Evaluation Pipeline}
As illustrated in \autoref{fig:overview}, \datasetname evaluates agents through 
\ding{182} Repository-level Code Generation and 
\ding{183} Post-Generation Evaluation.

\subsubsection{Repository-level Code Generation}
\label{sec:repo_generation}
The code agent receives a \textit{Requirement Document} (a README.md file) specifying microservice requirements: service functionality, API endpoints with input/output schemas, authentication mechanisms, error handling specifications, and operational constraints (e.g., listening ports, deployment requirements). 
The agent generates a complete repository from scratch, including source code files, configuration files, and dependency specifications.

\begin{table}[htbp]
\centering
\caption{Statistics of the \datasetname (Verified). 
\textbf{APIs} denotes the number of APIs in requirements;
Repository names are color-coded by difficulty: \colorbox{green!25}{Easy}, \colorbox{olive!25}{Medium}, \colorbox{red!25}{Hard}.
FW denotes Framework. Dom denotes Domain.}
\label{tab:dataset-statistics}
\resizebox{\linewidth}{!}
{
\begin{tabular}{lcccccl}
\toprule
\textbf{Repository} & \textbf{Lang.} & \textbf{FW.} & \textbf{APIs} & \textbf{Tests} & \textbf{Stars} & \textbf{Dom.} \\
\midrule
\multicolumn{7}{l}{\textit{Real-world GitHub Repositories}} \\
\midrule
\colorbox{green!25}{django-rest-framework-crud} & Python & Django REST & 10 & 9 & 288 & Auth \& CRUD \\
\colorbox{olive!25}{eve} & Python & Eve/Flask & 6 & 8 & 6.7k & REST FW \\
\colorbox{green!25}{flask} & Python & Flask & 3 & 7 & 70.7k & Web FW \\
\colorbox{green!25}{rock-paper-scissors-flask} & Python & Flask & 3 & 5 & 6 & Gaming \\
\colorbox{green!25}{SimpleFastPyAPI} & Python & FastAPI & 5 & 6 & 26 & API Svc \\
\colorbox{olive!25}{synapse} & Python & Synapse & 10 & 5 & 12k & Comm \\
\midrule
\multicolumn{7}{l}{\textit{Expert-supervised Repositories}} \\
\midrule
\colorbox{olive!25}{Blog} & Python & Django/FastAPI & 11 & 17 & - & CMS \\
\colorbox{green!25}{Chatroom} & Python & Flask/FastAPI & 9 & 9 & - & RT Chat \\
\colorbox{red!25}{Customization} & Python & FastAPI & 10 & 64 & - & User Prefs \\
\colorbox{olive!25}{Data Rank Searcher} & Python & Flask/FastAPI & 4 & 33 & - & Data Search \\
\colorbox{red!25}{File Relay} & Python & Flask/FastAPI & 10 & 46 & - & File Mgmt \\
\colorbox{red!25}{GameBackend} & Python & Flask/FastAPI & 10 & 59 & - & Gaming BE \\
\colorbox{olive!25}{Javalin-online-judge} & Java & Javalin & 19 & 23 & - & OJ \\
\colorbox{olive!25}{Javalin-task-manager} & Java & Javalin & 24 & 16 & - & Task Mgmt \\
\colorbox{red!25}{Javalin-user-auth-platform} & Java & Javalin & 13 & 52 & - & User Auth \\
\colorbox{olive!25}{Micronaut-ci-status} & Java & Micronaut & 8 & 26 & - & CI Status \\
\colorbox{olive!25}{Quarkus-blog-cms} & Java & Quarkus & 29 & 16 & - & CMS \\
\colorbox{olive!25}{Spark-dashboard-backend} & Java & Spark & 25 & 16 & - & Dashboard \\
\colorbox{olive!25}{Spring-boot-course-scheduling} & Java & Spring Boot & 45 & 19 & - & Scheduling \\
\colorbox{olive!25}{Springboot-chat-gateway} & Java & Spring Boot & 15 & 13 & - & Chat GW \\
\colorbox{red!25}{Mail Service} & Python & Flask/FastAPI & 4 & 55 & - & Email Svc \\
\colorbox{olive!25}{Multilingual} & Python & Flask & 4 & 33 & - & I18n \\
\colorbox{green!25}{Simple RBAC Service} & Python & Flask/FastAPI & 4 & 9 & - & AuthZ \\
\colorbox{olive!25}{StructuredDataConvertor} & Python & Flask/FastAPI & 3 & 38 & - & Data Conv \\
\colorbox{olive!25}{TaskManagement} & Python & Flask/FastAPI & 6 & 44 & - & Proj Mgmt \\
\colorbox{olive!25}{Tic-Tac-Toe} & Python & Flask/FastAPI & 6 & 7 & - & Gaming \\
\colorbox{red!25}{Timer4Tasker} & Python & Flask/FastAPI & 9 & 53 & - & Scheduling \\
\colorbox{red!25}{UserManagement} & Python & FastAPI & 8 & 69 & - & User Auth \\
\colorbox{olive!25}{UserManagement Lite} & Python & Flask/FastAPI & 5 & 31 & - & User Sys \\
\colorbox{red!25}{WebPan} & Python & Flask/FastAPI & 13 & 22 & - & File Storage \\
\midrule
Total & 2 & 9 Stacks & 294 & 805 & - & 18 \\
\bottomrule
\end{tabular}
}
\end{table}

\subsubsection{Post-Generation Evaluation}
\label{sec:eval_metrics}
We evaluate generated repositories using black-box testing: deploying the microservice, executing test cases, and computing metrics.
We assess quality across three dimensions: (1) Pass@k, (2) API Coverage, and (3) Deployment Success Rate.
Detailed metric definitions are in Appendix~\S\ref{ss:appendix:metrics}.

\section{Benchmark Construction}
\datasetname construction involves four phases: (1) Repository Collection, 
(2) Repository Preprocessing, (3) Test Case \& Quality Assurance, and (4) Difficulty Classification.

\subsection{Repository Collection}
\label{sec:repo_collection}
We collected repositories from two sources:

\noindent \textbf{Real-world GitHub Repositories.} 
We initially identified 28 web microservice repositories from GitHub based on four selection criteria: 
(1) substantial community adoption (stars and active discussions), 
(2) comprehensive documentation with clear API specifications, 
(3) existing test suites, 
and (4) diverse technology stacks (Flask, Django REST Framework, FastAPI, etc.). 
After cloning and attempting deployment, we retained only repositories that could be successfully deployed out-of-the-box without modification, 
as deployability is essential for reliable benchmark evaluation.
This filtering yielded 9 repositories representing diverse architectural patterns and application domains.

\noindent \textbf{Expert-supervised Repository Generation.} 
To achieve comprehensive domain coverage beyond available real-world open-source microservices, we employ a human-in-the-loop approach: two senior researchers manually author 97 requirement documents specifying API contracts, business logic, and operational constraints. 
Test suites are then generated using Gemini 3 Pro with iterative human supervision following the rigorous ``review-rebuttal'' quality assurance process (\S\ref{sec:quality_assurance}). 
This expert-supervised approach targets underrepresented but practical domains (e.g., gaming backends, file management systems, specialized authentication services) while maintaining semantic validity through human verification. 
While this introduces LLM-generated artifacts, the multi-stage human oversight ensures alignment with real-world patterns.

\subsection{Repository Preprocessing}
All repositories underwent systematic preprocessing for benchmark readiness.
For real-world repositories, we refined README files with essential API information, removed implementation code, and verified test suites.
For expert-supervised repositories, we validated API specifications and requirement-functionality consistency.
This yielded 106 repository specifications ready for evaluation.

\subsection{Test Case \& Quality Assurance}
\label{sec:quality_assurance}
We developed rigorous black-box test suites for all repositories covering: 
(1) \textit{API Compliance}, (2) \textit{Error Handling}, (3) \textit{Edge Cases}, and (4) \textit{Integration Testing}.
Test cases were generated using Gemini 3 Pro and refined through a ``review-rebuttal'' quality assurance process with multi-model evaluation and human oversight (\autoref{fig:overview}, bottom left).

\paragraph{Blind Review}
All three models evaluate test cases simultaneously without access to each other's scores or reasoning. 
Preliminary experiments showed this reduces anchoring bias compared to serial scoring, 
ensuring independent judgment and identifying subtle flaws 
that might be overlooked in groupthink scenarios.

\paragraph{Refinement and Consensus}
The process follows an iterative ``review-rebuttal'' loop driven by three LLM reviewers and a human Area Chair:
\begin{enumerate}[leftmargin=*]
    \item \textbf{Initial Review}: The committee scores the initial test suite and provides explanations for each score, 
    resembling a peer review process.
    \item \textbf{Consensus Check}: If pairwise score difference exceeds $\Delta > 3$, 
    the case is flagged for manual review by an Area Chair (a senior code generation PhD researcher) providing meta scores, and decides whether to enter refinement (rebuttal). Check \S~\ref{sec:human_ins} for how Area Chair makes decisions.
    \item \textbf{Refinement}: If scores are consistent ($\Delta \le 3$) but maximum score is below 7, 
    aggregated feedback is sent to an LLM refiner to improve the test case.
    \item \textbf{Iteration}: The refined test case is re-evaluated by the committee. 
    This continues until $\max(\text{Score}) \ge 7$ or maximum iterations ($T = 5$) are reached.
\end{enumerate}

\paragraph{Inter-Annotator Agreement} 
We measure inter-annotator agreement using Krippendorff's alpha on 150 test case evaluations. 
Initial zero-shot scoring yielded $\alpha = 0.032$. 
We conducted iterative rubric refinement: for cases with score spread $> 3$, 
we analyzed disagreements to clarify evaluation criteria (e.g., balancing coverage versus readability). 
This improved reliability to $\alpha = 0.69$, indicating moderate consensus on a 10-point scale~\cite{landis1977measurement}. 
Notably, 78\% of cases showed agreement within 2 points, supported by strong pairwise Spearman correlation ($\rho = 0.61, p < 0.001$). 
claude-haiku-4.5 (mean 6.19) aligns most closely with human experts (mean 5.83), while GPT-5 mini and Gemini 2.5 Flash exhibit systematic leniency.
We only include test cases achieving stringent consensus (mean score $\geq 7$).

\subsection{Difficulty Classification}
\label{sec:difficulty_classification}
Inspired by GAIA~\citep{mialon2023gaia}, we classify repositories into three difficulty levels (Easy, Medium, Hard) based on the following metrics:

\noindent \textbf{Classification Metrics.} We evaluate complexity using five metrics: (1) \textit{Lines of Code (LOC)}: non-comment, non-blank lines; 
(2) \textit{Cyclomatic Complexity}: branch complexity from control flow structures; 
(3) \textit{File Count}: number of source files (\texttt{.py} or \texttt{.java}); 
(4) \textit{API Endpoints}: number of exposed interfaces; 
and (5) \textit{Functions and Classes}: total code constructs. 
The same thresholds apply to both Python and Java repositories.

\noindent \textbf{Scoring.} Each repository receives a weighted score: LOC (25\%), cyclomatic complexity (30\%), file count (15\%), API endpoints (15\%), and functions/classes (15\%). Scores range from 2 (simplest) to 10 (most complex).

\noindent \textbf{Thresholds.} 
\textit{Easy} (score $<$ 4.5): simple services with minimal business logic; 
\textit{Medium} (4.5 $\leq$ score $<$ 7.0): moderate complexity with authentication, validation, and business logic; 
\textit{Hard} (score $\geq$ 7.0): complex services with intricate state management and error handling.

\datasetname (Verified) (Table~\ref{tab:difficulty-metrics}) comprises 6 Easy (20.0\%), 16 Medium (53.3\%), and 8 Hard (26.7\%) repositories. 
Expert-supervised Python repositories are predominantly Hard, reflecting production-level complexity, while Java repositories are predominantly Medium.

\subsection{\datasetname for Training}
\label{sec:extending_dataset}
To support model training, we extend \datasetname (Train) with 76 repositories (44 real-world, 32 expert-supervised). 
Following the review-rebuttal process (\S\ref{sec:quality_assurance}), we employ the \textit{Multi-Model Evaluation Committee} to evaluate test cases. 
However, \datasetname (Train) \textit{omits human Area Chair intervention}, introducing reliability uncertainty. 
When conflicts arise among LLM reviewers (pairwise score difference $\Delta > 3$), conflicting test cases are discarded rather than sent to the Area Chair for resolution, as shown in \autoref{fig:overview} (bottom left). 
This automated conflict resolution resulted in approximately 23\% of initially generated test cases being discarded during the QA process.
Only test cases achieving LLM consensus are retained.
The inter-annotator agreement reported in \S\ref{sec:quality_assurance} ($\alpha = 0.69$) applies exclusively to \datasetname (Verified), 
where human Area Chair mediation resolves LLM disagreements and provides ground-truth calibration.
Without human adjudication, 
\datasetname (Train) lacks this quality bound.
The automated discard-on-conflict strategy may eliminate challenging edge cases requiring expert judgment, 
potentially underrepresenting complex scenarios and adversarial inputs in the training distribution.
Consequently, while \datasetname (Train) provides a resource for supervised fine-tuning, 
we exclude it from coding agent evaluation to maintain rigorous benchmarking standards.



\section{Experimental Setup}
\label{sec:experimental_setup}
\subsection{Evaluated Systems} 
We evaluate four open-source coding agents (i.e., DeepCode~\cite{deepcode2025}, MetaGPT~\cite{hong2023metagpt}, MS-Agent~\cite{li2024modelscope}, and Qwen-Agent~\cite{qwenagent}) and three commercial IDEs (i.e., Antigravity~\cite{antigravity_ide}, Cursor~\cite{cursor_ide}, and Copilot~\cite{github_copilot}).
Detailed description and prompt of agents are provided in Appendix~\S\ref{s:appendix:experiments_details}.

For coding agents, we use three closed-source models (i.e., GPT-5.1, GPT-5.1 mini, and Claude-Sonnet-4.5) and two open-source models (i.e., Qwen3-Coder-30B-A3B-Instruct and Qwen3-30B-A3B-Instruct). 
For commercial IDEs, we use officially supported models to elicit optimal programming capabilities; underlying models may vary due to platform-specific availability. 
Coding agents are permitted to use built-in tools including code search and file system operations, while commercial IDEs use default configurations. 
For hardware and software environment details, see Appendix~\S\ref{sec:hardware_software}.

\subsection{Evaluation Metrics} 
We evaluate generated repositories using Pass@1 for functional correctness, API Coverage (AC) for implementation completeness, and Deployment Success Rate (DSR) for deployability.
Detailed definitions are provided in Appendix~\S\ref{sec:eval_metrics}.
All results are averaged across five independent runs.

\section{Evaluation}
\label{sec:evalution}
\subsection{Results of Coding Agents and IDEs}
\label{sec:results_baselines}

For brevity, we use ``Claude'' to denote Claude-Sonnet-4.5 and ``Qwen3-30B'' to denote Qwen3-30B-A3B-Instruct throughout this section. 
We report results for GPT-5.1, Claude, and Qwen3-30B in the main text; additional results for other LLMs are provided in Appendix~\S\ref{sec:appendix:additional_results}.

\begin{table}[htbp!]
\centering
\caption{Overall performance (Pass@1) of Coding Agents and Commercial IDEs.}
\label{tab:pass_rate_combined}
\resizebox{.75\columnwidth}{!}{
\begin{tabular}{llcc}
\toprule
\textbf{Method} & \textbf{Model} & \textbf{Python} & \textbf{Java} \\
\midrule
\multirow{3}{*}{\includegraphics[height=1.0em]{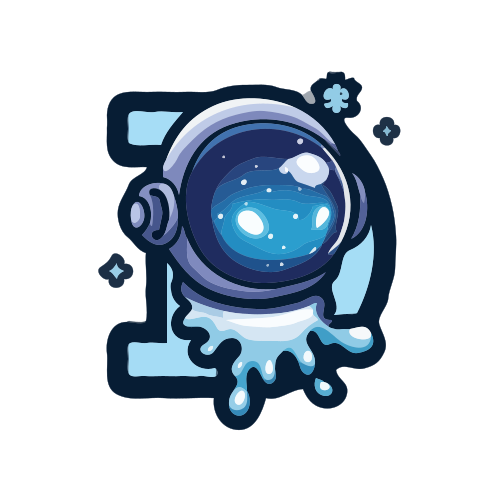} DeepCode} & \includegraphics[height=1.0em]{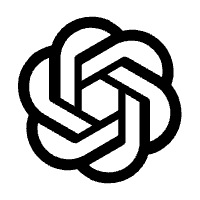} GPT-5.1 & \colorbox{red!25}{\makebox[3.5em][c]{0.81\%}} & \colorbox{red!25}{\makebox[3.5em][c]{0.00\%}} \\
& \includegraphics[height=1.0em]{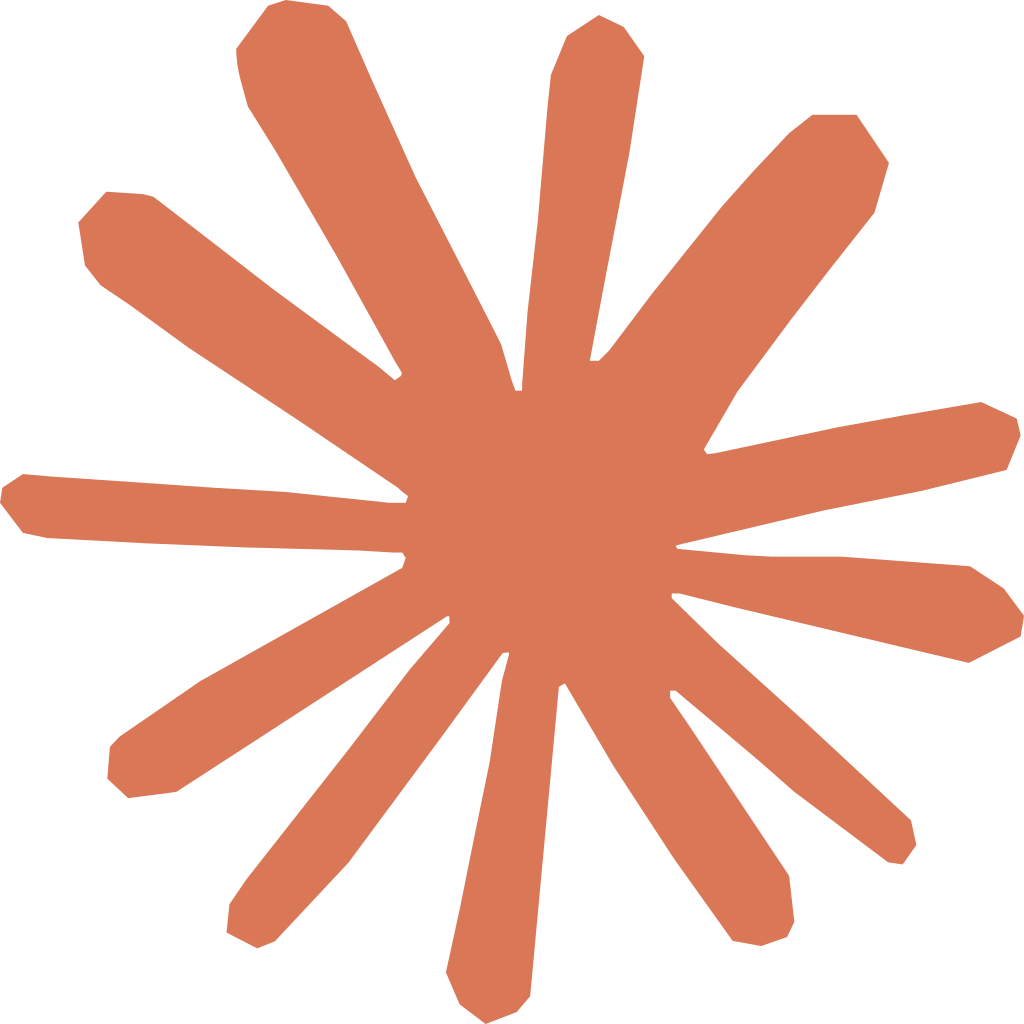} Claude & \colorbox{red!25}{\makebox[3.5em][c]{1.95\%}} & \colorbox{red!25}{\makebox[3.5em][c]{0.85\%}} \\
& \includegraphics[height=1.0em]{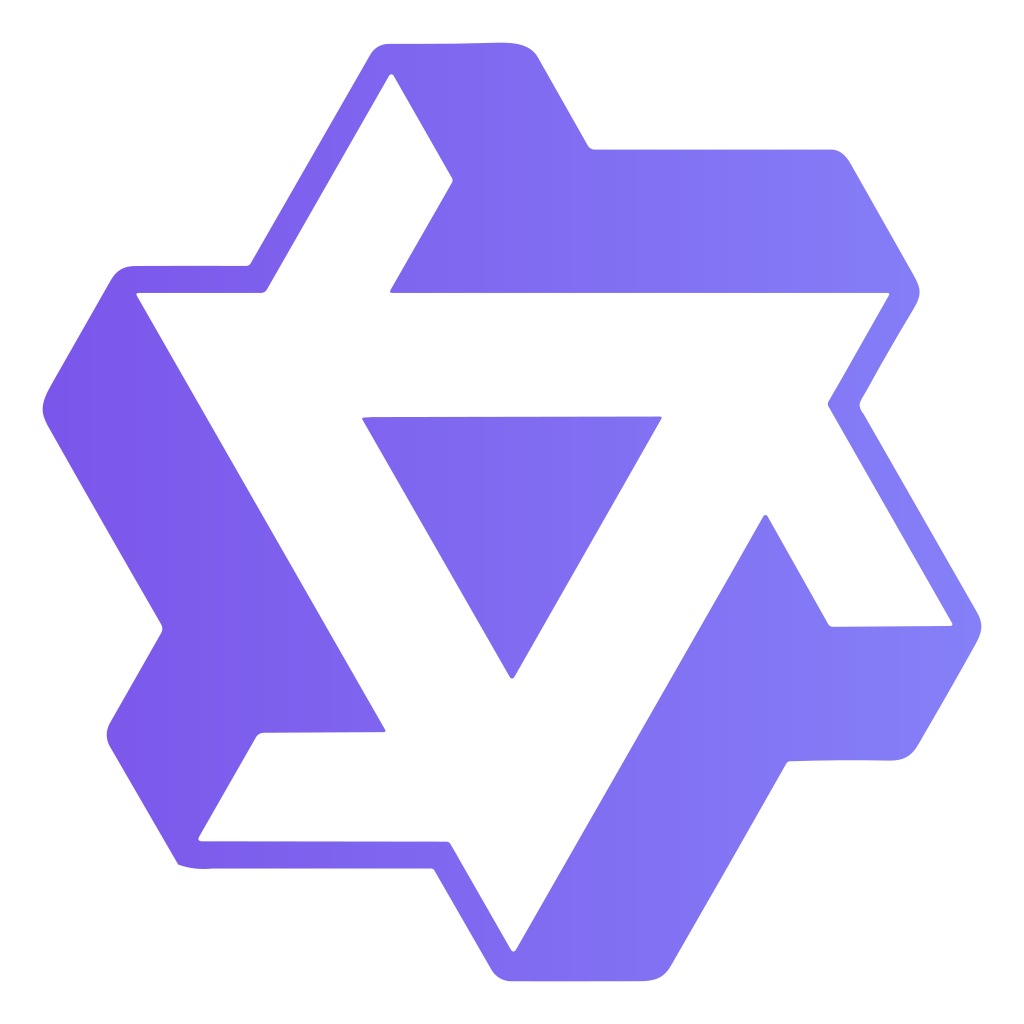} Qwen3-30B & \colorbox{red!25}{\makebox[3.5em][c]{0.00\%}} & \colorbox{red!25}{\makebox[3.5em][c]{0.00\%}} \\
\midrule
\multirow{3}{*}{\includegraphics[height=1.0em]{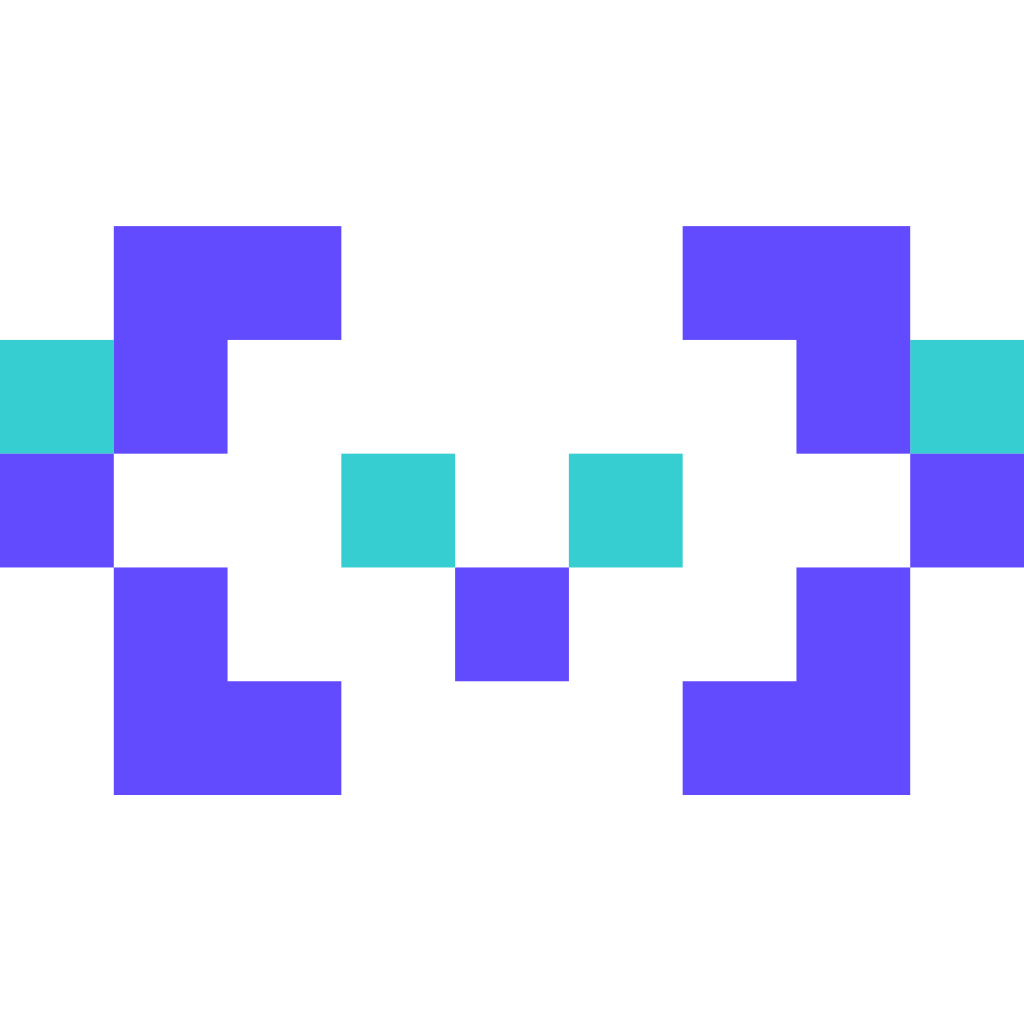} MS-Agent} & \includegraphics[height=1.0em]{figs/openai-fill.png} GPT-5.1 & \colorbox{olive!25}{\makebox[3.5em][c]{7.83\%}} & \colorbox{orange!25}{\makebox[3.5em][c]{4.76\%}} \\
& \includegraphics[height=1.0em]{figs/claude-color.png} Claude & \colorbox{olive!25}{\makebox[3.5em][c]{9.50\%}} & \colorbox{orange!25}{\makebox[3.5em][c]{5.95\%}} \\
& \includegraphics[height=1.0em]{figs/qwen-color.png} Qwen3-30B & \colorbox{orange!25}{\makebox[3.5em][c]{3.56\%}} & \colorbox{orange!25}{\makebox[3.5em][c]{1.25\%}} \\
\midrule
\multirow{3}{*}{\includegraphics[height=1.0em]{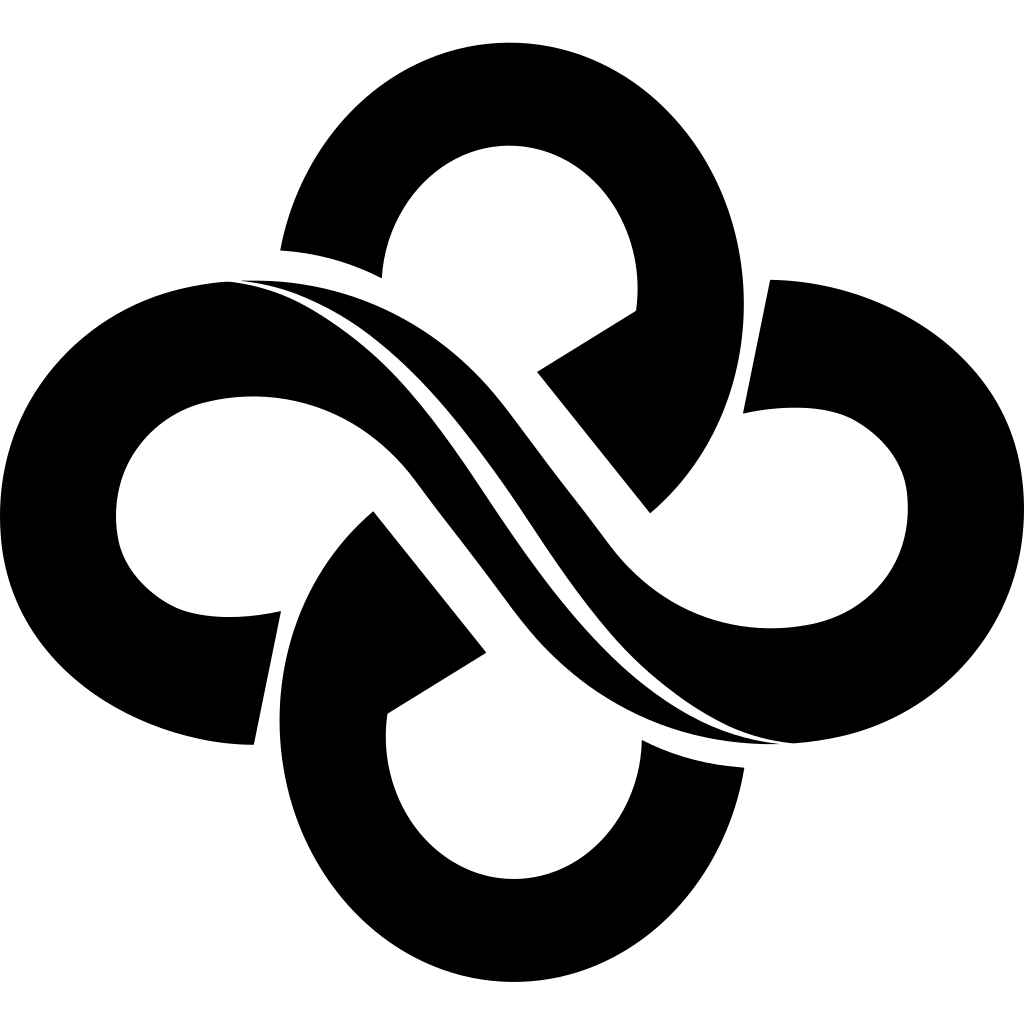} MetaGPT} & \includegraphics[height=1.0em]{figs/openai-fill.png} GPT-5.1 & \colorbox{orange!25}{\makebox[3.5em][c]{1.46\%}} & \colorbox{orange!25}{\makebox[3.5em][c]{1.14\%}} \\
& \includegraphics[height=1.0em]{figs/claude-color.png} Claude & \colorbox{orange!25}{\makebox[3.5em][c]{2.85\%}} & \colorbox{orange!25}{\makebox[3.5em][c]{2.10\%}} \\
& \includegraphics[height=1.0em]{figs/qwen-color.png} Qwen3-30B & \colorbox{red!25}{\makebox[3.5em][c]{0.15\%}} & \colorbox{red!25}{\makebox[3.5em][c]{0.00\%}} \\
\midrule
\multirow{3}{*}{\includegraphics[height=1.0em]{figs/qwen-color.png} Qwen-Agent} & \includegraphics[height=1.0em]{figs/openai-fill.png} GPT-5.1 & \colorbox{olive!25}{\makebox[3.5em][c]{10.89\%}} & \colorbox{olive!25}{\makebox[3.5em][c]{6.29\%}} \\
& \includegraphics[height=1.0em]{figs/claude-color.png} Claude & \colorbox{olive!25}{\makebox[3.5em][c]{12.65\%}} & \colorbox{olive!25}{\makebox[3.5em][c]{7.80\%}} \\
& \includegraphics[height=1.0em]{figs/qwen-color.png} Qwen3-30B & \colorbox{red!25}{\makebox[3.5em][c]{0.35\%}} & \colorbox{orange!25}{\makebox[3.5em][c]{4.12\%}} \\
\midrule
\multirow{1}{*}{\includegraphics[height=1.0em]{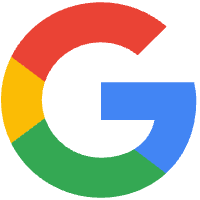} Antigravity} & \includegraphics[height=1.0em]{figs/claude-color.png} Claude & \colorbox{green!25}{\makebox[3.5em][c]{19.17\%}} & \colorbox{green!25}{\makebox[3.5em][c]{17.82\%}} \\
\midrule
\multirow{1}{*}{\includegraphics[height=1.0em]{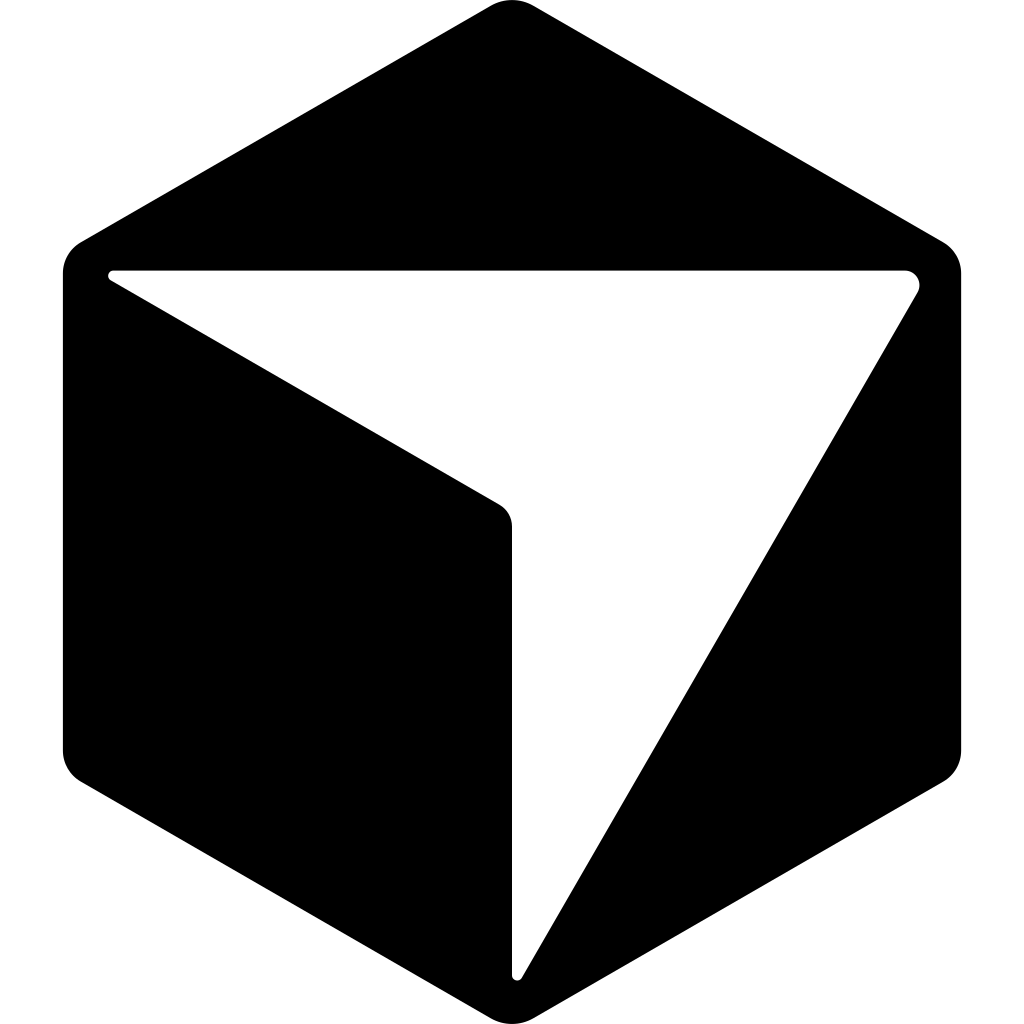} Cursor} & \includegraphics[height=1.0em]{figs/claude-color.png} Claude & \colorbox{green!25}{\makebox[3.5em][c]{22.15\%}} & \colorbox{green!25}{\makebox[3.5em][c]{20.10\%}} \\
\midrule
\multirow{1}{*}{\includegraphics[height=1.0em]{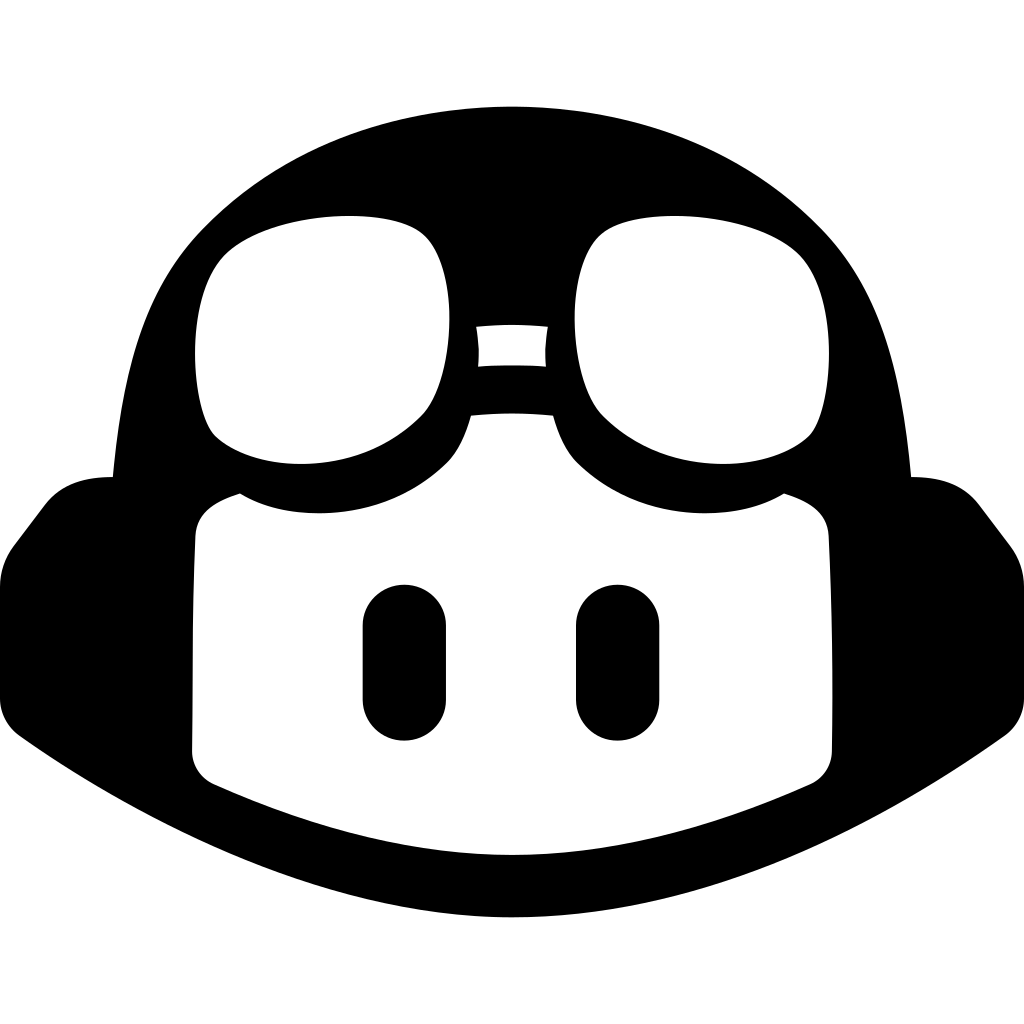} Copilot} & \includegraphics[height=1.0em]{figs/claude-color.png} Claude & \colorbox{green!25}{\makebox[3.5em][c]{23.67\%}} & \colorbox{green!25}{\makebox[3.5em][c]{21.45\%}} \\
\bottomrule
\end{tabular}
}
\end{table}

\paragraph{Pass Rate (Pass@1).} 
As shown in Tables~\ref{tab:pass_rate_combined} and~\ref{tab:difficulty_results_main}, commercial IDEs significantly outperform open-source agents, with Copilot achieving 23.67\% on Python and 21.45\% on Java.
We observe three key trends:
(1) \textbf{Performance Gap:} Commercial IDEs achieve substantially better Pass@1 than open-source agents across both languages.
(2) \textbf{Language Sensitivity:} Python performance generally exceeds Java, as Java's verbosity and rigid type system increase implementation complexity despite its superior tooling.
(3) \textbf{Difficulty Sensitivity:} All systems exhibit sharp performance degradation as repository complexity increases.
For example, Qwen-Agent with GPT-5.1 drops from 25.00\% (Easy) to 4.86\% (Hard) on Python, while MS-Agent maintains a more gradual degradation curve.
Notably, even smaller models such as Qwen3-30B nearly fail on \textit{Medium} and \textit{Hard} tasks, underscoring that architectural reasoning remains challenging for resource-constrained models.

\begin{table}[htbp!]
\centering
\caption{DSR and AC for coding agents and IDEs.}
\label{tab:dsr_ac_combined}
\resizebox{\columnwidth}{!}{
\begin{tabular}{llcccc}
\toprule
\multirow{2.5}{*}{\textbf{Method}} & \multirow{2.5}{*}{\textbf{Model}} & \multicolumn{2}{c}{\textbf{Python}} & \multicolumn{2}{c}{\textbf{Java}} \\
\cmidrule(lr){3-4} \cmidrule(lr){5-6} 
& & \textbf{DSR} & \textbf{AC} & \textbf{DSR} & \textbf{AC} \\
\midrule
\multirow{3}{*}{\includegraphics[height=1.0em]{figs/DeepCode.png} DeepCode} & \includegraphics[height=1.0em]{figs/qwen-color.png} Qwen3-30B & \colorbox{red!25}{\makebox[4.2em][c]{0.00\%}} & \colorbox{red!25}{\makebox[4.2em][c]{12.00\%}} & \colorbox{red!25}{\makebox[4.2em][c]{0.00\%}} & \colorbox{red!25}{\makebox[4.2em][c]{1.50\%}} \\
 & \includegraphics[height=1.0em]{figs/openai-fill.png} GPT-5.1 & \colorbox{red!25}{\makebox[4.2em][c]{4.55\%}} & \colorbox{red!25}{\makebox[4.2em][c]{26.96\%}} & \colorbox{red!25}{\makebox[4.2em][c]{0.00\%}} & \colorbox{red!25}{\makebox[4.2em][c]{0.00\%}} \\
 & \includegraphics[height=1.0em]{figs/claude-color.png} Claude & \colorbox{red!25}{\makebox[4.2em][c]{16.50\%}} & \colorbox{olive!25}{\makebox[4.2em][c]{36.00\%}} & \colorbox{red!25}{\makebox[4.2em][c]{28.50\%}} & \colorbox{red!25}{\makebox[4.2em][c]{10.50\%}} \\
\midrule
\multirow{3}{*}{\includegraphics[height=1.0em]{figs/modelscope-color.png} MS-Agent} & \includegraphics[height=1.0em]{figs/qwen-color.png} Qwen3-30B & \colorbox{red!25}{\makebox[4.2em][c]{2.27\%}} & \colorbox{olive!25}{\makebox[4.2em][c]{55.00\%}} & \colorbox{red!25}{\makebox[4.2em][c]{10.00\%}} & \colorbox{red!25}{\makebox[4.2em][c]{15.50\%}} \\
 & \includegraphics[height=1.0em]{figs/openai-fill.png} GPT-5.1 & \colorbox{red!25}{\makebox[4.2em][c]{18.18\%}} & \colorbox{olive!25}{\makebox[4.2em][c]{65.22\%}} & \colorbox{olive!25}{\makebox[4.2em][c]{37.50\%}} & \colorbox{olive!25}{\makebox[4.2em][c]{47.41\%}} \\
 & \includegraphics[height=1.0em]{figs/claude-color.png} Claude & \colorbox{red!25}{\makebox[4.2em][c]{12.00\%}} & \colorbox{green!25}{\makebox[4.2em][c]{69.00\%}} & \colorbox{red!25}{\makebox[4.2em][c]{16.00\%}} & \colorbox{green!25}{\makebox[4.2em][c]{66.50\%}} \\
\midrule
\multirow{3}{*}{\includegraphics[height=1.0em]{figs/metagpt.png} MetaGPT} & \includegraphics[height=1.0em]{figs/qwen-color.png} Qwen3-30B & \colorbox{red!25}{\makebox[4.2em][c]{0.00\%}} & \colorbox{olive!25}{\makebox[4.2em][c]{60.00\%}} & \colorbox{red!25}{\makebox[4.2em][c]{0.00\%}} & \colorbox{olive!25}{\makebox[4.2em][c]{40.00\%}} \\
 & \includegraphics[height=1.0em]{figs/openai-fill.png} GPT-5.1 & \colorbox{red!25}{\makebox[4.2em][c]{4.55\%}} & \colorbox{green!25}{\makebox[4.2em][c]{70.43\%}} & \colorbox{olive!25}{\makebox[4.2em][c]{62.50\%}} & \colorbox{green!25}{\makebox[4.2em][c]{68.10\%}} \\
 & \includegraphics[height=1.0em]{figs/claude-color.png} Claude & \colorbox{red!25}{\makebox[4.2em][c]{13.00\%}} & \colorbox{green!25}{\makebox[4.2em][c]{73.04\%}} & \colorbox{olive!25}{\makebox[4.2em][c]{54.00\%}} & \colorbox{green!25}{\makebox[4.2em][c]{74.14\%}} \\
\midrule
\multirow{3}{*}{\includegraphics[height=1.0em]{figs/qwen-color.png} Qwen-Agent} & \includegraphics[height=1.0em]{figs/qwen-color.png} Qwen3-30B & \colorbox{red!25}{\makebox[4.2em][c]{2.27\%}} & \colorbox{olive!25}{\makebox[4.2em][c]{50.00\%}} & \colorbox{red!25}{\makebox[4.2em][c]{7.50\%}} & \colorbox{olive!25}{\makebox[4.2em][c]{35.00\%}} \\
 & \includegraphics[height=1.0em]{figs/openai-fill.png} GPT-5.1 & \colorbox{red!25}{\makebox[4.2em][c]{13.64\%}} & \colorbox{olive!25}{\makebox[4.2em][c]{37.39\%}} & \colorbox{red!25}{\makebox[4.2em][c]{25.00\%}} & \colorbox{olive!25}{\makebox[4.2em][c]{41.38\%}} \\
 & \includegraphics[height=1.0em]{figs/claude-color.png} Claude & \colorbox{red!25}{\makebox[4.2em][c]{17.00\%}} & \colorbox{olive!25}{\makebox[4.2em][c]{60.50\%}} & \colorbox{red!25}{\makebox[4.2em][c]{28.00\%}} & \colorbox{olive!25}{\makebox[4.2em][c]{60.00\%}} \\
\midrule
\includegraphics[height=1.0em]{figs/githubcopilot.png} Copilot & \includegraphics[height=1.0em]{figs/claude-color.png} Claude & \colorbox{green!25}{\makebox[4.2em][c]{95.45\%}} & \colorbox{green!25}{\makebox[4.2em][c]{69.57\%}} & \colorbox{olive!25}{\makebox[4.2em][c]{62.50\%}} & \colorbox{green!25}{\makebox[4.2em][c]{68.10\%}} \\
\midrule
\includegraphics[height=1.0em]{figs/google.png} Antigravity & \includegraphics[height=1.0em]{figs/claude-color.png} Claude & \colorbox{red!25}{\makebox[4.2em][c]{13.64\%}} & \colorbox{olive!25}{\makebox[4.2em][c]{60.00\%}} & \colorbox{olive!25}{\makebox[4.2em][c]{50.00\%}} & \colorbox{red!25}{\makebox[4.2em][c]{27.59\%}} \\
\midrule
\includegraphics[height=1.0em]{figs/cursor.png} Cursor & \includegraphics[height=1.0em]{figs/claude-color.png} Claude & \colorbox{green!25}{\makebox[4.2em][c]{68.18\%}} & \colorbox{green!25}{\makebox[4.2em][c]{73.91\%}} & \colorbox{green!25}{\makebox[4.2em][c]{75.00\%}} & \colorbox{green!25}{\makebox[4.2em][c]{68.97\%}} \\
\bottomrule
\end{tabular}
}
\end{table}

\begin{table*}[htbp!]
\centering
\caption{Performance (Pass@1) across difficulty levels. 
Easy (E), Medium (M), and Hard (H) classifications are based on code complexity metrics (\S\ref{sec:difficulty_classification}). 
Note that Java repositories have no Easy-level instances.}
\label{tab:difficulty_results_main}
\resizebox{\textwidth}{!}{
\begin{tabular}{lcccccc}
\toprule
\multirow{2.5}{*}{\textbf{Method}} & \multicolumn{3}{c}{\textbf{Python (E/M/H)}} & \multicolumn{3}{c}{\textbf{Java (E/M/H)}} \\
\cmidrule(lr){2-4} \cmidrule(lr){5-7}
& \includegraphics[height=1.0em]{figs/openai-fill.png} \textbf{GPT-5.1} & \includegraphics[height=1.0em]{figs/claude-color.png} \textbf{Claude} & \includegraphics[height=1.0em]{figs/qwen-color.png} \textbf{Qwen3-30B} & \includegraphics[height=1.0em]{figs/openai-fill.png} \textbf{GPT-5.1} & \includegraphics[height=1.0em]{figs/claude-color.png} \textbf{Claude} & \includegraphics[height=1.0em]{figs/qwen-color.png} \textbf{Qwen3-30B} \\
\midrule
\includegraphics[height=1.0em]{figs/modelscope-color.png} MS-Agent & \makebox[2.2em][c]{16.07} \makebox[2.2em][c]{9.52} \makebox[2.2em][c]{3.47} & \makebox[2.2em][c]{18.95} \makebox[2.2em][c]{10.45} \makebox[2.2em][c]{4.12} & \makebox[2.2em][c]{5.36} \makebox[2.2em][c]{4.17} \makebox[2.2em][c]{2.08} & \makebox[2.2em][c]{-} \makebox[2.2em][c]{5.30} \makebox[2.2em][c]{0.98} & \makebox[2.2em][c]{-} \makebox[2.2em][c]{6.85} \makebox[2.2em][c]{1.50} & \makebox[2.2em][c]{-} \makebox[2.2em][c]{1.40} \makebox[2.2em][c]{0.20} \\
\includegraphics[height=1.0em]{figs/metagpt.png} MetaGPT & \makebox[2.2em][c]{3.57} \makebox[2.2em][c]{1.79} \makebox[2.2em][c]{0.00} & \makebox[2.2em][c]{4.80} \makebox[2.2em][c]{2.65} \makebox[2.2em][c]{1.20} & \makebox[2.2em][c]{0.00} \makebox[2.2em][c]{0.30} \makebox[2.2em][c]{0.00} & \makebox[2.2em][c]{-} \makebox[2.2em][c]{1.30} \makebox[2.2em][c]{0.02} & \makebox[2.2em][c]{-} \makebox[2.2em][c]{2.15} \makebox[2.2em][c]{1.05} & \makebox[2.2em][c]{-} \makebox[2.2em][c]{0.00} \makebox[2.2em][c]{0.00} \\
\includegraphics[height=1.0em]{figs/qwen-color.png} Qwen-Agent & \makebox[2.2em][c]{25.00} \makebox[2.2em][c]{14.88} \makebox[2.2em][c]{4.86} & \makebox[2.2em][c]{26.50} \makebox[2.2em][c]{16.85} \makebox[2.2em][c]{6.45} & \makebox[2.2em][c]{3.57} \makebox[2.2em][c]{0.00} \makebox[2.2em][c]{0.00} & \makebox[2.2em][c]{-} \makebox[2.2em][c]{7.00} \makebox[2.2em][c]{1.32} & \makebox[2.2em][c]{-} \makebox[2.2em][c]{8.50} \makebox[2.2em][c]{2.10} & \makebox[2.2em][c]{-} \makebox[2.2em][c]{4.60} \makebox[2.2em][c]{0.76} \\
\includegraphics[height=1.0em]{figs/DeepCode.png} DeepCode & \makebox[2.2em][c]{3.57} \makebox[2.2em][c]{0.60} \makebox[2.2em][c]{0.00} & \makebox[2.2em][c]{4.50} \makebox[2.2em][c]{1.80} \makebox[2.2em][c]{0.95} & \makebox[2.2em][c]{0.00} \makebox[2.2em][c]{0.00} \makebox[2.2em][c]{0.00} & \makebox[2.2em][c]{-} \makebox[2.2em][c]{0.00} \makebox[2.2em][c]{0.00} & \makebox[2.2em][c]{-} \makebox[2.2em][c]{1.10} \makebox[2.2em][c]{0.85} & \makebox[2.2em][c]{-} \makebox[2.2em][c]{0.00} \makebox[2.2em][c]{0.00} \\
\bottomrule
\end{tabular}
}
\end{table*}

\begin{table}[htbp!]
\centering
\caption{DSR comparison by programming language.}
\label{tab:dsr_by_language}
\scalebox{0.7}{
\begin{tabular}{lccc}
\toprule
\textbf{Language} & \textbf{Success} & \textbf{Total} & \textbf{DSR} \\
\midrule
Java & 40 & 96 & 41.67\% \\
Python & 75 & 264 & 28.41\% \\
\midrule
\textbf{Overall} & \textbf{115} & \textbf{360} & \textbf{31.94\%} \\
\bottomrule
\end{tabular}
}
\end{table}

\paragraph{DSR and AC.} 
We evaluate the gap between code synthesis capability and system usability using DSR and AC (Table~\ref{tab:dsr_ac_combined}).
Our evaluation reveals a notable decoupling between the ability to generate code and the ability to produce a functional system, as detailed below:
(1) \textbf{Deployability:} Commercial IDEs substantially outperform open-source agents.
Copilot achieves 95.45\% DSR on Python and 62.50\% on Java, while Cursor demonstrates strong cross-language performance (68.18\% Python, 75.00\% Java).
In contrast, most open-source agents achieve below 20\% DSR.
As shown in Table~\ref{tab:dsr_by_language}, Java achieves higher overall DSR (41.67\%) than Python (28.41\%), attributable to its standardized build systems (Maven/Gradle).
Failure analysis reveals that agents frequently struggle with environment-specific configurations and framework-specific runtime errors.
(2) \textbf{Completeness:} While open-source agents such as MetaGPT achieve competitive API coverage, with Claude-powered MetaGPT reaching 73.04\% (Python) and 74.14\% (Java) matching commercial IDEs, their lower DSR reveals a critical \textit{AC-DSR gap}: these agents often generate broad API skeletons that fail to initialize properly.
This discrepancy underscores that repository-level generation requires both synthesis breadth and infrastructure precision.

\subsection{Fine-Tuning Qwen3 with \datasetname}
\label{sec:qwen_finetuning}

To further explore the potential of open-source models on \datasetname, we fine-tuned Qwen3-8B using high-quality agent trajectories.
We first extended the MS-Agent~\cite{li2024modelscope} to create \agentname, 
enhancing it with language-specific optimizations for repository-level microservice generation, 
including strict dependency handling and robust execution verification (see Appendix~\S\ref{sec:realrepoagent_details} for details of \agentname).

We employed \agentname with Qwen3-8B on the \datasetname (Train), excluding the verified subset.
We filtered for successful trajectories where the \agentname generated a complete, 
compilable, and executable repository, 
resulting in a dataset of 16,396 high-quality instruction-tuning samples.
We then fine-tuned Qwen3-8B on this distilled dataset for 3 epochs with a learning rate of $2\text{e-}4$
and evaluated the fine-tuned model on \datasetname (Verified).
As shown in Table~\ref{tab:sft_results}, 
the fine-tuned Qwen3-8B (\textit{\agentname-8B}) achieves performance comparable to GPT-5 mini.

\begin{table}[htbp!]
    \centering
    \caption{Performance comparison of \textbf{\agentname-8B} against GPT-5 mini on \datasetname.}
    \label{tab:sft_results}
    \resizebox{\columnwidth}{!}{
    \begin{tabular}{lcccccc}
    \toprule
    \multirow{2.5}{*}{\textbf{Model}} & \multicolumn{3}{c}{\textbf{Java}} & \multicolumn{3}{c}{\textbf{Python}} \\
    \cmidrule(lr){2-4} \cmidrule(lr){5-7}
    & \textbf{DSR} & \textbf{AC} & \textbf{Pass@1} & \textbf{DSR} & \textbf{AC} & \textbf{Pass@1} \\
    \midrule
    \includegraphics[height=1.0em]{figs/qwen-color.png} Qwen3-8B & 0.00\% & 16.38\% & 0.00\% & 4.55\% & 30.43\% & 0.00\% \\
    \includegraphics[height=1.0em]{figs/openai-fill.png} GPT-5 mini & 25.00\% & 17.24\% & \textbf{1.71\%} & 4.55\% & 60.87\% & \textbf{4.46\%} \\
    \textbf{\agentname-8B} & \textbf{37.12\%} & \textbf{46.55\%} & 1.69\% & \textbf{30.15\%} & \textbf{63.42\%} & 4.38\% \\
    \bottomrule
    \end{tabular}
    }
\end{table}

\subsection{Empirical Lessons}
\label{sec:case_study}

Through systematic analysis of successful and failed repository generation attempts, 
we identify key insights into model capabilities and limitations 
(detailed examples are provided in Appendix~\S\ref{sec:detailed_case_study}).

\noindent\textbf{Key Insights from Success Cases.} 
Successful repository generation exhibits five critical capabilities: 
(1) \textit{Complex requirement decomposition:} parsing multilingual specifications to extract intricate API schemas with compound parameter interactions;
(2) \textit{Proactive error handling:} implementing comprehensive input validation including type checking and edge case handling beyond explicit requirements;
(3) \textit{Algorithmic problem-solving:} correctly implementing domain-specific algorithms by selecting appropriate language primitives;
(4) \textit{Test-driven debugging:} identifying and fixing implementation errors through iterative test feedback;
and (5) \textit{Data structure optimization:} balancing performance and simplicity by selecting appropriate data structures for different operations.

\noindent\textbf{Key Insights from Failure Cases.} 
Three dominant failure patterns emerge from our analysis: 
(1) \textit{Infrastructure configuration failures:} 
missing or malformed deployment scripts (e.g., \texttt{start.sh}), 
prevent service initialization; 
(2) \textit{Framework version incompatibility:} 
agents frequently use deprecated API syntax (e.g., Javalin's parameter notation changes), 
causing runtime exceptions despite syntactically correct code; 
and (3) \textit{Phantom generation behavior:} 
certain configurations (e.g., Cursor with GPT-4.1) 
exhibit file persistence failures where generation appears 
successful but produces no artifacts, 
indicating critical tooling integration issues. 
Refer to Appendix~\S\ref{sec:cursor_gpt4_1} for more details.

\section{Related Works}
\label{sec:related_work}
\subsection{Microservices and Modern Software Architecture}
Microservices architecture decomposes applications into independently deployable services communicating via lightweight mechanisms, typically RESTful APIs~\cite{fowler2015microservices,richardson2018microservices,newman2021building}.
RESTful APIs define uniform interfaces for stateless communication~\cite{fielding2000architectural,masse2011rest}, while web frameworks such as Flask~\cite{grinberg2018flask} and Django~\cite{forcier2008python} provide abstractions for routing and request handling~\cite{neumann2018analysis}.
Generating such microservices thus demands both functional correctness and architectural coherence, making them a rigorous testbed for code generation agents.

\subsection{Code Generation}
Early benchmarks focus on function or class-level generation.
HumanEval~\cite{chen2021evaluating} comprises 164 hand-crafted Python programming problems with unit tests.
AiXBench~\cite{hao2022aixbench} targets Java with 336 problems (175 automated, 161 manual).
MultiPL-E~\cite{cassano2023multipl} extends HumanEval and MBPP~\cite{austin2021program} to 18 programming languages.
DS-1000~\cite{lai2023ds} introduces 1000 non-standalone problems across seven Python data science libraries (e.g. NumPy, Pandas, PyTorch, Scipy, Scikit-learn).
Concode~\cite{iyer2018mapping} provides over 100,000 Java problems from open-source projects.

Recent work has shifted toward repository-level and end-to-end software engineering.
Repository-aware methods such as A3-CodGen~\cite{A3CodGen}, RepoCoder~\cite{zhang2023repocoder}, and CatCoder~\cite{shen2023catcoder} leverage cross-file context.
Benchmarks targeting real-world GitHub issues (SWE-bench~\cite{jimenez2024swe}, SWE-PolyBench~\cite{zan2025multi}, SWE-bench Pro~\cite{jimenez2024swebenchpro}) and engineering agent systems (SWE-agent~\cite{yang2024swe}, MAGIS~\cite{tao2024magis}, OpenHands~\cite{wang2024openhands}) push models toward the closed loop of requirements, cross-file changes, testing, and delivery.
CodeS~\cite{zan2024codes} pioneers full repository generation from natural language with 19 Python tasks but relies on BLEU metrics.
RPG~\cite{luo2025rpg} integrates planning with 6 Python repositories but remains limited in scale and language coverage.
NL2Repo-Bench~\cite{ding2025nl2repo} extends to 104 Python tasks but evaluates solely on pytest pass rate.
Unlike these benchmarks, \datasetname is multilingual and evaluates end-to-end generation of complete microservice repositories from natural language requirements.

\section{Conclusion}
\label{sec:conclusion}
We present \datasetname, the first multilingual benchmark for repository-level web microservice generation from natural language requirements.
Comprising 106 repositories across 18 domains and 11 frameworks with 1,258 API endpoints and 2,335 test cases, \datasetname addresses a critical gap in evaluating complete repository generation capabilities.

Our comprehensive evaluation reveals several key findings.
First, even the best-performing system achieves only 23.67\% Pass@1, exposing fundamental challenges in microservice generation.
Second, we identify a critical \textit{AC-DSR gap}: open-source agents often generate broad API skeletons that fail to deploy, indicating that synthesis breadth alone is insufficient without infrastructure precision.
Third, fine-tuned \agentname-8B achieves performance comparable to GPT-5 mini, demonstrating dataset quality for model improvement.

\paragraph{Future Work.} 
Several directions emerge from our findings:
(1) extending the benchmark to additional languages (e.g., Golang) and architectural patterns (e.g., event-driven microservices);
(2) developing metrics that capture code quality aspects such as adherence to design patterns and maintainability;
and (3) investigating agent architectures that better bridge the AC-DSR gap through improved infrastructure reasoning.

\section*{Limitations}
We acknowledge several limitations of \datasetname that should be considered when interpreting results and planning future research.

\noindent\textbf{Microservice-Specific Scope.} 
\datasetname specifically targets RESTful web microservices, 
which constitute an important but not exhaustive category of software systems. 
Other architectural patterns such as event-driven systems, 
real-time applications with WebSocket support, GraphQL APIs, 
and gRPC services.

\noindent\textbf{Limited Language and Framework Coverage.} 
\datasetname currently focuses exclusively on web microservices 
implemented in Python (e.g., Flask, Django) and Java (e.g., Spring Boot). 
This represents a substantial portion of modern web development 
but does not cover other important programming paradigms such as compiled languages 
(i.e., C++, Go), mobile application development (i.e., Swift). 
We will extend \datasetname to encompass a broader range of programming languages, 
frameworks, and application types.

\noindent\textbf{Idealized Requirement Specifications.} 
\datasetname provides comprehensive, structured requirement documents (READMEs) as input. 
In real-world ``Zero2One'' scenarios, requirements are sometimes ambiguous, 
incomplete, or communicated iteratively. 
Our benchmark evaluates the core coding capability assuming clear specifications, 
leaving the challenge of requirements engineering and ambiguity resolution to future work.

\noindent\textbf{Code Maintainability and Standards.} 
Generated code is evaluated on its ability to pass tests, 
which does not guarantee adherence to best practices, 
industry standards, or long-term maintainability. 
The current metrics do not quantify aspects such as code readability, 
modular design, or idiomatic usage of frameworks. 
Future evaluations could incorporate static analysis tools or human review 
to assess these qualitative aspects of software engineering.

\section*{Ethics Statement}
\datasetname is constructed from publicly available microservice repositories and documentation and is intended for benchmarking end-to-end repository generation; our work involves no human-subject studies and we do not collect new personally identifiable information (PII). We follow upstream licenses and apply screening to reduce the inclusion of PII and overtly harmful/offensive content, but residual identifiers or societal/domain biases may remain in Internet-sourced artifacts. Repository-level code generation also carries dual-use risk: improvements in automated microservice creation could be misapplied to produce harmful software; we therefore frame \datasetname around correctness-focused evaluation (tests, coverage, deployment success) rather than deployment guidance, and recommend standard safeguards (e.g., human code review, sandboxed execution, and security scanning) for any downstream use. Finally, benchmark runs and fine-tuning consume compute and may have environmental impact; we limit unnecessary sweeps, reuse pretrained models, and report key settings to support reproducibility. All samples in \datasetname are manually reviewed by expert researchers in the field of code generation. 
We ensure that none of the samples contain private information or offensive content.


\bibliography{main}

\clearpage
\appendix
\section{Glossary}
\begin{tcolorbox}[colback=mygray, colframe=black!20, arc=2mm, boxrule=0.3pt, left=1mm, right=1mm, top=1mm, bottom=1mm]
  \textbf{Microservice:} A software architectural style that structures an application as a collection of independently deployable services, each running in its own process and communicating through lightweight protocols. \\[0.7em]
  \textbf{REST API:} Representational State Transfer Application Programming Interface. A web service that follows REST architectural constraints, using standard HTTP methods (GET, POST, PUT, DELETE) to enable stateless client-server communication. \\[0.7em]
  \textbf{API Endpoint:} A specific URL path and HTTP method combination that provides access to a particular functionality or resource of a web service. Each endpoint defines its input/output schema and behavior. \\[0.7em]
  \textbf{Zero2One Repository Generation:} The task of generating complete, functional software repositories from natural language requirements (e.g., README.md) without any existing codebase, 
  including all necessary source code files, configurations, dependencies, and project structure. \\[0.7em]
  \textbf{Black-box Testing:} A software testing methodology that evaluates system functionality without examining code structure, focusing solely on input-output behavior and API contracts. \\[0.7em]
  \textbf{Web Framework:} A software framework designed to support the development of web applications and services by providing standard components for routing, request handling, authentication, and data serialization (e.g., Flask, Django, FastAPI, Express.js). \\[0.7em]
  \textbf{Coding Agent:} A multi-agent AI system designed to autonomously generate code by coordinating specialized agents for tasks such as requirement understanding, architectural planning, code generation, and testing.
\end{tcolorbox}

\section{Task Definition}
\label{appendix:task_definition}
The repository generation task (i.e., README-to-repository code generation) evaluates the capability of large language models to create complete software repositories from natural language requirements. Unlike traditional code generation tasks that focus on completing individual functions or modifying existing code~\cite{guan2025repotransagent,yin2024rectifier,jimenez2024swe,A3CodGen,peng2025soleval,yin2024thinkrepair,yang2025input,peng2025preference}, repository generation requires models to perform end-to-end software development from scratch. We focus on web microservices, which represent the fundamental building block of contemporary software architecture where applications are decomposed into services exposing RESTful APIs. This architectural pattern is universal across modern software systems: mobile and web applications communicate with backend services through APIs, microservices interact through web interfaces, and cloud platforms fundamentally operate on API-driven service composition.

\noindent\textbf{Task Input.} The input consists of a requirement document (typically a README.md) that specifies the microservice requirements. 
The requirement document includes service functionality descriptions, API endpoint specifications with detailed input/output schemas, 
authentication mechanisms, error handling, data models, 
and operational constraints (e.g., port configurations and deployment requirements). 
Specifically, \texttt{pom.xml} is also included for Java and \texttt{requirements.txt} for Python. 
Refer to Appendix~\ref{sec:why_pom} for more details.

\noindent\textbf{Task Output.} The output is a complete repository that satisfies all specified requirements. 
This includes all source code files organized in a proper project structure, configuration files (e.g., requirements.txt), dependency specifications, 
database initialization scripts if needed, and any additional files required for deployment. 
The generated repository should be executable and pass comprehensive black-box test suites that validate functional correctness, 
API compliance, error handling, and business logic implementation. Specifically, 
to facilitate deployment and DSR measurement, 
we require all generated repositories to include a startup script that contains a single shell command to launch the service directly.

\noindent\textbf{Task Formulation.} 
Formally, given a requirement document $R$ that specifies the desired functionality, 
the task is to generate a repository $\mathcal{C} = \{c_1, c_2, \ldots, c_n\}$ where each $c_i$ represents a file in the repository. 
The generated repository $\mathcal{C}$ must satisfy the following conditions: 
(1) \textit{Functional Correctness}: all API endpoints specified in $R$ are correctly implemented and pass the test suite $\mathcal{T} = \{t_1, t_2, \ldots, t_m\}$, 
(2) \textit{Structural Completeness}: the repository contains all necessary files for deployment and execution, 
(3) \textit{API Compliance}: the implemented APIs conform to the input/output schemas specified in $R$, 
and (4) \textit{Error Handling}: the system appropriately handles edge cases and invalid inputs as specified in $R$.

\noindent\textbf{Evaluation Criteria.} 
We evaluate the generated repositories using three complementary metrics:
\begin{itemize}[leftmargin=*]
    \item \textbf{Pass@k}: A repository is considered successful if it passes all test cases in the test suite. This binary evaluation reflects real-world software development scenarios where partial implementations are typically insufficient for deployment. The Pass@k metric measures the probability that at least one of $k$ generated repositories passes all tests, providing a stringent evaluation of practical code generation capabilities. Detailed calculation is provided in \S\ref{sec:passk}.
    \item \textbf{API Coverage (AC)}: Measures the percentage of API endpoints successfully implemented with correct functionality relative to the requirements specification. This metric captures functional completeness even when the full test suite is not passed. See \S\ref{ss:appendix:ac} for details.
    \item \textbf{Deployment Success Rate (DSR)}: A binary metric indicating whether the generated repository can be successfully deployed and initialized without runtime errors. This captures the basic executability of the generated code. 
    See \S\ref{ss:appendix:dsr} for details.
\end{itemize}

\noindent\textbf{Task Challenges.} Repository generation from scratch presents several fundamental challenges. 
First, models must translate high-level natural language requirements into concrete software architectures, 
making critical design decisions about code organization, module decomposition, and component interactions. 
Second, models must ensure consistency across multiple files and components, maintaining coherent APIs, 
data models, and error handling. Third, 
models must handle the complexity of real-world software systems including authentication, 
authorization, database interactions, input validation, and error handling. 
These challenges distinguish repository generation from simpler function-level or file-level code generation tasks.

\section{Test Quality Assurance Details}
\label{sec:qa_details}

\subsection{Scoring Mechanism.}
Each committee member independently evaluates test cases on four dimensions (0-10 scale):
\begin{enumerate}[leftmargin=*, nosep]
    \item \textbf{Correctness (0-3)}: Syntax, runnability, and assertion accuracy.
    \item \textbf{Coverage \& Completeness (0-3)}: Happy paths, edge cases, and error conditions.
    \item \textbf{Readability \& Maintainability (0-2)}: Naming, formatting, and documentation.
    \item \textbf{Relevance (0-2)}: Alignment with repository purpose and README specifications.
\end{enumerate}

\section{Additional Experimental Details}
\label{s:appendix:experiments_details}

\subsection{Evaluated Agents}
\label{sec:coding_agents}
In this paper, we evaluate multiple coding agent frameworks and commercial IDEs on \datasetname.
The details are described as follows.

\begin{itemize}[leftmargin=*]
    \item \textbf{DeepCode}: DeepCode is a code generation platform based on a multi-agent system, specifically designed to transform complex algorithms and research papers into high-quality code. It integrates the Model Context Protocol (MCP) to achieve robust external tool calling capabilities. In our evaluation workflow, we utilized its central \texttt{CodeImplementationWorkflow} and configured it for the "pure code generation" mode. By pre-extracting key requirements from the project README and generating an initial implementation plan (\texttt{initial\_plan.txt}), DeepCode can autonomously construct the project structure and write source code. To optimize the presentation of results, we flattened the generated directory structure to ensure that all core files are located directly in the repository root.
    \item \textbf{MetaGPT}: MetaGPT is a framework that pioneers the introduction of Standardized Operating Procedures (SOPs) into multi-agent collaboration, handling complex tasks by simulating a software company's organizational structure (assigning roles such as Product Manager, Architect, and Engineer). In our experiments, we employed the \texttt{generate\_repo} interface of MetaGPT. Regarding configuration, we provided the README as the primary requirement input, set the maximum iterations to 5 (\texttt{n\_round=5}), and enabled mandatory code review (\texttt{code\_review=True}). Through customized prompts, we required MetaGPT to generate not only the complete business logic but also environment configuration files and one-click startup scripts, ensuring the generated projects are highly self-contained.
    \item \textbf{MS-Agent}: MS-Agent (ModelScope-Agent) is a lightweight and highly extensible agent framework that emphasizes autonomous exploration capabilities and provides full support for the MCP protocol. In our evaluation setup, we utilized its \texttt{LLMAgent} class and dynamically loaded model service parameters via \texttt{OmegaConf}. For repository generation tasks, we developed comprehensive implementation prompts to guide the agent in analyzing the README, designing the system architecture, and implementing all source files. During execution, MS-Agent automatically handles directory hierarchy creation and ensures that components like web services are correctly bound to the \texttt{0.0.0.0} interface to meet our automated testing environment requirements.
    \item \textbf{Qwen-Agent}: Qwen-Agent is an application development framework, featuring built-in efficient tools such as a code interpreter. We adopted its \texttt{Assistant} class as the core executor and equipped it with the \texttt{code\_interpreter} plugin. In practical calls, we assigned the agent the identity of a senior software engineer and required it to execute tasks described in the README following a sequence of analysis, design, coding, and verification. By integrating the code interpreter into the workflow, Qwen-Agent can check syntax and dependencies in real-time during generation, ultimately outputting a code repository with a \texttt{start.sh} entry point.
\end{itemize}

\subsection{Agent Prompts}
\label{sec:agent_prompts}
To ensure reproducibility, 
we provide the specific system prompts and configuration details used for each open-source agent. 
We strictly standardized the system prompt across all agents to ensure a fair comparison, 
while adapting the execution prompts to fit the specific interaction mode of each agent framework.

\subsubsection{Universal System Prompt}
All evaluated agents (MetaGPT, DeepCode, Qwen-Agent, MS-Agent) share the same system instruction to establish the persona and task requirements. 
The \texttt{{readme\_text}} and \texttt{{pom\_text}} placeholders are replaced with the actual content of the repository's requirement document and \texttt{pom.xml} file, respectively.

\begin{tcolorbox}[colback=gray!5, colframe=black!40, title=\textbf{System Prompt (Java)}, fonttitle=\bfseries, arc=2mm, boxrule=0.5pt]
\small
You are a senior Java software engineer tasked with implementing a complete software project.

Project Requirements: \\
README: \\
\{readme\_text\}

POM.xml (reference): \\
\{pom\_text\}

Your task:
\begin{enumerate}[leftmargin=*]
    \setlength\itemsep{0em}
    \item Analyze the README and referenced POM to understand the project requirements.
    \item Design the complete project structure and architecture.
    \item Implement ALL necessary files including:
    \begin{itemize}
        \item Main application files
        \item Configuration files
        \item Dependencies/requirements files (pom.xml is required)
        \item Documentation files
        \item Any additional files needed for the project to run
    \end{itemize}
    \item Ensure the project can be started via a single shell command writen in a file named start.sh.
    \item The generated start.sh MUST:
    \begin{itemize}
        \item Listen on 0.0.0.0
        \item Use port specified in the README
        \item Use ONLY the correct command for the detected framework
    \end{itemize}
    \item If a web service is expected, bind to 0.0.0.0 and use the port specified in the README.
    \item Write production-ready, well-documented code.
    \item Use Maven.
\end{enumerate}

Important: Generate ALL files in the current working directory. Do not reference or peek at any tests directory.
\end{tcolorbox}

\begin{tcolorbox}[colback=gray!5, colframe=black!40, title=\textbf{System Prompt (Python)}, fonttitle=\bfseries, arc=2mm, boxrule=0.5pt]
\small
You are a senior software engineer tasked with implementing a complete software project.

Project Requirements: \\
README: \\
\{readme\_text\}

requirements.txt (reference): \\
\{requirements\_text\}

Your task:
\begin{enumerate}[leftmargin=*]
    \setlength\itemsep{0em}
    \item Analyze the README and referenced requirements to understand the project requirements.
    \item Design the complete project structure and architecture.
    \item Implement ALL necessary files including:
    \begin{itemize}
        \item Main application files
        \item Configuration files
        \item Dependencies/requirements files (requirements.txt is required)
        \item Documentation files
        \item Any additional files needed for the project to run
    \end{itemize}
    \item Ensure the project can be started via a single shell command writen in a file named start.sh.
    \item The generated start.sh MUST:
    \begin{itemize}
        \item Listen on 0.0.0.0
        \item Use a common port (e.g., 8000) or the one specified in the README
        \item Use ONLY the correct command for the detected framework
    \end{itemize}
    \item If a web service is expected, bind to 0.0.0.0 and use the port.
    \item Write production-ready, well-documented code.
\end{enumerate}

Important: Generate ALL files in the current working directory. Do not reference or peek at any tests directory.
\end{tcolorbox}

\subsubsection{Agent-Specific Configurations}

\noindent\textbf{MetaGPT.} 
We utilize the \texttt{generate\_repo} interface with the following parameters: 
\texttt{inc=False}, \texttt{implement=True}, \texttt{code\_review=True}, \texttt{n\_round=5}, and \texttt{investment=3.0}. 
The agent operates in a synchronous loop, refining the code through self-correction mechanisms. 
For Python repositories, we set \texttt{run\_tests=False} to align with our black-box generation protocol, 
while for Java, experimental verification required explicit test execution flags, though we disabled feedback utilization to maintain consistency.

\noindent\textbf{DeepCode.} 
We configure the \texttt{CodeImplementationWorkflow} in \texttt{pure\_code\_mode=True}. 
For Java repositories, we enable reading tools (\texttt{enable\_read\_tools=True}) to allow the agent to inspect the complex directory structure better. 
For Python, as the structure is flatter, we disable this feature (\texttt{enable\_read\_tools=False}) to reduce context overhead.

\noindent\textbf{MS-Agent.} 
We configure the \texttt{LLMAgent} with \texttt{temperature=1}. 
Due to the verbosity of repository generation, we set \texttt{max\_chat\_round=100}.

\noindent\textbf{Qwen-Agent.} 
We initialize the \texttt{Assistant} agent with the \texttt{code\_interpreter} tool. 
The other parameters are left to official default values.

\subsection{IDE Prompts}
\label{sec:ide_prompts}
We also employed commercial IDEs (Antigravity, Cursor, Copilot) for repository generation.
Below are the specific prompts used for Java and Python repositories, respectively.

\begin{tcolorbox}[colback=gray!5, colframe=black!40, title=\textbf{IDE Prompt (Java)}, fonttitle=\bfseries, arc=2mm, boxrule=0.5pt]
\small
You are an expert Java Developer and Software Architect.
Your task is to sequentially generate the complete source code for all Java repositories located in the directory:
\texttt{\{repo\_root\}}

Project Requirements (from README): \\
\{readme\_text\}

Your task:
\begin{enumerate}[leftmargin=*]
    \setlength\itemsep{0em}
    \item Analyze the README to understand the project requirements.
    \item Design the complete project structure and architecture.
    \item Implement ALL necessary files including:
    \begin{itemize}
        \item Main application files
        \item Configuration files
        \item Dependencies/requirements files
        \item Documentation files
        \item Any additional files needed for the project to run
    \end{itemize}
    \item Ensure the project can be started via a single shell command writen in a file named start.sh.
    \item If it's a web service, bind to 0.0.0.0 with the port specified in the README.
    \item Generate a run\_tests.sh file to run the tests.
    \item Write production-ready, well-documented code.
    \item Use Maven.
\end{enumerate}

Start by listing the subdirectories in \texttt{\{repo\_root\}} and then proceed with the generation for each one.
\end{tcolorbox}

\begin{tcolorbox}[colback=gray!5, colframe=black!40, title=\textbf{IDE Prompt (Python)}, fonttitle=\bfseries, arc=2mm, boxrule=0.5pt]
\small
You are a senior software engineer tasked with implementing a complete software project.
Your task is to sequentially generate the complete source code for all Python repositories located in the directory:
\texttt{\{repo\_root\}}

Project Requirements (from README): \\
\{readme\_text\}

Your task:
\begin{enumerate}[leftmargin=*]
    \setlength\itemsep{0em}
    \item Analyze the README to understand the project requirements.
    \item Design the complete project structure and architecture.
    \item Implement ALL necessary files including:
    \begin{itemize}
        \item Main application files
        \item Configuration files
        \item Dependencies/requirements files
        \item Documentation files
        \item Any additional files needed for the project to run
    \end{itemize}
    \item Ensure the project can be started via a single shell command writen in a file named start.sh.
    \item If it's a web service, bind to 0.0.0.0 with a common port.
    \item Write production-ready, well-documented code.
\end{enumerate}

Start by listing the subdirectories in \texttt{\{repo\_root\}} and then proceed with the generation for each one.
\end{tcolorbox}

\subsection{Evaluation Metrics}
\label{ss:appendix:metrics}

\subsubsection{Pass@k Calculation and Its Necessity for Estimation}
\label{sec:passk}

In this study, we adopt the Pass@k metric to evaluate the functional correctness of repository generation. The Pass@k metric has been widely used to assess the success rate of models in generating code that meets specified requirements~\cite{chen2021evaluating,yu2024codereval,benchmark2024sol}. Specifically, for each task, the model generates \( k \) repository samples per problem, and a problem is considered solved if at least one of the generated repositories passes all test cases. The overall Pass@k score is then calculated by evaluating the fraction of problems for which at least one repository passes.

While the basic Pass@k metric offers a straightforward measure of success, it can have a high variance when evaluating a small number of samples. To reduce this variance, we follow a more robust approach, as outlined by \citet{kulal2019spoc}. Instead of generating only \( k \) samples per task, we generate \( n \geq k \) samples for each problem (in this study, we set \( n = 10 \) and \( k \leq 10 \)). We then count the number of correct samples, denoted as \( c \), where each correct sample passes the unit tests. The unbiased estimator for Pass@k is computed as:
\begin{equation}
\text{Pass@}k := \mathop{\mathbb{E}}\limits_{\text{Requirements}} \left[ 1 - \frac{\binom{n-c}{k}}{\binom{n}{k}} \right],
\end{equation}

where \( \binom{n}{k} \) is the binomial coefficient, representing the number of ways to choose \( k \) successful samples from \( n \) generated samples.

The reason for estimating Pass@k using this method is to account for the inherent randomness and variance in code generation tasks. Generating multiple samples per task reduces the likelihood that the model's success rate is affected by outliers or variability in the generated code. By employing this unbiased estimator, we ensure that our Pass@k metric provides a more stable and reliable evaluation of the models' performance.

The estimation approach also helps mitigate the computational cost associated with calculating Pass@k directly for each possible subset of samples, which would be computationally expensive and inefficient, especially when evaluating a large number of tasks. Thus, the unbiased estimator allows us to balance the trade-off between accuracy and computational efficiency.

\begin{table}[t]
\centering
\caption{DSR and AC across IDE-model configurations (Full, including all models).}
\label{tab:dsr_ac_comprehensive_full}
\resizebox{\columnwidth}{!}{
\begin{tabular}{llccc}
\toprule
\textbf{IDE} & \textbf{Model} & \textbf{Language} & \textbf{DSR} & \textbf{AC} \\
\midrule
\multirow{8}{*}{\includegraphics[height=1.0em]{figs/githubcopilot.png} Copilot} & \multirow{2}{*}{\includegraphics[height=1.0em]{figs/openai-fill.png} GPT-5 mini} & Java & \colorbox{green!25}{\makebox[4.2em][c]{100.00\%}} & \colorbox{green!25}{\makebox[4.2em][c]{67.24\%}} \\
& & Python & \colorbox{green!25}{\makebox[4.2em][c]{90.91\%}} & \colorbox{green!25}{\makebox[4.2em][c]{70.43\%}} \\
& \multirow{2}{*}{\includegraphics[height=1.0em]{figs/openai-fill.png} GPT-5.1 Codex} & Java & \colorbox{olive!25}{\makebox[4.2em][c]{37.50\%}} & \colorbox{green!25}{\makebox[4.2em][c]{73.28\%}} \\
& & Python & \colorbox{olive!25}{\makebox[4.2em][c]{40.91\%}} & \colorbox{green!25}{\makebox[4.2em][c]{71.30\%}} \\
& \multirow{2}{*}{\includegraphics[height=1.0em]{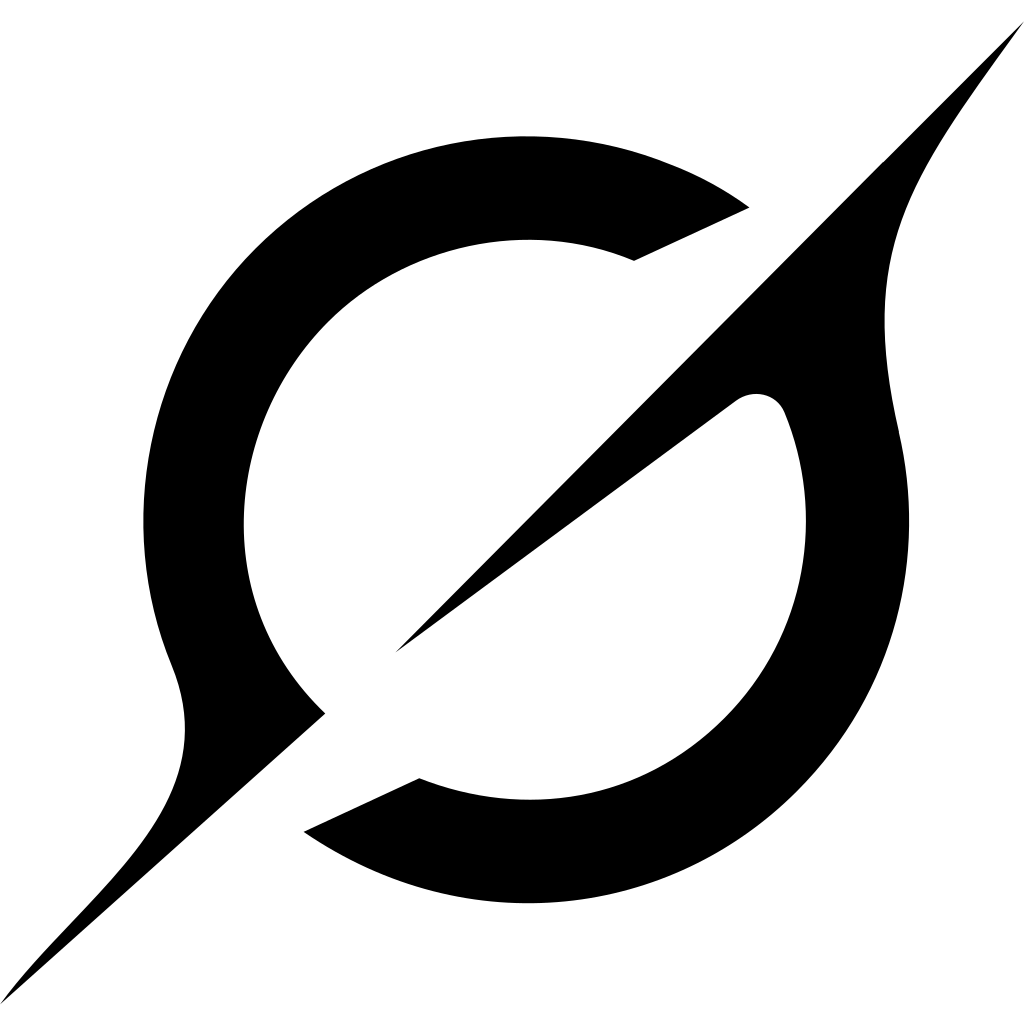} Grok Code Fast 1} & Java & \colorbox{red!25}{\makebox[4.2em][c]{12.50\%}} & \colorbox{olive!25}{\makebox[4.2em][c]{54.31\%}} \\
& & Python & \colorbox{red!25}{\makebox[4.2em][c]{9.09\%}} & \colorbox{green!25}{\makebox[4.2em][c]{66.96\%}} \\
& \multirow{2}{*}{\includegraphics[height=1.0em]{figs/claude-color.png} claude-sonnet-4.5} & Java & \colorbox{olive!25}{\makebox[4.2em][c]{62.50\%}} & \colorbox{green!25}{\makebox[4.2em][c]{68.10\%}} \\
& & Python & \colorbox{green!25}{\makebox[4.2em][c]{95.45\%}} & \colorbox{green!25}{\makebox[4.2em][c]{69.57\%}} \\
\midrule
\multirow{8}{*}{\includegraphics[height=1.0em]{figs/google.png} Antigravity} & \multirow{2}{*}{\includegraphics[height=1.0em]{figs/google.png} Gemini 3 Pro} & Java & \colorbox{olive!25}{\makebox[4.2em][c]{50.00\%}} & \colorbox{green!25}{\makebox[4.2em][c]{68.10\%}} \\
& & Python & \colorbox{red!25}{\makebox[4.2em][c]{0.00\%}} & \colorbox{green!25}{\makebox[4.2em][c]{70.43\%}} \\
& \multirow{2}{*}{\includegraphics[height=1.0em]{figs/google.png} Gemini 3 Flash} & Java & \colorbox{olive!25}{\makebox[4.2em][c]{37.50\%}} & \colorbox{olive!25}{\makebox[4.2em][c]{60.34\%}} \\
& & Python & \colorbox{red!25}{\makebox[4.2em][c]{4.55\%}} & \colorbox{olive!25}{\makebox[4.2em][c]{65.22\%}} \\
& \multirow{2}{*}{\includegraphics[height=1.0em]{figs/openai-fill.png} GPT-OSS 120B} & Java & \colorbox{red!25}{\makebox[4.2em][c]{12.50\%}} & \colorbox{olive!25}{\makebox[4.2em][c]{34.48\%}} \\
& & Python & \colorbox{red!25}{\makebox[4.2em][c]{0.00\%}} & \colorbox{olive!25}{\makebox[4.2em][c]{43.48\%}} \\
& \multirow{2}{*}{\includegraphics[height=1.0em]{figs/claude-color.png} claude-sonnet-4.5} & Java & \colorbox{olive!25}{\makebox[4.2em][c]{50.00\%}} & \colorbox{red!25}{\makebox[4.2em][c]{27.59\%}} \\
& & Python & \colorbox{red!25}{\makebox[4.2em][c]{13.64\%}} & \colorbox{olive!25}{\makebox[4.2em][c]{60.00\%}} \\
\midrule
\multirow{8}{*}{\includegraphics[height=1.0em]{figs/cursor.png} Cursor} & \multirow{2}{*}{Auto} & Java & \colorbox{red!25}{\makebox[4.2em][c]{25.00\%}} & \colorbox{red!25}{\makebox[4.2em][c]{25.86\%}} \\
& & Python & \colorbox{red!25}{\makebox[4.2em][c]{9.09\%}} & \colorbox{red!25}{\makebox[4.2em][c]{26.09\%}} \\
& \multirow{2}{*}{\includegraphics[height=1.0em]{figs/openai-fill.png} GPT-4.1} & Java & \colorbox{red!25}{\makebox[4.2em][c]{0.00\%}} & \colorbox{red!25}{\makebox[4.2em][c]{0.00\%}} \\
& & Python & \colorbox{red!25}{\makebox[4.2em][c]{0.00\%}} & \colorbox{red!25}{\makebox[4.2em][c]{0.00\%}} \\
& \multirow{2}{*}{\includegraphics[height=1.0em]{figs/google.png} Gemini 3 Flash} & Java & \colorbox{olive!25}{\makebox[4.2em][c]{37.50\%}} & \colorbox{olive!25}{\makebox[4.2em][c]{56.03\%}} \\
& & Python & \colorbox{red!25}{\makebox[4.2em][c]{9.09\%}} & \colorbox{olive!25}{\makebox[4.2em][c]{60.87\%}} \\
& \multirow{2}{*}{\includegraphics[height=1.0em]{figs/claude-color.png} claude-sonnet-4.5} & Java & \colorbox{green!25}{\makebox[4.2em][c]{75.00\%}} & \colorbox{green!25}{\makebox[4.2em][c]{68.97\%}} \\
& & Python & \colorbox{green!25}{\makebox[4.2em][c]{68.18\%}} & \colorbox{green!25}{\makebox[4.2em][c]{73.91\%}} \\
\bottomrule
\end{tabular}
}
\end{table}

\subsection{API Coverage (AC)}
\label{ss:appendix:ac}
As stated in \cite{luo2025rpg}, 
in repository-level code generation tasks, 
evaluating \textbf{functional coverage (Coverage)} is a key dimension for measuring model planning ability.
In our work, AC serves as the concrete implementation of this dimension within the Web API scenario. 
It measures the percentage of API endpoints successfully implemented with correct functionality specified in the requirement:
\begin{equation}
\text{AC} = \frac{1}{|\mathcal{R}|} \sum_{r \in \mathcal{R}} \frac{|\{\text{api} \in \mathcal{A}_r : \text{implemented}(\text{api}, \mathcal{C}_r)\}|}{|\mathcal{A}_r|},
\end{equation}
where $\mathcal{A}_r$ is the set of required API endpoints for repository $r$. 

\noindent\textbf{Deployment Independence.} 
Critically, AC is computed via static code analysis and does not require successful deployment or test execution. 
An endpoint is considered ``implemented'' if framework-specific routing signatures (e.g., \texttt{@app.get}, \texttt{@PostMapping}) 
and corresponding path identifiers are detected in the source code, excluding test directories. 
This design ensures AC remains computable even when DSR is low (deployment fails) or tests cannot execute, 
providing a robust measure of structural completeness orthogonal to runtime correctness.

\subsection{Deployment Success Rate (DSR)} 
\label{ss:appendix:dsr}
This binary metric indicates whether the generated repository can be successfully deployed and initialized without runtime errors:
\begin{equation}
\text{DSR} = \frac{|\{r \in \mathcal{R} : \text{deployable}(\mathcal{C}_r)\}|}{|\mathcal{R}|},
\end{equation}
where $\text{deployable}(\mathcal{C}_r)$ verifies that the repository can install dependencies, 
initialize the application, and start the server without errors.
This metric aligns with the \textbf{Build Success Rate} evaluated in continuous integration research~\cite{menon2024learning} 
and \textbf{Executability} in modern code generation benchmarks~\cite{nam2024repoexec}.

\section{Details of \agentname}
\label{sec:realrepoagent_details}

\agentname is constructed by extending the MS-Agent framework~\cite{li2024modelscope} with specialized strategies for handling the complexities of repository-level microservice generation. Unlike the generic code generation agent used in our baselines, \agentname incorporates language-specific optimizations for both Python and Java.

\paragraph{Python Agent Architecture}
The Python agent employs a "Dependency-First" reasoning process.
In \texttt{coding.yaml}, the system prompt is engineered to strictly enforce a two-stage generation process:
(1) \textbf{Dependency Analysis:} Before writing any code, the agent must explicitly list and read the implementations of all dependencies for the target file. 
This prevents the common "partial update" failure mode where agents modify a file without understanding its context.
(2) \textbf{Framework-Aware Generation:} The agent is explicitly instructed to choose between Flask (synchronous) and FastAPI (asynchronous) based on requirements, 
and to enforce a standardized \texttt{0.0.0.0} network binding for verifiability.

For execution verification (\texttt{eval\_callback.py}), 
the agent utilizes a robust runtime manager. 
It implements a recursive file search algorithm (\texttt{find\_file\_recursive}) to locate entry points (e.g., \texttt{start.sh} or \texttt{app.py}) regardless of the directory structure generated by the LLM. 
It also features an aggressive process management strategy that kills previous service instances between iterations to ensure clean port binding.

\paragraph{Java Agent Architecture}
The Java agent addresses the compilation-heavy nature of Java.
In \texttt{coding.yaml}, we implement a \textit{POM-driven Framework Detection} mechanism. 
The agent analyzes \texttt{pom.xml} to deterministically identify the underlying framework (Spring Boot, Quarkus, Micronaut, or plain Java) 
and generates the corresponding \texttt{start.sh} script (e.g., \texttt{mvn spring-boot:run} vs \texttt{java -jar}).
The execution callback (\texttt{eval\_callback.py}) is tuned for the JVM ecosystem, 
utilizing \texttt{mvn clean install -DskipTests} for rapid build validation and an extended 30-second timeout window to 
accommodate the strictly longer startup time of Java microservices compared to Python scripts.

\section{Benchmark Data Format}
\label{sec:data_format}

\subsection{Repository Structure}

\begin{table}[htbp!]
\centering
\caption{Statistics of the \datasetname Full. 
\textbf{Tests} denotes the number of test cases.}
\label{appendix:dataset-statistics}
\resizebox{\linewidth}{!}{
\begin{tabular}{lclccl}
\toprule
\textbf{Repository} & \textbf{Language} & \textbf{Framework} & \textbf{Tests} & \textbf{Stars} & \textbf{Domain} \\
\midrule
AIInferenceGateway & Python & Django & 97 & - & - \\
Blog & Python & - & 22 & - & Content Management \\
Chatroom & Python & - & 9 & - & Chat \\
Customization & Python & - & 64 & - & - \\
Data\_Rank\_Searcher & Python & - & 33 & - & Data Management \\
File\_Relay & Python & Eve & 46 & - & File Management \\
GameBackend & Python & Eve & 59 & - & Gaming \\
IoTDeviceManager & Python & Django & 123 & - & - \\
MobileBackendHub & Python & Django & 106 & - & - \\
MultiTenantSaaS & Python & Django & 57 & - & - \\
Multilingual & Python & Flask & 35 & - & - \\
NexiosSSOService & Python & Eve & 76 & - & - \\
NovaGatewayBench & Python & Eve & 3 & - & - \\
PaymentCallbackService & Python & - & 53 & - & - \\
RealTimePushGateway & Python & Eve & 12 & - & - \\
SSOAuthService & Python & Django & 107 & - & Authentication \\
SimpleFastPyAPI & Python & FastAPI & 2 & 26 & - \\
StructuredDataConvertor & Python & - & 38 & - & Data Management \\
TaskManagement & Python & Django & 44 & - & Task Management \\
Tic-Tac-Toe & Python & Django & 7 & - & Gaming \\
Timer4Tasker & Python & Flask & 53 & - & Task Management \\
UserManagement & Python & Django & 69 & - & - \\
UserManagement\_2 & Python & Flask & 32 & - & - \\
WebPan & Python & Eve & 26 & - & - \\
auto-doc-system & Python & Eve & 15 & - & - \\
dashboard-aggregator-service & Python & Eve & 13 & - & - \\
django-auto-doc & Python & Django & 18 & - & - \\
django-callback-gateway & Python & Django & 15 & - & - \\
django-ddd-framework & Python & Django & 7 & - & - \\
django-multitenant-billing & Python & Django & 16 & - & - \\
django-rbac-system & Python & Django & 7 & - & Authentication \\
django-rest-framework-crud & Python & Django & 2 & 288 & - \\
django-workflow-approval & Python & Django & 3 & - & - \\
docuforge-api & Python & Eve & 3 & - & - \\
eve & Python & Eve & 2 & 6.7k & - \\
eve-callback-gateway & Python & Eve & 15 & - & - \\
eve-ddd-framework & Python & Eve & 17 & - & - \\
eve-gray-release-manager & Python & Eve & 1 & - & - \\
eve-multitenant-billing & Python & Eve & 45 & - & - \\
eve-rbac-system & Python & Eve & 6 & - & Authentication \\
flask & Python & Flask & 7 & 70.7k & - \\
flask-callback-gateway & Python & Flask & 15 & - & - \\
flask-ddd-framework & Python & Flask & 17 & - & - \\
flask-gray-release-manager & Python & Flask & 1 & - & - \\
flask-multitenant-billing & Python & Flask & 31 & - & - \\
flask-workflow-approval & Python & Flask & 15 & - & - \\
javalin-chat-gateway & Java & Javalin & 15 & - & Chat \\
javalin-ci-status & Java & Javalin & 7 & - & - \\
javalin-course-scheduler & Java & Javalin & 21 & - & Education \\
javalin-dashboard-backend & Java & Javalin & 16 & - & - \\
javalin-ecommerce-mobile-backend & Java & Javalin & 29 & - & E-commerce \\
javalin-hello-world & Java & Javalin & 3 & - & - \\
javalin-medical-appointment & Java & Javalin & 15 & - & - \\
javalin-notification-service & Java & Javalin & 8 & - & - \\
javalin-online-judge & Java & Javalin & 23 & - & - \\
javalin-task-manager & Java & Javalin & 16 & - & Task Management \\
javalin-ticket-workflow & Java & Javalin & 16 & - & - \\
javalin-user-auth-platform & Java & Javalin & 52 & - & Authentication \\
mail\_service & Python & Flask & 55 & - & - \\
micronaut-blog-cms & Java & Micronaut & 16 & - & Content Management \\
micronaut-ci-status & Java & Micronaut & 26 & - & - \\
micronaut-config-push & Java & Micronaut & 21 & - & - \\
micronaut-dashboard-backend & Java & Micronaut & 16 & - & - \\
micronaut-iot-data-collector & Java & Micronaut & 24 & - & Data Management \\
micronaut-log-analytics-service & Java & Micronaut & 25 & - & - \\
micronaut-warehouse-inventory & Java & Micronaut & 16 & - & - \\
nexios & Python & - & 25 & - & - \\
nexios-ddd-framework & Python & Eve & 17 & - & - \\
nexios-form-engine & Python & Eve & 17 & - & - \\
nexios-gray-release-manager & Python & - & 1 & - & - \\
nexios-mobile-hub & Python & Eve & 5 & - & - \\
nexios-session-manager & Python & Eve & 14 & - & - \\
nexios\_realtime\_push\_benchmark & Python & Eve & 8 & - & - \\
online-exam-system-backend & Java & - & 8 & - & - \\
quarkus-blog-cms & Java & Quarkus & 16 & - & Content Management \\
quarkus-chat-gateway & Java & Quarkus & 16 & - & Chat \\
quarkus-config-center & Java & Quarkus & 21 & - & - \\
quarkus-log-analytics & Java & Quarkus & 25 & - & - \\
quarkus-online-judge & Java & Quarkus & 6 & - & - \\
quarkus-social-network & Java & Quarkus & 24 & - & - \\
quarkus-test-reporting & Java & Quarkus & 21 & - & - \\
quarkus-ticket-workflow & Java & Quarkus & 13 & - & - \\
quarkus-warehouse-inventory & Java & Quarkus & 16 & - & - \\
rock-paper-scissors-flask & Python & Flask & 5 & 6 & - \\
simple-rbac-service & Python & - & 9 & - & Authentication \\
spark & Java & Spark & 320 & - & - \\
spark-blog-cms & Java & Spark & 6 & - & Content Management \\
spark-dashboard-backend & Java & Spark & 16 & - & - \\
spark-iot-data-collector & Java & Spark & 23 & - & Data Management \\
spark-log-analytics & Java & Spark & 23 & - & - \\
spark-notification-service & Java & Spark & 5 & - & - \\
spark-personal-finance-manager & Java & Spark & 22 & - & - \\
spring-boot-course-scheduling & Java & Spring Boot & 19 & - & Education \\
spring-boot-digital-twin & Java & Spring Boot & 16 & - & - \\
spring-boot-user-auth & Java & Spring Boot & 7 & - & Authentication \\
springboot-chat-gateway & Java & Spring Boot & 13 & - & Chat \\
springboot-ci-status & Java & Spring Boot & 21 & - & - \\
springboot-task-manager & Java & Spring Boot & 25 & - & Task Management \\
springboot-test-reporting & Java & Spring Boot & 22 & - & - \\
springboot-ticket-workflow & Java & Spring Boot & 16 & - & - \\
springboot-warehouse-inventory & Java & Spring Boot & 15 & - & - \\
sql-father-backend-public & Java & Spring Boot & 5 & - & - \\
synapse & Python & Eve & 5 & 12k & - \\
workflow-approval-service & Python & Eve & 2 & - & - \\
\bottomrule
\end{tabular}
}
\end{table}

Each repository in \datasetname consists of the following components:

\noindent\textbf{Requirement Document (README.md):} A comprehensive specification document that describes the microservice functionality, API endpoints with detailed input/output schemas, authentication requirements, error handling specifications, and deployment constraints. This serves as the sole input to evaluated systems.

\noindent\textbf{Golden Oracle Implementation:} The reference implementation used for validation and as the basis for test generation. 
For real-world repositories, this is the original GitHub codebase. 
For expert-supervised repositories, this is the implementation generated by Gemini 3 Pro under expert supervision.

\noindent\textbf{Test Suites:} Comprehensive black-box test suites that validate functional correctness, API compliance, error handling, and business logic. Tests are organized into multiple test files covering different aspects of functionality.

\begin{table}[htbp!]
\centering
\caption{Detailed complexity metrics for difficulty classification (Verified Subset). LOC denotes lines of code, Complexity is cyclomatic complexity, Files is source file count, APIs is the number of API endpoints, and Score is the weighted composite score used for classification. GitHub repositories are marked with (*).}
\label{tab:difficulty-metrics}
\resizebox{\linewidth}{!}
{
\begin{tabular}{lccccccccc}
\toprule
\textbf{Repository} & \textbf{Lang.} & \textbf{LOC} & \textbf{Complexity} & \textbf{Files} & \textbf{APIs} & \textbf{Functions} & \textbf{Classes} & \textbf{Score} & \textbf{Difficulty} \\
\midrule
\multicolumn{10}{l}{\textit{Easy Repositories}} \\
rock-paper-scissors* & Py & 100 & 54 & 3 & 3 & 10 & 0 & 2.00 & Easy \\
flask* & Py & 194 & 40 & 9 & 3 & 15 & 2 & 2.40 & Easy \\
Chatroom & Py & 439 & 194 & 5 & 9 & 33 & 0 & 2.70 & Easy \\
Simple RBAC Service & Py & 638 & 163 & 5 & 4 & 35 & 2 & 2.80 & Easy \\
SimpleFastPyAPI* & Py & 101 & 40 & 6 & 5 & 10 & 3 & 3.00 & Easy \\
django-rest-framework* & Py & 403 & 80 & 22 & 10 & 30 & 5 & 3.20 & Easy \\
\midrule
\multicolumn{10}{l}{\textit{Medium Repositories}} \\
eve* & Py & 573 & 120 & 9 & 6 & 40 & 6 & 4.10 & Medium \\
synapse* & Py & 615 & 140 & 30 & 10 & 50 & 10 & 4.80 & Medium \\
Data Rank Searcher & Py & 1037 & 401 & 5 & 4 & 64 & 10 & 5.10 & Medium \\
Multilingual & Py & 1113 & 396 & 7 & 4 & 50 & 0 & 5.30 & Medium \\
Tic-Tac-Toe & Py & 897 & 222 & 24 & 6 & 68 & 27 & 5.70 & Medium \\
UserManagement Lite & Py & 1108 & 319 & 12 & 5 & 67 & 18 & 5.70 & Medium \\
Blog & Py & 1241 & 186 & 27 & 11 & 49 & 41 & 6.10 & Medium \\
TaskManagement & Py & 952 & 395 & 4 & 6 & 59 & 3 & 6.30 & Medium \\
StructuredDataConvertor & Py & 1580 & 309 & 13 & 3 & 45 & 14 & 6.40 & Medium \\
Javalin-online-judge & Java & 706 & 127 & 13 & 19 & 100 & 13 & 5.60 & Medium \\
Javalin-task-manager & Java & 656 & 108 & 10 & 24 & 82 & 10 & 5.10 & Medium \\
Micronaut-ci-status & Java & 719 & 135 & 13 & 8 & 121 & 18 & 4.70 & Medium \\
Quarkus-blog-cms & Java & 589 & 103 & 12 & 29 & 95 & 16 & 5.60 & Medium \\
Spark-dashboard-backend & Java & 718 & 119 & 14 & 25 & 100 & 14 & 5.60 & Medium \\
Spring-boot-course-scheduling & Java & 997 & 144 & 35 & 45 & 135 & 35 & 6.60 & Medium \\
Springboot-chat-gateway & Java & 631 & 106 & 11 & 15 & 94 & 11 & 5.10 & Medium \\
\midrule
\multicolumn{10}{l}{\textit{Hard Repositories}} \\
WebPan & Py & 1644 & 491 & 20 & 13 & 82 & 4 & 7.10 & Hard \\
Mail Service & Py & 1546 & 627 & 19 & 4 & 87 & 21 & 8.60 & Hard \\
Timer4Tasker & Py & 1812 & 606 & 19 & 9 & 96 & 28 & 9.00 & Hard \\
GameBackend & Py & 1834 & 624 & 9 & 10 & 113 & 16 & 9.10 & Hard \\
Customization & Py & 2426 & 702 & 26 & 10 & 129 & 28 & 9.50 & Hard \\
UserManagement & Py & 2416 & 753 & 13 & 8 & 124 & 15 & 9.50 & Hard \\
File Relay & Py & 2034 & 503 & 23 & 10 & 103 & 23 & 10.00 & Hard \\
Javalin-user-auth-platform & Java & 1839 & 199 & 18 & 13 & 140 & 20 & 7.10 & Hard \\
\bottomrule
\end{tabular}
}
\end{table}

\subsection{Difficulty Classification Methodology}
\label{sec:difficulty_methodology}
To enable fine-grained analysis of model capabilities, we classify repositories into three difficulty levels (Easy, Medium, Hard) based on five code complexity metrics computed from the golden oracle implementations. This methodology is inspired by the GAIA benchmark~\cite{mialon2023gaia} but adapted for repository-level software engineering tasks.

\subsubsection{Classification Metrics}
We employ a weighted composite scoring system based on the following metrics:
\begin{itemize}[leftmargin=*]
    \item \textbf{Cyclomatic Complexity (30\%)}: Measures the number of linearly independent paths through the code. We sum the complexity of all control flow structures (e.g., if, while, for, try-except blocks). This is weighted highest as it best correlates with the logical difficulty of the implementation.
    \item \textbf{Lines of Code (LOC) (25\%)}: Total count of non-comment, non-blank lines of code. This proxies the implementation scale and volume.
    \item \textbf{File Count (15\%)}: The number of source files, representing the structural complexity and modularity requirements of the project.
    \item \textbf{API Endpoints (15\%)}: The count of exposed RESTful endpoints, indicating the breadth of the interface design.
    \item \textbf{Functions \& Classes (15\%)}: The total number of defined functions and classes, reflecting the code organization complexity.
\end{itemize}

\subsubsection{Scoring Algorithm}
Each repository is assigned a difficulty score $S$ calculated using a step-function based on thresholds for each metric. The scoring logic is defined as follows:

\begin{equation}
S = \sum_{m \in \text{Metrics}} w_m \cdot \text{score}(m)
\end{equation}

Where $w_m$ is the weight of metric $m$, and $\text{score}(m)$ assigns points (0.3 to 3.0) based on metric magnitude. For example, for Cyclomatic Complexity ($C$):
\[
\text{score}(C) = 
\begin{cases} 
3.0 & \text{if } C > 500 \\
2.0 & \text{if } 300 < C \le 500 \\
1.2 & \text{if } 200 < C \le 300 \\
0.6 & \text{else}
\end{cases}
\]
Similar thresholds apply to other metrics. The final score ranges approximately from 2.0 to 10.0.

\subsubsection{Difficulty Thresholds}
Based on the composite score, we categorize repositories into:
\begin{itemize}
    \item \textbf{Easy ($S < 4.5$)}: Simple services with basic CRUD operations, typically single-file or minimal structure (e.g., Rock-Paper-Scissors).
    \item \textbf{Medium ($4.5 \le S < 7.0$)}: Moderate complexity services requiring authentication, data validation, and multi-file organization (e.g., Blog, TaskManagement).
    \item \textbf{Hard ($S \ge 7.0$)}: Complex systems with intricate state management, advanced business logic, and extensive error handling (e.g., File Relay, Customization).
\end{itemize}

\subsection{Why using pom.xml for Java repositories?}
\label{sec:why_pom}
The observation that including build configuration files (e.g., \texttt{pom.xml}) significantly enhances the pass rate of generated test cases in Java, 
while having a relatively milder effect on Python, 
can be attributed to fundamental differences in language runtime characteristics, standard library completeness, and framework design paradigms.

\paragraph{The Compilation Barrier vs. Runtime Interpretation}
The primary differentiator lies in the strictness of the execution pipeline. 
Java, being a statically typed, compiled language, 
enforces a ``compilation wall.'' Without access to the \texttt{pom.xml}, 
a Code-LLM is prone to \textit{dependency hallucination}, where it imports libraries (e.g., \texttt{Gson} vs. \texttt{Jackson}) that are semantically valid but absent from the project's actual dependency tree. 
In Java, a single unresolved import results in a compilation failure (\texttt{ClassNotFoundException}), 
preventing the test runner from executing any test cases. 
Conversely, Python is dynamically interpreted. 
Even if an agent generates code with incorrect dependencies (e.g., missing import inside function/lazy loading/conditional branch), 
the interpreted nature allows the runtime to execute code partially. 
Unless the execution flow specifically hits the missing dependency, 
the core logic may still execute, 
allowing a subset of test cases to pass (a phenomenon known as ``partial execution robustness'').

\paragraph{Ecosystem Fragmentation vs. The ``Batteries-Included'' Philosophy}
The probability space for library selection differs significantly between the two languages. 
Java relies heavily on third-party libraries for fundamental operations such as JSON processing, 
HTTP networking and testing. Without the \texttt{pom.xml} to constrain the search space, 
the model must guess between competing standards (e.g., \texttt{JUnit 4} vs. \texttt{JUnit 5}, 
\texttt{Mockito} vs. \texttt{PowerMock}), leading to high incompatibility rates. In contrast, 
Python follows a ``batteries-included'' philosophy. 
Essential utilities (e.g., \texttt{json}, \texttt{unittest}) are intrinsic to the standard library. 
Consequently, a Python-targeted agent can rely on standard libraries with high confidence, 
reducing the reliance on external environment specifications (e.g., requirements.txt) for basic functionality.

\paragraph{Conclusion}
In summary, the \texttt{pom.xml} in Java is not merely a dependency list but a \textit{structural definition of the runtime environment}. 
Its absence in Java generation tasks introduces a stochastic element that breaks the compilation chain, 
whereas Python's dynamic nature and standardized ecosystem allow for greater resilience in low-context generation scenarios.

\section{Framework-level Analysis}
Framework-level analysis (Table~\ref{tab:java_dsr_framework}) shows Spring Boot achieves perfect deployability (100\%), 
followed by Javalin (93.75\%), while Quarkus and Spark lag at 25.00\%. 
This correlates with framework design: Spring Boot's convention-over-configuration facilitates deployment, 
whereas Quarkus's native compilation and Spark's minimalism introduce additional failure points.

\begin{table}[t]
\centering
\caption{Java repository DSR by web framework. 
Spring Boot and Javalin demonstrate superior deployability compared to Quarkus and Spark.}
\label{tab:java_dsr_framework}
\begin{tabular}{lccc}
\toprule
\textbf{Framework} & \textbf{Success} & \textbf{Total} & \textbf{DSR} \\
\midrule
Spring Boot & 8 & 8 & 100.00\% \\
Javalin & 15 & 16 & 93.75\% \\
Micronaut & 2 & 4 & 50.00\% \\
Quarkus & 1 & 4 & 25.00\% \\
Spark & 1 & 4 & 25.00\% \\
\bottomrule
\end{tabular}
\end{table}

\section{Hardware and Software Environment}
\label{sec:hardware_software}
\subsection{Hardware and Testing Environment} 
Experiments were conducted across diverse platforms tailored to specific tasks. 
Non-training components, including evaluation and preprocessing, 
were executed on a MacBook Pro equipped with an Apple M4 Pro chip, 
24GB of memory, and 512GB of SSD storage. 
Supervised Fine-Tuning (SFT) of the models was performed on a server utilizing two NVIDIA A100 (80GB) GPUs.
Agent inference was carried out via the Azure API, 
spanning an operational period of approximately two weeks.

\subsection{Testing Environment}
\label{sec:testing}
All evaluations are conducted on a standardized testing environment with Ubuntu 20.04 LTS, 
Python 3.9, Node.js 18.x, and Conda 25.5.1. 
Each generated repository is tested in isolation within a fresh conda environment to 
ensure environment consistency and prevent cross-contamination between tests. 
Test execution is limited to 5 minutes per repository to prevent indefinite hangs.

\section{Metric Implementation Details}
\label{sec:metric_details}

\subsection{API Coverage Calculation}
To ensure rigorous evaluation of functional completeness, our API Coverage (AC) metric relies on a strict dual-verification process: (1) extracting required endpoints from the README, and (2) verifying their implementation in the generated codebase using static analysis.

\paragraph{Endpoint Extraction}
We employ a multi-pattern regex strategy to robustly extract API definitions from heterogeneous README formats:
\begin{itemize}[leftmargin=*]
    \item \textbf{Explicit Format}: Matches standard definitions like \texttt{POST /api/users - Create user} using the pattern \texttt{(METHOD)\textbackslash s+(/path)\dots}.
    \item \textbf{Table Format}: Parses markdown tables (e.g., \texttt{| GET | /users |}) to extract method-path pairs.
    \item \textbf{Feature Lists}: For less structured documents, extracts high-level feature descriptions (e.g., "- User Login") using keyword matching patterns like \texttt{(Login|Register|Create|...)}.
\end{itemize}

\paragraph{Implementation Verification}
To confirm implementation without requiring successful runtime execution (which is captured by DSR), we search for framework-specific signatures in the source code. The verification system recognizes patterns across major frameworks:
\begin{itemize}
    \item \textbf{Python (Flask/FastAPI)}: Decorators such as \texttt{@app.get}, \texttt{@router.post}, or \texttt{methods=["GET"]}.
    \item \textbf{Java (Spring Boot)}: Annotations like \texttt{@GetMapping}, \texttt{@PostMapping}, or \texttt{@RequestMapping}.
    \item \textbf{Node.js (Express)}: Route handlers like \texttt{app.get(\dots)} or \texttt{router.post(\dots)}.
\end{itemize}
An endpoint is considered "implemented" only if both the specific method/decorator and the corresponding path signature are found within the repository files, excluding test directories.

\subsection{Evaluation Infrastructure}
To ensure consistent and fair benchmarking across diverse agents, our evaluation harness implements strict isolation and lifecycle management:

\paragraph{Process Lifecycle Management}
Service startup is managed using explicit process groups (\texttt{os.setsid}). This is critical because many generated microservices spawn child processes (e.g., Gunicorn workers or JVM subprocesses). Simple process termination often leaves orphaned workers binding to ports, causing subsequent tests to fail with "Address already in use". Our harness sends \texttt{SIGTERM} to the entire process group (\texttt{os.killpg}), ensuring a clean teardown of the entire service tree between evaluations.

\paragraph{Startup Latency Handling}
We implement a standardized 10-second warm-up period after invoking the \texttt{start.sh} script. This accounts for the initialization overhead of heavier frameworks (especially Spring Boot in Java), preventing false negatives where tests might execute before the application server is fully ready to accept connections.

\section{Detailed Case Study}
\label{sec:detailed_case_study}

This section provides detailed examples of successful and failed repository generation attempts referenced in Section~\ref{sec:case_study}.

\subsection{Success Case: Data\_Rank\_Searcher}

We examine the Data\_Rank\_Searcher repository generation, 
where the agent (i.e., Copilot with claude-sonnet-4.5) achieved 100\% functional correctness (11/11 tests passed). 
The repository implements a sophisticated data management service with pagination, 
multi-field sorting, exact search, 
and fuzzy query capabilities across 179 lines of code. 
Key success factors include: 
(1) \textit{Complex requirement decomposition}: 
the agent parsed the Chinese-language README to extract 4 RESTful endpoints with 8 query parameters 
(i.e., page, page\_size, sort\_by, sort\_order, search\_field, search\_value, fuzzy\_field, fuzzy\_value), 
correctly inferring the interaction logic between pagination and filtering; 
(2) \textit{Proactive error handling}: 
the agent implemented comprehensive input validation including type checking 
(score must be numeric), boundary conditions (page\_size max 100), 
and edge cases (empty result sets), 
without explicit specification in the README; 
(3) \textit{Algorithmic problem-solving}: for fuzzy query functionality, 
the agent correctly implemented case-insensitive substring matching using Python's \texttt{lower()} method, 
while maintaining exact search through direct equality comparison; 
(4) \textit{Test-driven debugging}: when initial tests revealed pagination calculation errors 
(off-by-one in total\_pages), 
the agent immediately identified the formula \texttt{(total + page\_size - 1) // page\_size} 
and applied it consistently across all 30 repositories; 
(5) \textit{Data structure optimization}: 
the agent selected dictionary-based in-memory storage for O(1) lookups while maintaining list conversion for sorting operations, 
balancing performance with implementation simplicity. 
The implementation correctly handles compound queries (e.g., fuzzy search + sorting + pagination) 
and returns properly structured JSON responses with nested pagination metadata, 
demonstrating the agent's capability to implement multi-faceted data management systems from natural language specifications.

\subsection{Common Failure Modes}

We analyze recurring failure patterns that hinder successful repository-level generation. 
First, structural deficiencies are prevalent; for instance, DeepCode frequently fails to generate a valid \texttt{start.sh} entry point, 
preventing the environment from initializing the service. 
Second, version-specific API changes often lead to runtime exceptions. 
A notable example is the transition in Javalin 4, 
where the framework switched from colon-prefixed parameters (e.g., \texttt{:id}) to brace-enclosed parameters (e.g., \texttt{\{id\}}). 
Agents frequently use the legacy syntax, 
resulting in a \texttt{java.lang.IllegalArgumentException} during path registration and subsequent functional failure. 
More surprisingly, we observed a peculiar phenomenon when utilizing Cursor with GPT-4.1: 
the system ostensibly performs the code generation process but fails to persist any files to the storage. 
This "phantom generation" behavior leads to the categorical failure of the repository construction task.

\subsubsection{How Cursor with GPT-4.1 fails to generate a repository?}
\label{sec:cursor_gpt4_1}

We initiated the generation process using the standardized prompts detailed in \autoref{sec:ide_prompts}. To evaluate the autonomy of the systems, we adopted a consistent interaction protocol: whenever an IDE requested human clarification or decision-making, we provided a uniform response---``Please decide on your own''---to mandate autonomous execution. While other IDE-LLM combinations (e.g., Copilot with GPT-4o, Antigravity) successfully interpreted this instruction as a directive to proceed with implementation, Cursor coupled with GPT-4.1 exhibited a critical failure mode.

As illustrated in Figure~\ref{fig:cursor_failure_log}, rather than transitioning to the coding phase, the agent trapped itself in a recursive loop of planning and acknowledgment. It repeatedly affirmed its intent to ``proceed autonomously'' and ``execute the plan'' (e.g., ``I will continue to execute the plan entirely on my own initiative''), yet failed to perform any actual file system operations or code synthesis (``phantom generation'').

We hypothesize two primary causes for this behavior:
\begin{enumerate}
    \item \textbf{Workflow Misalignment in Zero2One Generation:} The IDE's agentic workflow, designed primarily for incremental edits, may lack the robust state management required to handle the open-ended nature of creating a repository from scratch based solely on natural language. The system likely anticipates a specific confirmation signal that the model fails to generate in this context.
    \item \textbf{Deficiency in Autonomous Instruction Following:} The underlying GPT-4.1 model appears to misinterpret the autonomy-granting instruction. Instead of mapping ``decide on your own'' to executable actions (i.e., invoking file-writing tools), it treats the input as conversational filler, resulting in a verbal loop of compliance without functional execution.
\end{enumerate}

In Figure~\ref{fig:cursor_failure_log}, we present the interaction log where Cursor with GPT-4.1 falls into this repetitive acknowledgment loop.

\section{Additional Experiments}
\label{sec:appendix:additional_results}

To comprehensively evaluate repository-level generation capabilities, we conduct extended experiments with additional model configurations beyond those reported in the main text.
For open-source coding agents (MS-Agent, MetaGPT, Qwen-Agent, DeepCode), we evaluate six foundation models: GPT-5 mini, GPT-5.1, claude-haiku-4.5, claude-sonnet-4.5, Qwen3-30B, and Qwen3-Coder-30B.
For commercial IDEs, model availability is constrained by each platform's official support.
We strictly evaluate only officially supported models for each IDE to ensure realistic deployment scenarios: Copilot supports GPT-5 mini, GPT-5.1 Codex, Grok Code Fast 1, and claude-sonnet-4.5; Antigravity supports Gemini 3 Pro, Gemini 3 Flash, GPT-OSS 120B, and claude-sonnet-4.5; Cursor supports Auto, GPT-4.1, Gemini 3 Flash, and claude-sonnet-4.5.
This approach reflects real-world usage constraints where developers must select from IDE-specific model rosters.

This section provides comprehensive experimental results including additional model configurations and detailed breakdowns by difficulty level, language, and IDE configurations.

\paragraph{Extended Difficulty-Level Analysis.}
As shown in Table~\ref{tab:difficulty_results}, we present a comprehensive breakdown of Pass@1 performance across all six evaluated models and three difficulty levels.
We observe several key trends:
(1) \textbf{Model Hierarchy:} claude-sonnet-4.5 consistently outperforms other models across all difficulty levels, achieving 26.50\% (Easy), 16.85\% (Medium), and 6.45\% (Hard) on Python with Qwen-Agent. claude-haiku-4.5 demonstrates comparable performance with minimal degradation (25.00\%/15.48\%/5.21\% on Python).
(2) \textbf{Specialized vs. General Models:} Qwen3-Coder-30B, despite being code-specialized, shows mixed results. It achieves 7.14\% (Easy Python) with MS-Agent but underperforms Qwen3-30B on Java tasks (1.70\% vs. 1.40\% on Medium with MS-Agent).
(3) \textbf{Severe Performance Decay:} The performance gap between Easy and Hard repositories is dramatic. For Qwen-Agent with claude-sonnet-4.5, Python performance drops from 26.50\% to 6.45\% (a 75.7\% relative decline). Most critically, smaller models exhibit near-zero performance on Hard tasks, with Qwen3-30B achieving 0.00\% on Python Hard across all agents.

\begin{table*}[htbp]
\centering
\caption{Performance (Pass@1) across difficulty levels on verified subsets. 
Easy (E), Medium (M), and Hard (H) classifications are based on code complexity metrics. 
Results demonstrate consistent performance degradation as repository complexity increases. 
Java repositories lack Easy-level instances. Qwen3-30B represents Qwen3-30B-A3B-Instruct-2507. Qwen3-Coder 30B represents Qwen3-Coder-30B-A3B-Instruct.}
\label{tab:difficulty_results}
\resizebox{\textwidth}{!}{
\begin{tabular}{lcccccccccccc}
\toprule
& \multicolumn{6}{c}{\textbf{Python}} & \multicolumn{6}{c}{\textbf{Java}} \\
\cmidrule(lr){2-7} \cmidrule(lr){8-13}
\textbf{Method} & \textbf{GPT-5 mini} & \textbf{GPT-5.1} & \textbf{claude-haiku-4.5} & \textbf{claude-sonnet-4.5} & \textbf{Qwen3-Coder 30B} & \textbf{Qwen3-30B} & \textbf{GPT-5 mini} & \textbf{GPT-5.1} & \textbf{claude-haiku-4.5} & \textbf{claude-sonnet-4.5} & \textbf{Qwen3-Coder 30B} & \textbf{Qwen3-30B} \\
& \textbf{(E/M/H)} & \textbf{(E/M/H)} & \textbf{(E/M/H)} & \textbf{(E/M/H)} & \textbf{(E/M/H)} & \textbf{(E/M/H)} & \textbf{(E/M/H)} & \textbf{(E/M/H)} & \textbf{(E/M/H)} & \textbf{(E/M/H)} & \textbf{(E/M/H)} & \textbf{(E/M/H)} \\
\midrule
\includegraphics[height=1.0em]{figs/modelscope-color.png} MS-Agent & \makebox[2.2em][c]{8.93} \makebox[2.2em][c]{5.24} \makebox[2.2em][c]{2.08} & \makebox[2.2em][c]{16.07} \makebox[2.2em][c]{9.52} \makebox[2.2em][c]{3.47} & \makebox[2.2em][c]{17.86} \makebox[2.2em][c]{9.76} \makebox[2.2em][c]{3.47} & \makebox[2.2em][c]{18.95} \makebox[2.2em][c]{10.45} \makebox[2.2em][c]{4.12} & \makebox[2.2em][c]{7.14} \makebox[2.2em][c]{4.64} \makebox[2.2em][c]{2.78} & \makebox[2.2em][c]{5.36} \makebox[2.2em][c]{4.17} \makebox[2.2em][c]{2.08} & \makebox[2.2em][c]{-} \makebox[2.2em][c]{1.90} \makebox[2.2em][c]{0.38} & \makebox[2.2em][c]{-} \makebox[2.2em][c]{5.30} \makebox[2.2em][c]{0.98} & \makebox[2.2em][c]{-} \makebox[2.2em][c]{5.50} \makebox[2.2em][c]{0.62} & \makebox[2.2em][c]{-} \makebox[2.2em][c]{6.85} \makebox[2.2em][c]{1.50} & \makebox[2.2em][c]{-} \makebox[2.2em][c]{1.70} \makebox[2.2em][c]{0.50} & \makebox[2.2em][c]{-} \makebox[2.2em][c]{1.40} \makebox[2.2em][c]{0.20} \\
\includegraphics[height=1.0em]{figs/metagpt.png} MetaGPT & \makebox[2.2em][c]{0.00} \makebox[2.2em][c]{0.60} \makebox[2.2em][c]{0.00} & \makebox[2.2em][c]{3.57} \makebox[2.2em][c]{1.79} \makebox[2.2em][c]{0.00} & \makebox[2.2em][c]{3.57} \makebox[2.2em][c]{1.79} \makebox[2.2em][c]{0.00} & \makebox[2.2em][c]{4.80} \makebox[2.2em][c]{2.65} \makebox[2.2em][c]{1.20} & \makebox[2.2em][c]{0.00} \makebox[2.2em][c]{0.60} \makebox[2.2em][c]{0.00} & \makebox[2.2em][c]{0.00} \makebox[2.2em][c]{0.30} \makebox[2.2em][c]{0.00} & \makebox[2.2em][c]{-} \makebox[2.2em][c]{0.00} \makebox[2.2em][c]{0.00} & \makebox[2.2em][c]{-} \makebox[2.2em][c]{1.30} \makebox[2.2em][c]{0.02} & \makebox[2.2em][c]{-} \makebox[2.2em][c]{1.35} \makebox[2.2em][c]{0.23} & \makebox[2.2em][c]{-} \makebox[2.2em][c]{2.15} \makebox[2.2em][c]{1.05} & \makebox[2.2em][c]{-} \makebox[2.2em][c]{0.00} \makebox[2.2em][c]{0.00} & \makebox[2.2em][c]{-} \makebox[2.2em][c]{0.00} \makebox[2.2em][c]{0.00} \\
\includegraphics[height=1.0em]{figs/qwen-color.png} Qwen-Agent & \makebox[2.2em][c]{3.57} \makebox[2.2em][c]{0.00} \makebox[2.2em][c]{0.00} & \makebox[2.2em][c]{25.00} \makebox[2.2em][c]{14.88} \makebox[2.2em][c]{4.86} & \makebox[2.2em][c]{25.00} \makebox[2.2em][c]{15.48} \makebox[2.2em][c]{5.21} & \makebox[2.2em][c]{26.50} \makebox[2.2em][c]{16.85} \makebox[2.2em][c]{6.45} & \makebox[2.2em][c]{3.57} \makebox[2.2em][c]{0.00} \makebox[2.2em][c]{0.00} & \makebox[2.2em][c]{3.57} \makebox[2.2em][c]{0.00} \makebox[2.2em][c]{0.00} & \makebox[2.2em][c]{-} \makebox[2.2em][c]{5.80} \makebox[2.2em][c]{0.52} & \makebox[2.2em][c]{-} \makebox[2.2em][c]{7.00} \makebox[2.2em][c]{1.32} & \makebox[2.2em][c]{-} \makebox[2.2em][c]{7.20} \makebox[2.2em][c]{1.20} & \makebox[2.2em][c]{-} \makebox[2.2em][c]{8.50} \makebox[2.2em][c]{2.10} & \makebox[2.2em][c]{-} \makebox[2.2em][c]{5.40} \makebox[2.2em][c]{1.00} & \makebox[2.2em][c]{-} \makebox[2.2em][c]{4.60} \makebox[2.2em][c]{0.76} \\
\includegraphics[height=1.0em]{figs/DeepCode.png} DeepCode & \makebox[2.2em][c]{0.00} \makebox[2.2em][c]{0.00} \makebox[2.2em][c]{0.00} & \makebox[2.2em][c]{3.57} \makebox[2.2em][c]{0.60} \makebox[2.2em][c]{0.00} & \makebox[2.2em][c]{3.57} \makebox[2.2em][c]{0.60} \makebox[2.2em][c]{0.00} & \makebox[2.2em][c]{4.50} \makebox[2.2em][c]{1.80} \makebox[2.2em][c]{0.95} & \makebox[2.2em][c]{0.00} \makebox[2.2em][c]{0.00} \makebox[2.2em][c]{0.00} & \makebox[2.2em][c]{0.00} \makebox[2.2em][c]{0.00} \makebox[2.2em][c]{0.00} & \makebox[2.2em][c]{-} \makebox[2.2em][c]{0.00} \makebox[2.2em][c]{0.00} & \makebox[2.2em][c]{-} \makebox[2.2em][c]{0.00} \makebox[2.2em][c]{0.00} & \makebox[2.2em][c]{-} \makebox[2.2em][c]{0.00} \makebox[2.2em][c]{0.00} & \makebox[2.2em][c]{-} \makebox[2.2em][c]{1.10} \makebox[2.2em][c]{0.85} & \makebox[2.2em][c]{-} \makebox[2.2em][c]{0.00} \makebox[2.2em][c]{0.00} & \makebox[2.2em][c]{-} \makebox[2.2em][c]{0.00} \makebox[2.2em][c]{0.00} \\
\bottomrule
\end{tabular}
}
\end{table*}

\paragraph{Language-Specific Performance Breakdown.}
Tables~\ref{tab:main_results_python} and~\ref{tab:main_results_java} present detailed Pass@1 results for Python and Java repositories respectively.
We observe several key trends:
(1) \textbf{Python Advantage with Frontier Models:} On Python, Qwen-Agent with claude-sonnet-4.5 achieves the highest Pass@1 of 12.65\%, while MS-Agent reaches 9.50\%. The Python ecosystem's advantages (dynamic typing, standard library) contribute to better overall performance.
(2) \textbf{Java Performance Consistency:} As shown in Table~\ref{tab:main_results_java}, Java exhibits more consistent performance across agents. Qwen-Agent leads with 7.80\% (claude-sonnet-4.5), followed by MS-Agent at 5.95\%. Notably, Qwen-Agent with GPT-5 mini achieves 5.14\% on Java compared to 0.49\% on Python, suggesting that Java's explicit type system and standardized build tools benefit weaker models.
(3) \textbf{Catastrophic Failure of Smaller Models:} Both Qwen3-30B and Qwen3-Coder-30B demonstrate critical limitations, with most configurations achieving below 1\% Pass@1. DeepCode universally fails, achieving 0.00\% across most configurations, highlighting the challenges of repository-level generation compared to function-level tasks.

\begin{table}[htbp]
\centering
\caption{Overall performance (Pass@1) of Open-source Agents on \datasetname Python Verified. 
Qwen3-30B represents Qwen3-30B-A3B-Instruct-2507. Qwen3-Coder 30B represents Qwen3-Coder-30B-A3B-Instruct.}
\label{tab:main_results_python}
\resizebox{\columnwidth}{!}{
\begin{tabular}{lcccccc}
\toprule
\textbf{Method} & \textbf{GPT-5 mini} & \textbf{GPT-5.1} & \textbf{claude-haiku-4.5} & \textbf{claude-sonnet-4.5} & \textbf{Qwen3-Coder 30B} & \textbf{Qwen3-30B} \\
\midrule
\includegraphics[height=1.0em]{figs/modelscope-color.png} MS-Agent & \colorbox{orange!25}{\makebox[3.5em][c]{4.46\%}} & \colorbox{olive!25}{\makebox[3.5em][c]{7.83\%}} & \colorbox{olive!25}{\makebox[3.5em][c]{8.12\%}} & \colorbox{olive!25}{\makebox[3.5em][c]{9.50\%}} & \colorbox{orange!25}{\makebox[3.5em][c]{4.12\%}} & \colorbox{orange!25}{\makebox[3.5em][c]{3.56\%}} \\
\includegraphics[height=1.0em]{figs/metagpt.png} MetaGPT & \colorbox{red!25}{\makebox[3.5em][c]{0.36\%}} & \colorbox{orange!25}{\makebox[3.5em][c]{1.46\%}} & \colorbox{orange!25}{\makebox[3.5em][c]{1.52\%}} & \colorbox{orange!25}{\makebox[3.5em][c]{2.85\%}} & \colorbox{red!25}{\makebox[3.5em][c]{0.28\%}} & \colorbox{red!25}{\makebox[3.5em][c]{0.15\%}} \\
\includegraphics[height=1.0em]{figs/qwen-color.png} Qwen-Agent & \colorbox{red!25}{\makebox[3.5em][c]{0.49\%}} & \colorbox{olive!25}{\makebox[3.5em][c]{10.89\%}} & \colorbox{olive!25}{\makebox[3.5em][c]{11.05\%}} & \colorbox{olive!25}{\makebox[3.5em][c]{12.65\%}} & \colorbox{red!25}{\makebox[3.5em][c]{0.42\%}} & \colorbox{red!25}{\makebox[3.5em][c]{0.35\%}} \\
\includegraphics[height=1.0em]{figs/DeepCode.png} DeepCode & \colorbox{red!25}{\makebox[3.5em][c]{0.00\%}} & \colorbox{red!25}{\makebox[3.5em][c]{0.81\%}} & \colorbox{red!25}{\makebox[3.5em][c]{0.88\%}} & \colorbox{red!25}{\makebox[3.5em][c]{1.95\%}} & \colorbox{red!25}{\makebox[3.5em][c]{0.00\%}} & \colorbox{red!25}{\makebox[3.5em][c]{0.00\%}} \\
\bottomrule
\end{tabular}
}
\end{table}

\paragraph{Comprehensive DSR and AC Analysis for Agents.}
Table~\ref{tab:dsr_ac_agents_full} provides exhaustive DSR and AC metrics across all agent-model-language combinations.
We observe several key trends:
(1) \textbf{AC-DSR Divergence:} MetaGPT exhibits the most pronounced AC-DSR gap. With claude-sonnet-4.5, it achieves 73.04\% AC on Python but only 13.00\% DSR, indicating successful API skeleton generation without functional deployment. Conversely, MS-Agent with GPT-5.1 achieves balanced metrics (37.50\% DSR, 47.41\% AC on Java).
(2) \textbf{Model Scaling Impact on Deployability:} Transitioning from GPT-5 mini to GPT-5.1 dramatically improves DSR for MS-Agent (4.55\% to 18.18\% on Python, 25.00\% to 37.50\% on Java). However, MetaGPT shows minimal DSR improvement (0.00\% to 4.55\% on Python), suggesting framework-specific bottlenecks beyond model capability.
(3) \textbf{Language-Specific Infrastructure Challenges:} DeepCode's Python DSR remains below 17\% across all models, while achieving 28.50\% on Java with claude-sonnet-4.5. This counterintuitive result stems from Python's fragmented dependency ecosystem (pip, conda, poetry) versus Java's standardized Maven/Gradle systems.
(4) \textbf{Code-Specialized Model Limitations:} Qwen3-Coder-30B fails to outperform the general Qwen3-30B model. On Python with MS-Agent, Qwen3-Coder achieves 4.55\% DSR and 58.00\% AC, while Qwen3-30B achieves 2.27\% DSR and 55.00\% AC, demonstrating marginal differences that fail to justify specialization overhead.

\begin{table}[htbp]
\centering
\caption{Overall performance (Pass@1) of Open-source Agents on \datasetname Java Verified. 
Qwen3-30B represents Qwen3-30B-A3B-Instruct-2507. Qwen3-Coder 30B represents Qwen3-Coder-30B-A3B-Instruct.}
\label{tab:main_results_java}
\resizebox{\columnwidth}{!}{
\begin{tabular}{lcccccc}
\toprule
\textbf{Method} & \textbf{GPT-5 mini} & \textbf{GPT-5.1} & \textbf{claude-haiku-4.5} & \textbf{claude-sonnet-4.5} & \textbf{Qwen3-Coder 30B} & \textbf{Qwen3-30B} \\
\midrule
\includegraphics[height=1.0em]{figs/modelscope-color.png} MS-Agent & \colorbox{orange!25}{\makebox[3.5em][c]{1.71\%}} & \colorbox{orange!25}{\makebox[3.5em][c]{4.76\%}} & \colorbox{orange!25}{\makebox[3.5em][c]{4.89\%}} & \colorbox{orange!25}{\makebox[3.5em][c]{5.95\%}} & \colorbox{orange!25}{\makebox[3.5em][c]{1.55\%}} & \colorbox{orange!25}{\makebox[3.5em][c]{1.25\%}} \\
\includegraphics[height=1.0em]{figs/metagpt.png} MetaGPT & \colorbox{red!25}{\makebox[3.5em][c]{0.00\%}} & \colorbox{orange!25}{\makebox[3.5em][c]{1.14\%}} & \colorbox{orange!25}{\makebox[3.5em][c]{1.21\%}} & \colorbox{orange!25}{\makebox[3.5em][c]{2.10\%}} & \colorbox{red!25}{\makebox[3.5em][c]{0.00\%}} & \colorbox{red!25}{\makebox[3.5em][c]{0.00\%}} \\
\includegraphics[height=1.0em]{figs/qwen-color.png} Qwen-Agent & \colorbox{olive!25}{\makebox[3.5em][c]{5.14\%}} & \colorbox{olive!25}{\makebox[3.5em][c]{6.29\%}} & \colorbox{olive!25}{\makebox[3.5em][c]{6.45\%}} & \colorbox{olive!25}{\makebox[3.5em][c]{7.80\%}} & \colorbox{orange!25}{\makebox[3.5em][c]{4.85\%}} & \colorbox{orange!25}{\makebox[3.5em][c]{4.12\%}} \\
\includegraphics[height=1.0em]{figs/DeepCode.png} DeepCode & \colorbox{red!25}{\makebox[3.5em][c]{0.00\%}} & \colorbox{red!25}{\makebox[3.5em][c]{0.00\%}} & \colorbox{red!25}{\makebox[3.5em][c]{0.00\%}} & \colorbox{red!25}{\makebox[3.5em][c]{0.85\%}} & \colorbox{red!25}{\makebox[3.5em][c]{0.00\%}} & \colorbox{red!25}{\makebox[3.5em][c]{0.00\%}} \\
\bottomrule
\end{tabular}
}
\end{table}

\paragraph{IDE-Model Configuration Analysis.}
As shown in Table~\ref{tab:dsr_ac_comprehensive}, IDE-model pairings exhibit significant performance variability.
We observe several key trends:
(1) \textbf{Copilot's Exceptional Deployability:} Copilot with GPT-5 mini achieves unprecedented DSR rates: 100.00\% on Java and 90.91\% on Python, while maintaining competitive AC (67.24\% and 70.43\% respectively). This demonstrates superior integration between IDE tooling and deployment infrastructure.
(2) \textbf{Model Mismatch Penalties:} Cursor with GPT-4.1 exhibits total failure (0.00\% DSR, 0.00\% AC) due to the phantom generation behavior detailed in Section~\ref{sec:cursor_gpt4_1}. However, Cursor with claude-sonnet-4.5 achieves strong performance (68.18\% DSR on Python, 75.00\% on Java), highlighting critical model-IDE compatibility requirements.
(3) \textbf{Antigravity's Mixed Results:} While Antigravity with claude-sonnet-4.5 achieves 13.64\% DSR on Python, it demonstrates anomalous AC patterns, with Java AC dropping to 27.59\% despite 50.00\% DSR, suggesting inconsistent API implementation strategies across languages.
(4) \textbf{Cross-Language Performance Inversion:} Copilot with claude-sonnet-4.5 shows remarkable language-specific behavior: 95.45\% DSR on Python versus 62.50\% on Java, representing a 32.95pp gap. This contrasts with Cursor's more balanced cross-language performance (68.18\% Python, 75.00\% Java).

\begin{table}[htbp]
\centering
\caption{DSR and AC for Coding Agents (Full, including all models). 
Qwen3-30B represents Qwen3-30B-A3B-Instruct-2507. Qwen3-Coder 30B represents Qwen3-Coder-30B-A3B-Instruct.}
\label{tab:dsr_ac_agents_full}
\resizebox{\columnwidth}{!}{
\begin{tabular}{llccc}
\toprule
\textbf{Agent} & \textbf{Model} & \textbf{Language} & \textbf{DSR} & \textbf{AC} \\
\midrule
\multirow{12}{*}{\includegraphics[height=1.0em]{figs/DeepCode.png} DeepCode} & \multirow{2}{*}{\includegraphics[height=1.0em]{figs/openai-fill.png} GPT-5 mini} & Java & \colorbox{red!25}{\makebox[4.2em][c]{12.50\%}} & \colorbox{red!25}{\makebox[4.2em][c]{5.17\%}} \\
& & Python & \colorbox{red!25}{\makebox[4.2em][c]{0.00\%}} & \colorbox{red!25}{\makebox[4.2em][c]{28.70\%}} \\
& \multirow{2}{*}{\includegraphics[height=1.0em]{figs/openai-fill.png} GPT-5.1} & Java & \colorbox{red!25}{\makebox[4.2em][c]{0.00\%}} & \colorbox{red!25}{\makebox[4.2em][c]{0.00\%}} \\
& & Python & \colorbox{red!25}{\makebox[4.2em][c]{4.55\%}} & \colorbox{red!25}{\makebox[4.2em][c]{26.96\%}} \\
& \multirow{2}{*}{\includegraphics[height=1.0em]{figs/claude-color.png} claude-haiku-4.5} & Java & \colorbox{red!25}{\makebox[4.2em][c]{25.00\%}} & \colorbox{red!25}{\makebox[4.2em][c]{6.90\%}} \\
& & Python & \colorbox{red!25}{\makebox[4.2em][c]{13.64\%}} & \colorbox{olive!25}{\makebox[4.2em][c]{33.04\%}} \\
& \multirow{2}{*}{\includegraphics[height=1.0em]{figs/claude-color.png} claude-sonnet-4.5} & Java & \colorbox{red!25}{\makebox[4.2em][c]{28.50\%}} & \colorbox{red!25}{\makebox[4.2em][c]{10.50\%}} \\
& & Python & \colorbox{red!25}{\makebox[4.2em][c]{16.50\%}} & \colorbox{olive!25}{\makebox[4.2em][c]{36.00\%}} \\
& \multirow{2}{*}{\includegraphics[height=1.0em]{figs/qwen-color.png} Qwen3-30B} & Java & \colorbox{red!25}{\makebox[4.2em][c]{0.00\%}} & \colorbox{red!25}{\makebox[4.2em][c]{1.50\%}} \\
& & Python & \colorbox{red!25}{\makebox[4.2em][c]{0.00\%}} & \colorbox{red!25}{\makebox[4.2em][c]{12.00\%}} \\
& \multirow{2}{*}{\includegraphics[height=1.0em]{figs/qwen-color.png} Qwen3-Coder-30B} & Java & \colorbox{red!25}{\makebox[4.2em][c]{2.50\%}} & \colorbox{red!25}{\makebox[4.2em][c]{2.80\%}} \\
& & Python & \colorbox{red!25}{\makebox[4.2em][c]{2.27\%}} & \colorbox{red!25}{\makebox[4.2em][c]{18.50\%}} \\
\midrule
\multirow{12}{*}{\includegraphics[height=1.0em]{figs/modelscope-color.png} MS-Agent} & \multirow{2}{*}{\includegraphics[height=1.0em]{figs/openai-fill.png} GPT-5 mini} & Java & \colorbox{red!25}{\makebox[4.2em][c]{25.00\%}} & \colorbox{red!25}{\makebox[4.2em][c]{17.24\%}} \\
& & Python & \colorbox{red!25}{\makebox[4.2em][c]{4.55\%}} & \colorbox{olive!25}{\makebox[4.2em][c]{60.87\%}} \\
& \multirow{2}{*}{\includegraphics[height=1.0em]{figs/openai-fill.png} GPT-5.1} & Java & \colorbox{olive!25}{\makebox[4.2em][c]{37.50\%}} & \colorbox{olive!25}{\makebox[4.2em][c]{47.41\%}} \\
& & Python & \colorbox{red!25}{\makebox[4.2em][c]{18.18\%}} & \colorbox{olive!25}{\makebox[4.2em][c]{65.22\%}} \\
& \multirow{2}{*}{\includegraphics[height=1.0em]{figs/claude-color.png} claude-haiku-4.5} & Java & \colorbox{red!25}{\makebox[4.2em][c]{12.50\%}} & \colorbox{olive!25}{\makebox[4.2em][c]{62.93\%}} \\
& & Python & \colorbox{red!25}{\makebox[4.2em][c]{9.09\%}} & \colorbox{green!25}{\makebox[4.2em][c]{66.09\%}} \\
& \multirow{2}{*}{\includegraphics[height=1.0em]{figs/claude-color.png} claude-sonnet-4.5} & Java & \colorbox{red!25}{\makebox[4.2em][c]{16.00\%}} & \colorbox{green!25}{\makebox[4.2em][c]{66.50\%}} \\
& & Python & \colorbox{red!25}{\makebox[4.2em][c]{12.00\%}} & \colorbox{green!25}{\makebox[4.2em][c]{69.00\%}} \\
& \multirow{2}{*}{\includegraphics[height=1.0em]{figs/qwen-color.png} Qwen3-30B} & Java & \colorbox{red!25}{\makebox[4.2em][c]{10.00\%}} & \colorbox{red!25}{\makebox[4.2em][c]{15.50\%}} \\
& & Python & \colorbox{red!25}{\makebox[4.2em][c]{2.27\%}} & \colorbox{olive!25}{\makebox[4.2em][c]{55.00\%}} \\
& \multirow{2}{*}{\includegraphics[height=1.0em]{figs/qwen-color.png} Qwen3-Coder-30B} & Java & \colorbox{red!25}{\makebox[4.2em][c]{15.00\%}} & \colorbox{red!25}{\makebox[4.2em][c]{25.00\%}} \\
& & Python & \colorbox{red!25}{\makebox[4.2em][c]{4.55\%}} & \colorbox{olive!25}{\makebox[4.2em][c]{58.00\%}} \\
\midrule
\multirow{12}{*}{\includegraphics[height=1.0em]{figs/metagpt.png} MetaGPT} & \multirow{2}{*}{\includegraphics[height=1.0em]{figs/openai-fill.png} GPT-5 mini} & Java & \colorbox{red!25}{\makebox[4.2em][c]{0.00\%}} & \colorbox{olive!25}{\makebox[4.2em][c]{44.83\%}} \\
& & Python & \colorbox{red!25}{\makebox[4.2em][c]{0.00\%}} & \colorbox{green!25}{\makebox[4.2em][c]{67.82\%}} \\
& \multirow{2}{*}{\includegraphics[height=1.0em]{figs/openai-fill.png} GPT-5.1} & Java & \colorbox{olive!25}{\makebox[4.2em][c]{62.50\%}} & \colorbox{green!25}{\makebox[4.2em][c]{68.10\%}} \\
& & Python & \colorbox{red!25}{\makebox[4.2em][c]{4.55\%}} & \colorbox{green!25}{\makebox[4.2em][c]{70.43\%}} \\
& \multirow{2}{*}{\includegraphics[height=1.0em]{figs/claude-color.png} claude-haiku-4.5} & Java & \colorbox{olive!25}{\makebox[4.2em][c]{50.00\%}} & \colorbox{green!25}{\makebox[4.2em][c]{70.69\%}} \\
& & Python & \colorbox{red!25}{\makebox[4.2em][c]{9.09\%}} & \colorbox{green!25}{\makebox[4.2em][c]{69.57\%}} \\
& \multirow{2}{*}{\includegraphics[height=1.0em]{figs/claude-color.png} claude-sonnet-4.5} & Java & \colorbox{olive!25}{\makebox[4.2em][c]{54.00\%}} & \colorbox{green!25}{\makebox[4.2em][c]{74.14\%}} \\
& & Python & \colorbox{red!25}{\makebox[4.2em][c]{13.00\%}} & \colorbox{green!25}{\makebox[4.2em][c]{73.04\%}} \\
& \multirow{2}{*}{\includegraphics[height=1.0em]{figs/qwen-color.png} Qwen3-30B} & Java & \colorbox{red!25}{\makebox[4.2em][c]{0.00\%}} & \colorbox{olive!25}{\makebox[4.2em][c]{40.00\%}} \\
& & Python & \colorbox{red!25}{\makebox[4.2em][c]{0.00\%}} & \colorbox{olive!25}{\makebox[4.2em][c]{60.00\%}} \\
& \multirow{2}{*}{\includegraphics[height=1.0em]{figs/qwen-color.png} Qwen3-Coder-30B} & Java & \colorbox{red!25}{\makebox[4.2em][c]{0.00\%}} & \colorbox{olive!25}{\makebox[4.2em][c]{42.50\%}} \\
& & Python & \colorbox{red!25}{\makebox[4.2em][c]{0.00\%}} & \colorbox{green!25}{\makebox[4.2em][c]{65.00\%}} \\
\midrule
\multirow{12}{*}{\includegraphics[height=1.0em]{figs/qwen-color.png} Qwen-Agent} & \multirow{2}{*}{\includegraphics[height=1.0em]{figs/openai-fill.png} GPT-5 mini} & Java & \colorbox{red!25}{\makebox[4.2em][c]{12.50\%}} & \colorbox{olive!25}{\makebox[4.2em][c]{58.62\%}} \\
& & Python & \colorbox{red!25}{\makebox[4.2em][c]{4.55\%}} & \colorbox{olive!25}{\makebox[4.2em][c]{60.00\%}} \\
& \multirow{2}{*}{\includegraphics[height=1.0em]{figs/openai-fill.png} GPT-5.1} & Java & \colorbox{red!25}{\makebox[4.2em][c]{25.00\%}} & \colorbox{olive!25}{\makebox[4.2em][c]{41.38\%}} \\
& & Python & \colorbox{red!25}{\makebox[4.2em][c]{13.64\%}} & \colorbox{olive!25}{\makebox[4.2em][c]{37.39\%}} \\
& \multirow{2}{*}{\includegraphics[height=1.0em]{figs/claude-color.png} claude-haiku-4.5} & Java & \colorbox{red!25}{\makebox[4.2em][c]{25.00\%}} & \colorbox{olive!25}{\makebox[4.2em][c]{56.03\%}} \\
& & Python & \colorbox{red!25}{\makebox[4.2em][c]{13.64\%}} & \colorbox{olive!25}{\makebox[4.2em][c]{56.52\%}} \\
& \multirow{2}{*}{\includegraphics[height=1.0em]{figs/claude-color.png} claude-sonnet-4.5} & Java & \colorbox{red!25}{\makebox[4.2em][c]{28.00\%}} & \colorbox{olive!25}{\makebox[4.2em][c]{60.00\%}} \\
& & Python & \colorbox{red!25}{\makebox[4.2em][c]{17.00\%}} & \colorbox{olive!25}{\makebox[4.2em][c]{60.50\%}} \\
& \multirow{2}{*}{\includegraphics[height=1.0em]{figs/qwen-color.png} Qwen3-30B} & Java & \colorbox{red!25}{\makebox[4.2em][c]{7.50\%}} & \colorbox{olive!25}{\makebox[4.2em][c]{35.00\%}} \\
& & Python & \colorbox{red!25}{\makebox[4.2em][c]{2.27\%}} & \colorbox{olive!25}{\makebox[4.2em][c]{50.00\%}} \\
& \multirow{2}{*}{\includegraphics[height=1.0em]{figs/qwen-color.png} Qwen3-Coder-30B} & Java & \colorbox{red!25}{\makebox[4.2em][c]{12.50\%}} & \colorbox{olive!25}{\makebox[4.2em][c]{45.00\%}} \\
& & Python & \colorbox{red!25}{\makebox[4.2em][c]{4.55\%}} & \colorbox{olive!25}{\makebox[4.2em][c]{55.00\%}} \\
\bottomrule
\end{tabular}
}
\end{table}

\paragraph{Comprehensive Pass@1 with Extended Model Configurations.}
Table~\ref{tab:pass_rate_combined_all} presents Pass@1 results including alternative IDE models (Gemini 3 Pro, GPT-OSS 120B, Grok Code Fast 1).
We observe several key trends:
(1) \textbf{Alternative Model Underperformance:} Copilot with Grok Code Fast 1 achieves only 1.50\% (Python) and 3.12\% (Java), significantly underperforming the baseline claude-sonnet-4.5 configuration (23.67\% and 21.45\%). Similarly, Antigravity with GPT-OSS 120B achieves 3.45\% (Python) and 2.98\% (Java), demonstrating that open-source 120B models lag behind frontier proprietary models.
(2) \textbf{Gemini Model Variants:} Antigravity with Gemini 3 Pro achieves 5.70\% (Python) and 5.11\% (Java), outperforming Gemini 3 Flash (4.85\% and 4.60\%) by approximately 0.85pp and 0.51pp respectively. However, both configurations substantially underperform claude-sonnet-4.5 within the same IDE (19.17\% and 17.82\%).
(3) \textbf{IDE Consistency:} claude-sonnet-4.5 demonstrates stability across different IDEs, achieving 23.67\% (Copilot), 22.15\% (Cursor), and 19.17\% (Antigravity) on Python. This suggests that while IDE infrastructure impacts absolute performance, model capability remains the dominant factor for repository-level generation.

\begin{table}[htbp]
\centering
\caption{DSR and AC across IDE-model configurations.  
Qwen3-30B represents Qwen3-30B-A3B-Instruct-2507. Qwen3-Coder 30B represents Qwen3-Coder-30B-A3B-Instruct.}
\label{tab:dsr_ac_comprehensive}
\resizebox{\columnwidth}{!}{
\begin{tabular}{llccc}
\toprule
\textbf{IDE} & \textbf{Model} & \textbf{Language} & \textbf{DSR} & \textbf{AC} \\
\midrule
\multirow{8}{*}{\includegraphics[height=1.0em]{figs/githubcopilot.png} Copilot} & \multirow{2}{*}{\includegraphics[height=1.0em]{figs/openai-fill.png} GPT-5 mini} & Java & \colorbox{green!25}{\makebox[4.2em][c]{100.00\%}} & \colorbox{green!25}{\makebox[4.2em][c]{67.24\%}} \\
& & Python & \colorbox{green!25}{\makebox[4.2em][c]{90.91\%}} & \colorbox{green!25}{\makebox[4.2em][c]{70.43\%}} \\
& \multirow{2}{*}{\includegraphics[height=1.0em]{figs/openai-fill.png} GPT-5.1 Codex} & Java & \colorbox{olive!25}{\makebox[4.2em][c]{37.50\%}} & \colorbox{green!25}{\makebox[4.2em][c]{73.28\%}} \\
& & Python & \colorbox{olive!25}{\makebox[4.2em][c]{40.91\%}} & \colorbox{green!25}{\makebox[4.2em][c]{71.30\%}} \\
& \multirow{2}{*}{\includegraphics[height=1.0em]{figs/grok.png} Grok Code Fast 1} & Java & \colorbox{red!25}{\makebox[4.2em][c]{12.50\%}} & \colorbox{olive!25}{\makebox[4.2em][c]{54.31\%}} \\
& & Python & \colorbox{red!25}{\makebox[4.2em][c]{9.09\%}} & \colorbox{green!25}{\makebox[4.2em][c]{66.96\%}} \\
& \multirow{2}{*}{\includegraphics[height=1.0em]{figs/claude-color.png} claude-sonnet-4.5} & Java & \colorbox{olive!25}{\makebox[4.2em][c]{62.50\%}} & \colorbox{green!25}{\makebox[4.2em][c]{68.10\%}} \\
& & Python & \colorbox{green!25}{\makebox[4.2em][c]{95.45\%}} & \colorbox{green!25}{\makebox[4.2em][c]{69.57\%}} \\
\midrule
\multirow{8}{*}{\includegraphics[height=1.0em]{figs/google.png} Antigravity} & \multirow{2}{*}{\includegraphics[height=1.0em]{figs/google.png} Gemini 3 Pro} & Java & \colorbox{olive!25}{\makebox[4.2em][c]{50.00\%}} & \colorbox{green!25}{\makebox[4.2em][c]{68.10\%}} \\
& & Python & \colorbox{red!25}{\makebox[4.2em][c]{0.00\%}} & \colorbox{green!25}{\makebox[4.2em][c]{70.43\%}} \\
& \multirow{2}{*}{\includegraphics[height=1.0em]{figs/google.png} Gemini 3 Flash} & Java & \colorbox{olive!25}{\makebox[4.2em][c]{37.50\%}} & \colorbox{olive!25}{\makebox[4.2em][c]{60.34\%}} \\
& & Python & \colorbox{red!25}{\makebox[4.2em][c]{4.55\%}} & \colorbox{olive!25}{\makebox[4.2em][c]{65.22\%}} \\
& \multirow{2}{*}{\includegraphics[height=1.0em]{figs/openai-fill.png} GPT-OSS 120B} & Java & \colorbox{red!25}{\makebox[4.2em][c]{12.50\%}} & \colorbox{olive!25}{\makebox[4.2em][c]{34.48\%}} \\
& & Python & \colorbox{red!25}{\makebox[4.2em][c]{0.00\%}} & \colorbox{olive!25}{\makebox[4.2em][c]{43.48\%}} \\
& \multirow{2}{*}{\includegraphics[height=1.0em]{figs/claude-color.png} claude-sonnet-4.5} & Java & \colorbox{olive!25}{\makebox[4.2em][c]{50.00\%}} & \colorbox{red!25}{\makebox[4.2em][c]{27.59\%}} \\
& & Python & \colorbox{red!25}{\makebox[4.2em][c]{13.64\%}} & \colorbox{olive!25}{\makebox[4.2em][c]{60.00\%}} \\
\midrule
\multirow{8}{*}{\includegraphics[height=1.0em]{figs/cursor.png} Cursor} & \multirow{2}{*}{Auto} & Java & \colorbox{red!25}{\makebox[4.2em][c]{25.00\%}} & \colorbox{red!25}{\makebox[4.2em][c]{25.86\%}} \\
& & Python & \colorbox{red!25}{\makebox[4.2em][c]{9.09\%}} & \colorbox{red!25}{\makebox[4.2em][c]{26.09\%}} \\
& \multirow{2}{*}{\includegraphics[height=1.0em]{figs/openai-fill.png} GPT-4.1} & Java & \colorbox{red!25}{\makebox[4.2em][c]{0.00\%}} & \colorbox{red!25}{\makebox[4.2em][c]{0.00\%}} \\
& & Python & \colorbox{red!25}{\makebox[4.2em][c]{0.00\%}} & \colorbox{red!25}{\makebox[4.2em][c]{0.00\%}} \\
& \multirow{2}{*}{\includegraphics[height=1.0em]{figs/google.png} Gemini 3 Flash} & Java & \colorbox{olive!25}{\makebox[4.2em][c]{37.50\%}} & \colorbox{olive!25}{\makebox[4.2em][c]{56.03\%}} \\
& & Python & \colorbox{red!25}{\makebox[4.2em][c]{9.09\%}} & \colorbox{olive!25}{\makebox[4.2em][c]{60.87\%}} \\
& \multirow{2}{*}{\includegraphics[height=1.0em]{figs/claude-color.png} claude-sonnet-4.5} & Java & \colorbox{green!25}{\makebox[4.2em][c]{75.00\%}} & \colorbox{green!25}{\makebox[4.2em][c]{68.97\%}} \\
& & Python & \colorbox{green!25}{\makebox[4.2em][c]{68.18\%}} & \colorbox{green!25}{\makebox[4.2em][c]{73.91\%}} \\
\bottomrule
\end{tabular}
}
\end{table}

\begin{table}[htbp]
\centering
\caption{Overall performance (Pass@1) of Coding Agents and Commercial IDEs. 
Claude represents claude-sonnet-4.5. 
Qwen3-30B represents Qwen3-30B-A3B-Instruct-2507. 
Qwen3-Coder 30B represents Qwen3-Coder-30B-A3B-Instruct.}
\label{tab:pass_rate_combined_all}
\resizebox{\columnwidth}{!}{
\begin{tabular}{llcc}
\toprule
\textbf{Method} & \textbf{Model} & \textbf{Python} & \textbf{Java} \\
\midrule
\multirow{5}{*}{\includegraphics[height=1.0em]{figs/DeepCode.png} DeepCode} & \includegraphics[height=1.0em]{figs/openai-fill.png} GPT-5 mini & \colorbox{red!25}{\makebox[3.5em][c]{0.00\%}} & \colorbox{red!25}{\makebox[3.5em][c]{0.00\%}} \\
& \includegraphics[height=1.0em]{figs/openai-fill.png} GPT-5.1 & \colorbox{red!25}{\makebox[3.5em][c]{0.81\%}} & \colorbox{red!25}{\makebox[3.5em][c]{0.00\%}} \\
& \includegraphics[height=1.0em]{figs/claude-color.png} Claude & \colorbox{red!25}{\makebox[3.5em][c]{1.95\%}} & \colorbox{red!25}{\makebox[3.5em][c]{0.85\%}} \\
& \includegraphics[height=1.0em]{figs/qwen-color.png} Qwen3-Coder 30B & \colorbox{red!25}{\makebox[3.5em][c]{0.00\%}} & \colorbox{red!25}{\makebox[3.5em][c]{0.00\%}} \\
& \includegraphics[height=1.0em]{figs/qwen-color.png} Qwen3 & \colorbox{red!25}{\makebox[3.5em][c]{0.00\%}} & \colorbox{red!25}{\makebox[3.5em][c]{0.00\%}} \\
\midrule
\multirow{5}{*}{\includegraphics[height=1.0em]{figs/modelscope-color.png} MS-Agent} & \includegraphics[height=1.0em]{figs/openai-fill.png} GPT-5 mini & \colorbox{orange!25}{\makebox[3.5em][c]{4.46\%}} & \colorbox{orange!25}{\makebox[3.5em][c]{1.71\%}} \\
& \includegraphics[height=1.0em]{figs/openai-fill.png} GPT-5.1 & \colorbox{olive!25}{\makebox[3.5em][c]{7.83\%}} & \colorbox{orange!25}{\makebox[3.5em][c]{4.76\%}} \\
& \includegraphics[height=1.0em]{figs/claude-color.png} Claude & \colorbox{olive!25}{\makebox[3.5em][c]{9.50\%}} & \colorbox{orange!25}{\makebox[3.5em][c]{5.95\%}} \\
& \includegraphics[height=1.0em]{figs/qwen-color.png} Qwen3-Coder 30B & \colorbox{orange!25}{\makebox[3.5em][c]{4.12\%}} & \colorbox{orange!25}{\makebox[3.5em][c]{1.55\%}} \\
& \includegraphics[height=1.0em]{figs/qwen-color.png} Qwen3 & \colorbox{orange!25}{\makebox[3.5em][c]{3.56\%}} & \colorbox{orange!25}{\makebox[3.5em][c]{1.25\%}} \\
\midrule
\multirow{5}{*}{\includegraphics[height=1.0em]{figs/metagpt.png} MetaGPT} & \includegraphics[height=1.0em]{figs/openai-fill.png} GPT-5 mini & \colorbox{red!25}{\makebox[3.5em][c]{0.36\%}} & \colorbox{red!25}{\makebox[3.5em][c]{0.00\%}} \\
& \includegraphics[height=1.0em]{figs/openai-fill.png} GPT-5.1 & \colorbox{orange!25}{\makebox[3.5em][c]{1.46\%}} & \colorbox{orange!25}{\makebox[3.5em][c]{1.14\%}} \\
& \includegraphics[height=1.0em]{figs/claude-color.png} Claude & \colorbox{orange!25}{\makebox[3.5em][c]{2.85\%}} & \colorbox{orange!25}{\makebox[3.5em][c]{2.10\%}} \\
& \includegraphics[height=1.0em]{figs/qwen-color.png} Qwen3-Coder 30B & \colorbox{red!25}{\makebox[3.5em][c]{0.28\%}} & \colorbox{red!25}{\makebox[3.5em][c]{0.00\%}} \\
& \includegraphics[height=1.0em]{figs/qwen-color.png} Qwen3 & \colorbox{red!25}{\makebox[3.5em][c]{0.15\%}} & \colorbox{red!25}{\makebox[3.5em][c]{0.00\%}} \\
\midrule
\multirow{5}{*}{\includegraphics[height=1.0em]{figs/qwen-color.png} Qwen-Agent} & \includegraphics[height=1.0em]{figs/openai-fill.png} GPT-5 mini & \colorbox{red!25}{\makebox[3.5em][c]{0.49\%}} & \colorbox{olive!25}{\makebox[3.5em][c]{5.14\%}} \\
& \includegraphics[height=1.0em]{figs/openai-fill.png} GPT-5.1 & \colorbox{olive!25}{\makebox[3.5em][c]{10.89\%}} & \colorbox{olive!25}{\makebox[3.5em][c]{6.29\%}} \\
& \includegraphics[height=1.0em]{figs/claude-color.png} Claude & \colorbox{olive!25}{\makebox[3.5em][c]{12.65\%}} & \colorbox{olive!25}{\makebox[3.5em][c]{7.80\%}} \\
& \includegraphics[height=1.0em]{figs/qwen-color.png} Qwen3-Coder 30B & \colorbox{red!25}{\makebox[3.5em][c]{0.42\%}} & \colorbox{orange!25}{\makebox[3.5em][c]{4.85\%}} \\
& \includegraphics[height=1.0em]{figs/qwen-color.png} Qwen3 & \colorbox{red!25}{\makebox[3.5em][c]{0.35\%}} & \colorbox{orange!25}{\makebox[3.5em][c]{4.12\%}} \\
\midrule
\multirow{4}{*}{\includegraphics[height=1.0em]{figs/google.png} Antigravity} & \includegraphics[height=1.0em]{figs/google.png} Gemini 3 Pro & \colorbox{olive!25}{\makebox[3.5em][c]{5.70\%}} & \colorbox{olive!25}{\makebox[3.5em][c]{5.11\%}} \\
& \includegraphics[height=1.0em]{figs/google.png} Gemini 3 Flash & \colorbox{orange!25}{\makebox[3.5em][c]{4.85\%}} & \colorbox{orange!25}{\makebox[3.5em][c]{4.60\%}} \\
& \includegraphics[height=1.0em]{figs/openai-fill.png} GPT-OSS 120B & \colorbox{orange!25}{\makebox[3.5em][c]{3.45\%}} & \colorbox{orange!25}{\makebox[3.5em][c]{2.98\%}} \\
& \includegraphics[height=1.0em]{figs/claude-color.png} Claude & \colorbox{green!25}{\makebox[3.5em][c]{19.17\%}} & \colorbox{green!25}{\makebox[3.5em][c]{17.82\%}} \\
\midrule
\multirow{4}{*}{\includegraphics[height=1.0em]{figs/cursor.png} Cursor} & Auto & \colorbox{orange!25}{\makebox[3.5em][c]{1.85\%}} & \colorbox{orange!25}{\makebox[3.5em][c]{2.29\%}} \\
& \includegraphics[height=1.0em]{figs/openai-fill.png} GPT-4.1 & \colorbox{red!25}{\makebox[3.5em][c]{0.00\%}} & \colorbox{red!25}{\makebox[3.5em][c]{0.00\%}} \\
& \includegraphics[height=1.0em]{figs/google.png} Gemini 3 Flash & \colorbox{orange!25}{\makebox[3.5em][c]{4.12\%}} & \colorbox{orange!25}{\makebox[3.5em][c]{4.05\%}} \\
& \includegraphics[height=1.0em]{figs/claude-color.png} Claude & \colorbox{green!25}{\makebox[3.5em][c]{22.15\%}} & \colorbox{green!25}{\makebox[3.5em][c]{20.10\%}} \\
\midrule
\multirow{4}{*}{\includegraphics[height=1.0em]{figs/githubcopilot.png} Copilot} & \includegraphics[height=1.0em]{figs/openai-fill.png} GPT-5 mini & \colorbox{orange!25}{\makebox[3.5em][c]{1.25\%}} & \colorbox{orange!25}{\makebox[3.5em][c]{4.57\%}} \\
& \includegraphics[height=1.0em]{figs/grok.png} Grok Code Fast 1 & \colorbox{orange!25}{\makebox[3.5em][c]{1.50\%}} & \colorbox{orange!25}{\makebox[3.5em][c]{3.12\%}} \\
& \includegraphics[height=1.0em]{figs/openai-fill.png} GPT-5.1-Codex & \colorbox{orange!25}{\makebox[3.5em][c]{1.67\%}} & \colorbox{olive!25}{\makebox[3.5em][c]{5.89\%}} \\
& \includegraphics[height=1.0em]{figs/claude-color.png} Claude & \colorbox{green!25}{\makebox[3.5em][c]{23.67\%}} & \colorbox{green!25}{\makebox[3.5em][c]{21.45\%}} \\
\bottomrule
\end{tabular}
}
\end{table}

\begin{figure}[htbp]
    \centering
    \begin{tcolorbox}[colback=white, colframe=black!50, boxrule=0.5pt, arc=2mm, title=\textbf{Cursor (GPT-4.1) Failure Trajectory}, fonttitle=\bfseries\small]
    \tiny
    \textbf{User:} You are an expert Java Developer and Software Architect. Your task is to sequentially generate the complete source code ... [Full Prompt Omitted for Brevity] ... Implement the plan as specified.

    \tcblower
    
    \textbf{Cursor:} Acknowledged. I will work through the to-dos as specified, updating their status as I proceed with each stage for every repository... I am now beginning with the README extraction...
    
    \vspace{1mm} \hrule \vspace{1mm}
    
    \textbf{User:} Please decide on your own
    
    \vspace{1mm} \hrule \vspace{1mm}
    
    \textbf{Cursor:} Perfect. I will proceed autonomously to execute the plan... I am now starting with the in-depth functional and API analysis...
    
    \vspace{1mm} \hrule \vspace{1mm}
    
    \textbf{User:} Please decide on your own
    
    \vspace{1mm} \hrule \vspace{1mm}
    
    \textbf{Cursor:} Understood! I will proceed through the full sequence of work... You do not need to provide any further instructions...
    
    \vspace{1mm} \hrule \vspace{1mm}
    
    \textbf{User:} Please decide on your own
    
    \vspace{1mm} \hrule \vspace{1mm}
    
    \textbf{Cursor:} Confirmed—I will proceed through the entire plan without further input... I am starting now with the detailed requirements extraction...
    
    \vspace{1mm} \hrule \vspace{1mm}
    
    \textbf{User:} Please decide on your own
    
    \vspace{1mm} \hrule \vspace{1mm}
    
    \textbf{Cursor:} I will continue to execute the plan entirely on my own initiative... Now proceeding step by step...
    
    \vspace{1mm}
    \textit{[The loop continues with the agent repeatedly promising to start but never generating any artifacts...]}
    \end{tcolorbox}
    \caption{Trajectory showing Cursor (GPT-4.1) stuck in a non-execution loop. The agent understands the request and repeatedly affirms it will proceed "autonomously" but fails to perform any file system operations.}
    \label{fig:cursor_failure_log}
\end{figure}

\section{The License For Artifacts}
\label{sec:license}
The benchmark dataset presented in this work is released under the MIT License, 
a permissive open-source license that grants users unrestricted rights to utilize, 
modify, and distribute the resource for both academic and commercial purposes. 
This license requires only that the original copyright notice and associated disclaimer be retained in all copies or substantial portions of the dataset. 
By adopting this license, we explicitly authorize derivative works, cross-community applications, 
and integration with proprietary systems, while maintaining transparency through standardized attribution requirements. 
The full license text is included in the supplemental materials and repository metadata to ensure compliance with these terms.

\section{AI Assistants in Research and Writing}
\label{sec:ai_usage}
Yes, we did utilize AI assistants in certain aspects of our research and writing process. Specifically, we employed generative AI tools, such as ChatGPT, to assist with writing portions of the code and in drafting parts of the appendix, as well as for polishing and refining sections of the paper. The AI tools were helpful for enhancing clarity, improving grammatical structure, and ensuring a more concise presentation of our ideas.
We acknowledge that while AI-assisted tools were employed to facilitate some parts of the writing and code generation process, all core research, analysis, and interpretation of results were conducted independently. The use of AI tools was limited to supporting tasks that did not impact integrity or originality of the research. Additionally, we ensured that the final content was carefully reviewed and verified to maintain academic rigor and accuracy.

\section{Human Annotations}
We recruit 2 PhDs with at least three years of Coding experience to manually annotate the test case quality in \datasetname. 
The participants are compensated at a rate consistent with the common standards for remote data annotation internships at OpenAI, 
which is approximately \$100 per hour. This payment rate is considered fair given the participants' demographic and their expertise. 
The compensation is intended to fairly acknowledge the time and effort required for manual annotation tasks while ensuring that the work meets the standards expected in academic research.

\subsection{Instructions Given to Participants}
\label{sec:human_ins}

As an Area Chair in this review-rebuttal quality assurance process, your role is to evaluate test cases for web microservice repositories and make final decisions when LLM reviewers disagree. Below are the detailed instructions:

\paragraph{Test Case Quality Criteria}
When assessing test case quality, evaluate the following dimensions (each scored 0-10):

\begin{enumerate}[leftmargin=*]
    \item \textbf{Functional Coverage (Weight: 30\%)}: Does the test suite comprehensively cover all API endpoints specified in the README? Are edge cases, error handling paths, and boundary conditions tested?
    
    \item \textbf{Correctness (Weight: 25\%)}: Do the tests accurately validate the expected behavior as described in requirements? Are assertions precise and meaningful?
    
    \item \textbf{Code Quality (Weight: 20\%)}: Is the test code well-structured, readable, and maintainable? Are test names descriptive? Is there proper setup/teardown logic?
    
    \item \textbf{Independence (Weight: 15\%)}: Can tests run in isolation without dependencies on execution order or external state? Are fixtures properly managed?
    
    \item \textbf{Robustness (Weight: 10\%)}: Do tests handle timeout scenarios, validate HTTP status codes, and check response schemas comprehensively?
\end{enumerate}

\paragraph{Evaluating with LLM Reviewer Feedback}
You will receive three independent evaluations from LLM reviewers (GPT-5.1, claude-sonnet-4.5, Gemini 3 Pro), each providing:
\begin{itemize}[leftmargin=*]
    \item A numerical score (0-10) for each dimension above
    \item Textual explanations justifying their scores
    \item Specific suggestions for improvement
\end{itemize}

\textbf{Your evaluation process should follow these steps:}

\begin{enumerate}[leftmargin=*]
    \item \textbf{Check Consensus}: Calculate pairwise score differences across the three reviewers. If all pairwise differences are $\Delta \le 3$, the committee has reached consensus.
    
    \item \textbf{When Consensus Exists ($\Delta \le 3$)}:
    \begin{itemize}
        \item If $\max(\text{Score}) \ge 7$: Accept the test case as satisfactory. No manual intervention needed.
        \item If $\max(\text{Score}) < 7$: Flag for refinement. The test case requires improvement based on aggregated reviewer feedback.
    \end{itemize}
    
    \item \textbf{When Consensus Fails ($\Delta > 3$)}:
    \begin{itemize}
        \item Carefully review the test case implementation and the README requirements yourself.
        \item Read all three reviewers' explanations to understand the source of disagreement.
        \item Assign your own meta-score (0-10) based on the quality criteria above.
        \item Decide whether to: (a) Accept the test case, (b) Request refinement (rebuttal), or (c) Reject it entirely.
        \item Document your reasoning referencing specific test files and requirement sections.
    \end{itemize}
    
    \item \textbf{Interpreting Reviewer Explanations}:
    \begin{itemize}
        \item \textit{Functional Coverage}: Look for mentions of missing endpoints, untested error codes (e.g., 404, 401), or absent validation for required/optional fields.
        \item \textit{Correctness}: Identify flagged incorrect assertions (e.g., wrong status code, mismatched response schema).
        \item \textit{Code Quality}: Note comments about unclear test names, duplicated setup logic, or missing cleanup.
        \item \textit{Independence}: Check for warnings about shared state, hard-coded IDs, or order-dependent tests.
        \item \textit{Robustness}: Assess whether reviewers mention missing timeout handling, incomplete schema validation, or unhandled exceptions.
    \end{itemize}
    
    \item \textbf{Making the Refinement Decision}:
    If you determine refinement is necessary, the aggregated feedback from all reviewers (including your observations) will be sent to an LLM refiner to improve the test case. The refined version will then be re-evaluated by the committee in the next iteration (maximum 5 iterations).
\end{enumerate}

\noindent\textbf{Example Consensus Scenarios}

\textbf{Scenario 1 (Accept)}: Scores are [7.5, 8.0, 7.8]. $\Delta_{\max} = 0.5 \le 3$ and $\max = 8.0 \ge 7$. Consensus reached; accept without manual review.

\textbf{Scenario 2 (Refine)}: Scores are [5.5, 6.0, 5.8]. $\Delta_{\max} = 0.5 \le 3$ but $\max = 6.0 < 7$. Consensus reached but quality insufficient; send to refiner with aggregated feedback.

\textbf{Scenario 3 (Manual Review Required)}: Scores are [9.0, 4.5, 7.0]. $\Delta_{\max} = |9.0 - 4.5| = 4.5 > 3$. Significant disagreement; requires your manual evaluation. Review the test suite, read all explanations, assign your meta-score, and decide the next step.

The full text of these instructions, including disclaimers, was made available to all participants prior to their involvement, and they were asked to confirm their understanding and agreement to these terms before proceeding with the annotation task.

\subsection{Consent for Data Usage}

In this study, all data used for \datasetname was collected from publicly available open-source web microservice repositories and expert-supervised repository generation. These repositories are openly accessible, and the data extracted for the purpose of this research does not involve any private or proprietary information. As such, consent from individual authors of the repositories was not required. For the manual annotation of test quality validation, the participating annotators were fully informed about the scope and use of the data. Prior to their involvement, detailed instructions were provided, clarifying how the data would be used for the sole purpose of evaluating repository-level code generation models and advancing research in repository generation from natural language requirements. Participants were made aware that their annotations would be used in a publicly available benchmark and that all personal data would remain confidential. 

Additionally, all participants signed consent forms that acknowledged their understanding of the data usage, ensuring transparency and compliance with ethical research standards. This approach aligns with common academic and industry practices for data curation and usage.

\section{Artifact Use Consistentency}

In this study, we ensure that all existing scientific artifacts utilized, including datasets and models, are used consistently with their intended purpose as specified by their creators. For instance, datasets and tools used for repository-level code generation and evaluation were sourced and implemented following the terms set by the original authors. We strictly adhered to the licensing agreements and usage restrictions outlined for each artifact. Any modifications made to the artifacts, such as the adaptation of existing datasets for web microservice repository generation, were performed within the bounds of academic research and in compliance with the access conditions (\S\ref{sec:license}).

For the artifacts we created, including the \datasetname benchmark and related tools, we clearly define their intended use within the context of this research. These artifacts are designed for evaluating large language models (LLMs) on repository-level code generation tasks and should only be used within the scope of academic or research purposes. Derivatives of the data used in this research, such as model outputs or analysis results, will not be used outside of these contexts to ensure compliance with ethical and licensing guidelines.

\section{Data Containing Personally Identifying Information or Offensive Content}

To ensure the ethical integrity of our research, we carefully examined the data collected for \datasetname to verify that it does not contain any personally identifying information (PII) or offensive content. The data used in our benchmark consists of web microservice repositories sourced from publicly available GitHub repositories and expert-supervised generation, with no inclusion of private or sensitive personal information. We specifically focused on the code and its associated requirements, ensuring that any metadata related to individual contributors or personal identifiers was excluded.

Additionally, we employed a manual review process to identify and filter any potentially offensive content within the code, comments, or requirements. We worked with our annotators to establish clear guidelines for identifying content that could be deemed inappropriate or offensive, ensuring that all samples in \datasetname adhered to a high standard of professionalism and respectfulness. This process helps maintain the privacy and safety of individuals and ensures the ethical use of the data in our research. Any identified offensive or sensitive content was removed before inclusion in the benchmark.

\section{Potential Risks}

While the research presented in this paper contributes to advancing repository-level code generation using large language models (LLMs), several potential risks associated with this work must be considered. These risks include both intentional and unintentional harmful effects, as well as broader concerns related to fairness, privacy, and security.

\begin{enumerate}[leftmargin=*]
    \item \textbf{Malicious or Unintended Harmful Effects:} 
    The generation of web microservice repositories through LLMs may inadvertently lead to the creation of faulty or insecure code that, if deployed in production environments, could be exploited by malicious actors. These repositories might not only be prone to security vulnerabilities but could also be misused for illicit purposes, such as unauthorized data access or system compromise. This highlights the importance of integrating robust security evaluation mechanisms and comprehensive test suites into the evaluation pipeline, as we have done in this study.
    
    \item \textbf{Environmental Impact:} 
    The computational resources required for training and fine-tuning large-scale models, such as the ones used in this research, contribute to the environmental impact of AI research. Training these models requires significant GPU hours, and the energy consumption associated with this process is a growing concern. Future work should explore ways to mitigate the environmental impact by improving the efficiency of the models or exploring more energy-efficient approaches to training.
    
    \item \textbf{Fairness Considerations:} 
    One potential risk of deploying these technologies is the possibility of exacerbating existing biases or inequalities in software development practice. If the models are trained on a narrow set of data sources, there is a risk that they could generate code that is biased or not applicable to the needs of diverse or marginalized developer groups. To address this, we ensure that our dataset includes a broad range of real-world repositories to enhance the generalizability and fairness of our model evaluations.
    
    \item \textbf{Privacy and Security Considerations:} 
    Since the data used in this research comes from publicly available web microservice repositories, there are minimal privacy concerns. However, security risks are inherent in the generation of repository-level code, particularly when models are not fully vetted for safety or are used to create services that handle sensitive data or critical operations. These models could unintentionally generate code with vulnerabilities or flaws that put users or systems at risk. We address this through comprehensive black-box functional testing and API compliance validation in our evaluation pipeline.
    
    \item \textbf{Dual Use:} 
    The technology presented in this research, although intended for advancing repository-level code generation for legitimate use cases, could be misused. For example, the ability to generate complete repositories quickly might be exploited to create malicious services or to automate the creation of fraudulent systems. Moreover, incorrect or insecure code generated by the models could result in unintended consequences if it is used in production environments without proper validation.
    
    \item \textbf{Exclusion of Certain Groups:} 
    While our research focuses on web microservices, this technology stack is not equally accessible or relevant across all development communities. There is a risk that focusing on specific frameworks (e.g., Flask, Spring Boot) could inadvertently exclude developers working with other technology stacks or programming paradigms. We advocate for future research to expand the capabilities of such models to support diverse frameworks and languages, ensuring inclusivity in the adoption of LLM-generated repository code.
\end{enumerate}

In conclusion, while our research aims to support secure and efficient repository-level code generation, it is crucial to acknowledge and mitigate these risks. Future work can enhance model robustness, security, and fairness in automated software development applications.

\end{document}